%% file: main_final.tex
\documentclass[12pt]{article}
\usepackage{jheppub}

\pdfoutput=1

\usepackage{amsmath,bbm,array,amsfonts,graphicx,wrapfig,lscape,float,mathtools,multirow,longtable}
\usepackage{makecell}
\usepackage[dvipsnames,table]{xcolor}
\usepackage{array}
\usepackage{color}
\usepackage{physics}
\usepackage{subcaption}
\usepackage{tikz,tikz-3dplot}
\usepackage[colorlinks=true]{hyperref}
\newcolumntype{C}[1]{>{\centering\let\newline\\\arraybackslash\hspace{0pt}}m{#1}}

\input{pref}

\newcommand{\req}[1]{(\ref{#1})}
\newcommand{\ie}{{\it i.e.}}
\renewcommand{\t}{\tilde}

\newcommand{\Fcal}{\mathcal F}
\newcommand{\Gcal}{\mathcal G}
\newcommand{\Kcal}{\mathcal K}
\newcommand{\Ncal}{\mathcal N}
\newcommand{\Mcal}{\mathcal M}
\newcommand{\Lcal}{\mathcal L}
\newcommand{\Wcal}{\mathcal W}
\newcommand{\Ocal}{\mathcal O}
\newcommand{\Tcal}{\mathcal T}
\newcommand{\Ccal}{\mathcal C}
\newcommand{\Bcal}{\mathcal B}

\renewcommand{\P}{\mathbb P}
\newcommand{\F}{\mathbb F}
\newcommand{\Z}{\mathbb Z}
\newcommand{\R}{\mathbb R}
\newcommand{\T}{\mathbb T}
\newcommand{\C}{\mathbb C}

\newcommand{\m}{m}
\newcommand{\ep}{\epsilon}
\newcommand{\epar}{\cdot}
\newcommand\topa[2]{\genfrac{}{}{0pt}{2}{\scriptstyle #1}{\scriptstyle #2}}
\newcommand{\corr}[1]{\left<#1\right>}
\newcommand{\cf}{{\it cf.}}

\allowdisplaybreaks 

\vspace*{-1.5cm}\title{Relevant Deformations, Brane Brick Models and Triality}

\author[a,b]{Mario Carcamo,}
\author[a,b,c]{Sebasti\'an Franco,}
\author[d,e]{Dongwook Ghim,}
\author[a,b]{\\ Georgios P. Goulas,}
\author[f,g]{Rak-Kyeong Seong}

\affiliation[a]{Physics Department, The City College of the CUNY\\
	160 Convent Avenue, New York, NY 10031, USA}
	
\affiliation[b]{Physics Program and \textsuperscript{$c$}Initiative for the Theoretical Sciences\\
	The Graduate School and University Center, The City University of New York\\
	365 Fifth Avenue, New York NY 10016, USA}

\affiliation[d]{
Particle Theory and Cosmology Group, Center for Theoretical Physics of the Universe, \\
Institute for Basic Science (IBS), 55 Expo-ro, Yuseong-gu, Daejeon 34126, Korea
}

\affiliation[e]{
RIKEN Center for Interdisciplinary Theoretical and Mathematical Sciences (iTHEMS), \\
RIKEN, 2-1 Hirosawa, Wako, Saitama 351-0198, Japan
}
	
\affiliation[f]{
Department of Mathematical Sciences, and
\textsuperscript{$g$}Department of Physics,\\
Ulsan National Institute of Science and Technology,\\
50 UNIST-gil, Ulsan 44919, South Korea
}

\emailAdd{mcarcamo@ccny.cuny.edu}
\emailAdd{sfranco@ccny.cuny.edu}
\emailAdd{dongwook@ibs.re.kr}
\emailAdd{ggoulas@gradcenter.cuny.edu}
\emailAdd{seong@unist.ac.kr}

\preprint{
\begin{flushright}
UNIST-MTH-25-RS-04\\
CTPU-PTC-25-37\\
RIKEN-iTHEMS-Report-25 
\end{flushright}
}

\abstract{We extend the study of relevant deformations connecting $2d$ $(0,2)$ gauge theories on D1-branes probing toric Calabi-Yau 4-folds beyond pure mass deformations. The underlying geometry provides powerful insights when field-theoretic tools are still lacking. We observe that the volume of the Sasaki-Einstein base of the Calabi-Yau 4-fold grows towards the IR, signaling the relevance of deformations. We exploit the map between gauge theory fields and GLSM fields to compute scaling dimensions directly from divisor volumes, allowing for a sharper determination of whether terms in the Lagrangian are relevant or irrelevant. Moreover, this map provides a systematic way to determine the precise set of terms needed to realize a given deformation. We also explore the interplay between general relevant deformations and triality, studying cases where non-mass deformations are mapped to mass deformations in a dual theory, and resolving puzzles that seem to require non-holomorphic couplings in one of the dual phases. Finally, we present evidence that when the Hilbert series of the mesonic moduli space is refined only under the $U(1)$  $R$-symmetry, it becomes invariant even under non-mass relevant deformations of the brane brick models corresponding to toric Calabi-Yau 4-folds related by a birational transformation, extending previous results to a broader class of deformations.}

\begin{document}

\maketitle

\section{Introduction}

An infinite class of $2d$ $(0,2)$ gauge theories can be engineered on the worldvolume of D1-branes probing toric Calabi-Yau (CY) 4-folds. The understanding of these theories has significantly expanded with the introduction of {\it brane brick models} \cite{Franco:2015tna,Franco:2015tya}, a class of Type IIA brane configurations related to D1-branes at toric singularities via T-duality.

Brane brick models elegantly encode the structure of the $2d$ theory while simultaneously simplifying the connection between the gauge theories and their corresponding toric CY 4-folds, reducing it to a combinatorial problem. For additional developments on brane brick models, see e.g. \cite{Franco:2016nwv,Franco:2016qxh,Franco:2016fxm,Franco:2017cjj,Franco:2018qsc,Franco:2019bmx,Franco:2020avj,Franco:2021elb,Franco:2022iap,Franco:2022gvl,Franco:2022isw,Franco:2023tly,Franco:2024lxs}.

A systematic investigation of mass deformations of 2d (0,2) gauge theories realized on D1-branes at CY$_4$ singularities was carried out in \cite{Franco:2023tyf}, focusing on the class of deformations for which the initial and final theories are associated to toric CY 4-folds. This work examined various aspects of these deformations, including their combinatorial description, the associated global symmetry breaking patterns, and their impact on the volume of the Sasaki-Einstein 7-manifold forming the base of the CY 4-fold. The first example of this type of deformations was presented earlier in \cite{Franco:2016fxm}. These deformations are generalizations to $2d$ of the famous Klebanov-Witten deformation relating the $4d$ $\mathcal{N}=1$ gauge theories for D3-branes probing $\mathbb{C}^2/\mathbb{Z}_2 \times \mathbb{C}$ and the conifold \cite{Klebanov:1998hh}. More examples of mass deformations for $4d$ theories associated with toric CY 3-folds were studied using brane tilings \cite{Franco:2005rj,Franco:2005sm} in \cite{Bianchi:2014qma}.

In this paper, we extend this investigation to a broader class of relevant deformations that connect two toric theories without involving mass terms. The analogous problem involving $4d$ gauge theories described by brane tilings has been studied in \cite{Cremonesi:2023psg}. The first known example of a pair of brane tilings corresponding to different toric CY 3-folds connected by a non-mass relevant deformation can be traced back to the original paper introducing brane tilings \cite{Franco:2005rj}. In that work, a toric phase of $PdP_3$ and another one for $dP_3$ which share the same quiver were considered. The two theories differ in their superpotential. From a current point of view, they are related by a relevant deformation. These deformations were systematically studied in significantly greater detail in \cite{Cremonesi:2023psg}. 
Our primary strategy is to analyze the interplay between relevant deformations and triality, focusing on higher‑order relevant deformations that map to mass terms in a triality dual.

Interestingly, the effect of mass deformations on the underlying CY 4-fold can be understood as a subclass of combinatorial and algebraic polytope mutations acting on the corresponding toric diagrams \cite{Ghim:2024asj,Ghim:2025zhs}. These mutations are also referred to as birational transformations of the associated toric varieties. It was shown in \cite{Ghim:2024asj,Ghim:2025zhs} that when mass deformations of brane brick models realize such birational transformations of the toric CY 4‑fold, they preserve both the Hilbert series of the mesonic moduli space refined only by the $U(1)_R$ symmetry and the number of mesonic generators. This is consistent with previous studies on CY 3-folds \cite{Franco:2023flw,Franco:2023mkw,Arias-Tamargo:2024fjt,CarrenoBolla:2024fxy}. We further show that the relevant non-mass deformations explored in this paper are governed by the same class of polytope mutations.

This paper is organized as follows. Section \sref{sec:review} presents a quick review of $2d$ $(0,2)$ gauge theories on D1-branes on toric CY 4-folds, brane brick models and triality. Section \sref{sec:vol} discusses how divisor volumes can be exploited to determine the scaling dimensions of quiver fields, and how the volumes of the Sasaki–Einstein bases of CY 4-folds can be used to establish the UV$\to$IR direction of RG flows. In Section \sref{section_examples}, we study a non-mass relevant deformation and a deformation involving both mass terms and higher-order contributions. Section \sref{section_non_mass_to_mass_via_triality} initiates the study of the interplay between relevant deformations and triality, showing that non-mass deformations in a theory can map to mass deformations in a triality dual. In Section \sref{section_additional_masses}, we study deformations in triality-dual theories that appear to require non-holomorphic terms in one of the two descriptions, and we resolve this apparent puzzle. Section \sref{section_irrelevant} investigates deformations to theories containing additional irrelevant terms. In Section \sref{section_polytope_mutation}, we generalize the correspondence between birational transformations of toric Calabi-Yau 4-folds and mass deformations of brane brick models observed in \cite{Ghim:2024asj,Ghim:2025zhs}
to include non-mass relevant deformations. We present our conclusions in Section \sref{section_conclusions}. Various appendices collect explicit information about the (fast) forward algorithm applied to the theories we study in the paper.

\section{$2d$ $(0,2)$ Gauge Theory and Brane Brick Models \label{sec:review} } 

Brane brick models have considerably streamlined the correspondence between toric CY$_4$'s and $2d$ $(0,2)$ gauge theories, in both directions.
One key reason is that the {\it GLSM fields}, which provide a symplectic description of the mesonic moduli space of the gauge theory, partially admit a combinatorial realization known as {\it brick matchings} in the brane brick model \cite{Franco:2015tya}. Moreover, various practical methods have been developed to construct the $2d$ $(0,2)$ gauge theory and the brane brick model corresponding to a given toric CY$_4$ (see e.g. \cite{Franco:2015tna,Franco:2015tya,Franco:2016qxh,Franco:2016fxm,Franco:2018qsc,Franco:2020avj}).\footnote{More precisely, multiple brane brick models can correspond to the same underlying toric CY$_4$. Such models are related by triality \cite{Franco:2016nwv}.}

\subsection{GLSM Fields and the $P$- and $\bar{P}$-Matrices} \label{sec:pmat}

A distinguishing feature of the $2d$ $(0,2)$ gauge theories on the worldvolume of D1-branes probing toric CY 4-folds is that the $J$- and $E$-terms associated to each Fermi multiplet $\Lambda_{ij}$ have a binomial structure,
\begin{align}
\begin{split}
    & \Lambda_{ij} \, : \, J^+_{ji} - J^-_{ji} \,, \\
    & \overline{\Lambda}_{ij} \, : \, E^+_{ij} - E^-_{ij} \,,
\end{split}
\end{align}
where $J^{\pm}_{ji}$ and $E^{\pm}_{ij}$ are holomorphic monomials in chiral fields. This property is often referred to as the {\it toric condition} \cite{Franco:2015tna}.

When all gauge groups are $U(1)$, the classical mesonic moduli space $\mathcal{M}^{mes} $ of the brane brick model is defined by the quotient
\begin{align}
    \mathcal{M}^{mes} = \textrm{Spec}
    \left( \IC [ X_{ij} ] / \mathcal{I}^{\textrm{Irr}}_{JE} \right)  // U(1)^{G-1} \,,
\end{align}
where $\IC [ X_{ij} ] $ is the coordinate ring in terms of chiral fields $X_{ij}$ and $\mathcal{I}^{\textrm{Irr}}_{JE}$ is the irreducible component of the toric ideal formed by the $J$- and $E$-terms.
Only $G-1$ of the $U(1)$ gauge groups are independent to each other explaining the quotient by 
$U(1)^{G-1}$.

The \textit{forward algorithm} \cite{Franco:2015tna} allows us
to identify the mesonic moduli space $\mathcal{M}^{mes}$ in terms of GLSM fields $p_a$, 
\begin{align} \label{Mmes-def}
    \mathcal{M}^{mes} = \textrm{Spec} \left(
    \IC [ p_a ] // Q_{JE} \right) // Q_D \,,
\end{align}
where $a =1 , \cdots , c $ and $c$ is the total number of GLSM fields \cite{Franco:2015tna}.
The $J$- and $E$-terms as well as the $U(1)$ gauge charges are
expressed in terms of a collection of $U(1)$ charges on GLSM fields
which are given by
the $Q_{JE}$-  and $Q_D$-matrices, respectively. 
These charge matrices are obtained as part of the forward algorithm. They are defined as follows, 
\begin{align} \label{Mmes-sym}
    Q_{JE} = \ker P \,,  \quad \bar{d} = Q_D \cdot P^T \,,
\end{align}
where $\bar{d}$ is the reduced incidence matrix of the quiver. The $P$-matrix is an important piece of information coming from the $J$- and $E$-terms under the forward algorithm, since it summarizes the map between chiral fields in the quiver gauge theory and GLSM fields.

Defining the total charge matrix $Q_t$ as the concatenation of $Q_{JE}$ and $Q_D$, the toric diagram of the underlying CY 4-fold is given by
\beq
G= \ker Q_t \, ,
\eeq
where each of the columns of $G$ give the position of the point in the toric diagram associated to the corresponding GLSM field. In general, the forward algorithm also produces {\it extra GLSM fields}. These additional fields behave like an over-parameterization of the mesonic moduli space $\mathcal{M}^{mes}$ and are not part of the corresponding toric diagram. However, they play a crucial role in determining the correct $Q_{JE}$ and $Q_D$ charges in the symplectic quotient definition of the mesonic moduli space. We refer the interested reader to \cite{Franco:2015tna} for further details.

Brane brick models offer an alternative combinatorial approach to determining the toric CY 4-fold associated with a given gauge theory. 
This method circumvents some of the more computationally intensive steps of the forward algorithm. The central objects in this construction are the \textit{brick matchings}, defined as follows \cite{Franco:2015tya}. First, we define the gauge-invariant \textit{plaquettes} as the monomials in the $J$- and $E$-terms combined with the corresponding Fermi field or conjugate Fermi field, 
\beal{plaq}
\Lambda_{ij} \cdot J^+_{ji} \,, \quad \Lambda_{ij}
\cdot J^-_{ji} \,, \quad
\overline{\Lambda}_{ij} \cdot E^+_{ij} \,, \quad\overline{\Lambda}_{ij} \cdot E^-_{ij} \,.
\eea
Then, a brick matching is defined as a collection of chiral, Fermi and conjugate Fermi fields that covers exactly every plaquette in the brane brick model by satisfying the following three conditions:

\begin{itemize}
    \item The chiral fields in a brick matching cover either the plaquettes $ ( \Lambda_{ij} \cdot J^+_{ji} \,, \Lambda_{ij} \cdot J^-_{ji} ) $
    or the plaquettes $ ( \overline{\Lambda}_{ij} \cdot E^+_{ij} \,, \overline{\Lambda}_{ij} \cdot E^-_{ij} ) $ exactly once each.
    \item If the chiral fields in a brick matching cover the plaquettes
    $ ( \overline{\Lambda}_{ij} \cdot E^+_{ij} \,, \overline{\Lambda}_{ij} \cdot E^-_{ij} ) \,, $
    then the brick matching should include $ \Lambda_{ij} \,.$
    \item If the chiral fields in a brick matching cover the plaquettes $ ( \Lambda_{ij} \cdot J^+_{ji} \,, \Lambda_{ij} \cdot J^-_{ji} ) \,,$ then the brick matching should include $\overline{\Lambda}_{ij} \,. $
\end{itemize}
We refer to \cite{Franco:2019bmx,Franco:2021elb} for alternative, yet equivalent, definitions of brick matchings. 

Remarkably, brick matchings coincide exactly with the non-extra GLSM fields obtained via the forward algorithm. 
Together with a prescription for assigning coordinates in the toric diagram, they furnish an alternative and efficient procedure for determining the toric CY 4-fold from the gauge theory, known as the \textit{fast forward algorithm} \cite{Franco:2015tya}.

We will denote by $\bar{P}$ the matrix encoding the map between chiral fields in the quiver gauge theory and brick matchings.\footnote{Equivalently, $\bar{P}$ is obtained from the $P$-matrix introduced above by removing the columns corresponding to the extra GLSM fields.} In general, $\bar{P}$ is defined as follows,
\beal{def-P}
\bar{P}_{X_{ij},a} =
    \begin{cases}
    1 \quad \text{ if } \, X_{ij} \in p_a \,, \\
    0 \quad \text{ if } \, X_{ij} \notin p_a \,,
    \end{cases}
\eea
for $a = 1 , \cdots , \bar{c} $ and $\bar{c}$ being the total number of brick matchings. The $\bar{P}$-matrix can be extended to include the (conjugate) Fermi field content of brick matchings \cite{Franco:2015tya}, 
by including new rows associated with Fermi and conjugate Fermi fields as follows.
\beal{def-Pfermi}
\left( \bar{P}_{\Lambda} \right)_{\Lambda_{ij} \text{ or } \overline{\Lambda}_{ij},P} =
    \begin{cases}
    1 \quad \text{ if } \, \Lambda_{ij} \in p_a \\
    1 \quad \text{ if } \, \overline{\Lambda}_{ij} \in p_a \\
    0 \quad \text{ otherwise} 
    \end{cases}
    ~.~
\eea
In the following work, we refer to the above matrix as the extended $\bar{P}$-matrix.

For all the models studied in this paper, we have carried out both the forward and fast forward algorithms, confirming that they yield consistent results. For brevity, we will present only the extended $\bar{P}$-matrices. Including the extra GLSM fields, namely presenting the extended $P$-matrices, would be impractical, as these matrices are generally prohibitively large for the examples we will study.
\\

\subsection{Mass Deformations of Brane Brick Models} \label{sec:deform}

As shown in \cite{Franco:2023tyf}, mass deformations can relate brane brick models associated to different toric CY 4-folds. 
SUSY-preserving mass terms in $2d$ $(0,2)$ theories 
are gauge invariant quadratic couplings between a chiral and a Fermi field, hence constraining the 
brane brick models undergoing mass deformation in this way.
Given a pair of chiral and Fermi fields connecting to the same pair of nodes in the quiver, 
the choice of the mass term introduced in the $J$- and $E$-terms
depends on the type of chiral-Fermi pair in the quiver of the $2d$ $(0,2)$ theory.
In general, mass deformation
of a brane brick model involves the following mass terms, 
\begin{align}
\begin{split}
    (\Lambda_{ij}, X_{ij} ) \in Q : & \quad J'_{ji} = J_{ji} \,,
    \quad E'_{ij} = \pm m X_{ij} + E_{ij} \\
    (\overline{\Lambda}_{ij} , X_{ji} ) \in Q : & \quad J'_{ji} = \pm m X_{ji} + J_{ji} \,, \quad E'_{ij} = E_{ij} \,,
\end{split}
\end{align}
where $J_{ji}'$ and $E'_{ij}$ are the $J$- and $E$-terms corresponding to the Fermi $\Lambda_{ij}$ after the mass deformation,
and $J_{ji}$ and $E_{ij}$ indicate the original $J$- and $E$-terms. 
We can integrate out the massive pair of chiral and Fermi fields, 
leaving the following replacement of the chiral fields in the remaining $J$- and $E$-terms,
\beal{chi-repl}
X_{ij} = \mp \frac{1}{m} (E_{ij}^+ - E_{ij}^-) \quad \textrm{or} \quad X_{ji} = \mp \frac{1}{m} (J^+_{ji} - J^-_{ji} ) \,.
\eea
\\

\subsubsection*{Change of Variables}

Note that the replacement of chiral fields in \eqref{chi-repl} 
sometimes violates the binomial property of the $J$- and $E$-terms \cite{Franco:2015tna}. 
We rectify this issue by further deforming the interactions with higher-order couplings among the chiral fields.
In \cite{Franco:2023tyf}, this remedy deformation is referred to as a \textit{redefinition of holomorphic interactions}, 
which generally takes the form,
\beal{holo-redef}
\Lambda_{ij}' \cdot X'_{jk} = \Lambda_{ij} \cdot (X_{jk} + \sum_h c_h^{(jk)} X_{jh} X_{hk} ) \,,
\eea
where $X_{jk}'$ and $\Lambda_{ij}'$ are the new fields after the redefinition and $c_h^{(jk)}$ are the coefficients specific to the redefinition.

In this paper, we introduce a more general redefinition of holomorphic interactions, 
which we refer to as a \textit{change of field variables}, in order to recover the toric property of the $J$- and $E$-terms.
This more general redefinition of field variables involves 
the following redefinitions of chiral and Fermi fields, 
\beal{def-cov}
X_{ij} 
&\rightarrow&
 X'_{ij} = a X_{ij} + b X_{i \cdots j} \,, 
 \nn\\
\Lambda_{ij} 
&\rightarrow&
 \Lambda'_{ij} =  \, s \Lambda_{ij} + \alpha X_{i \cdots k} \cdot \Lambda_{kj}
+ \beta \Lambda_{ik} \cdot X_{k \cdots j} + \gamma X_{i \cdots k} \cdot \Lambda_{kl} \cdot X_{l \cdots j} \,,
\eea
where $a,b, s, \alpha, \beta$ and $\gamma$ are collective notations for numerical coefficients,
and $ X_{i \cdots j} $ denotes a holomorphic monomial consisting of chiral fields, namely an oriented chain of chiral fields in the quiver, 
whose start and end nodes are labelled $i$ and $j$, respectively.
The notation suppresses the node labels inside the chain.

The change of chiral fields in \eref{def-cov}
leads to the following transformation of $J$- and $E$-terms,
\beal{cov-chiral}
J (X'_{ij}) = J (aX_{ij} + b X_{i \cdots j}) \,, \quad E (X'_{ij}) = E ( aX_{ij} + b X_{i \cdots j}) \,,
\eea
where the gauge group indices of the $J$- and $E$-term are suppressed for clarity. 
The change of Fermi fields in \eref{def-cov} 
affects the $J$- and $E$-terms in a less straightforward way.
For the $E$-terms, 
the deformation affects the chirality constraint for the Fermi fields in the form $E_{ij} = \overline{D}_+ \Lambda_{ij} \,.$ 
This results in the following new $E$-term, 
\begin{align} \label{cov-E}
\begin{split}
E'_{ij} (X) =  \frac{1}{s}  & \Big[ E_{ij} (X') - \alpha X_{i \cdots k} \cdot E_{kj} (X') \\
& - \beta E_{ik} (X') \cdot X_{k \cdots j} - \gamma X_{i \cdots k} \cdot E_{kl} (X') \cdot X_{l \cdots j} \Big] \,.
\end{split}
\end{align}
For
the new $J$-terms, let us recall the Lagrangian of $2d$ $(0,2)$ gauge theories, where
the $J$-term is introduced through a holomorphic coupling with the Fermi field as follows,
\beal{lag-J}
\mathcal{L}_J = - \int d^2y ~d\theta^+ \Big[ \Lambda_{ij} J_{ji} (X) + \Lambda_{kl} J_{lk} (X) + \cdots \Big] - \textit{c.c} \,,
\eea
where $J_{lk}$ stands for the other $J$-terms coupled to Fermi fields $\Lambda_{kl}$.
Substituting $\Lambda_{ij}$ with $\Lambda'_{ij}$ as defined in \eqref{def-cov}, 
we have a change in the holomorphic coupling involving $\Lambda_{ij}\,, \Lambda_{jk} \,, \Lambda_{ik} \,,$ and $\Lambda_{kl} $ next to the $\Lambda_{ij} J_{ji} (X) $ coupling.
By rearranging the $J$-terms, we have,
\begin{align} \label{cov-J}
\begin{split}
J'_{ji} (X) & = s J_{ji} (X') \,, \\
J'_{jk} (X) & = J_{jk} (X') + \alpha J_{ji} (X') \cdot X_{i \cdots k} \,, \\
J'_{ki} (X) & = J_{ki} (X') + \beta X_{k \cdots j} \cdot J_{ji} (X') \,, \\
J'_{lk} (X) & = J_{lk} (X') + \gamma X_{l \cdots j} \cdot J_{ji} (X') \cdot X_{i \cdots k} \,,
\end{split}
\end{align}
where $J'_{jk} \,, J'_{kl} \,, \text{ and } J'_{lk} $ stand for the new $J$-terms coupled to
the redefined Fermi fields $\Lambda_{kj} \,, \Lambda_{ki} \,, \text { and } \Lambda_{kl} \,, $ respectively.

In summary, 
the change of field variables that we introduced in \eqref{def-cov} 
result in the introduction of 
higher-order interactions to the
holomorphic $J$- and $E$-terms as follows, 
\beal{cov-summary}
E_{ij} (X) &\rightarrow& s^{-1} \big[  E_{ij} - \alpha X_{i \cdots k} \cdot E_{kj} - \beta E_{ik} \cdot X_{k \cdots j} - \gamma X_{i \cdots k} \cdot E_{kl} \cdot X_{lj} \big] \,, \nn\\
J_{ji} (X) & \rightarrow& s J_{ji} \,, \nn\\
J_{jk} (X) &\rightarrow& J_{jk}  + \alpha J_{ji} \cdot X_{i \cdots k}  \,, \nn\\
J_{ki} (X) &\rightarrow& J_{ki}  + \beta X_{k \cdots j} \cdot J_{ji} \,, \nn\\
J_{lk} (X) &\rightarrow& J_{lk}  + \gamma X_{l \cdots j} \cdot J_{ji} \cdot X_{i \cdots k} \,,
\eea
where the holomorphic functions on the right-hand side above are to be understood as those 
resulting from the change of field variables  in \eqref{cov-chiral}.
\\

\subsubsection*{Brick Matchings and Mass Deformations}

Brick matchings correspond to vertices of the toric diagram of the toric Calabi-Yau 4-folds
as well as are combinatorially related to chiral fields of the brane brick model, 
forming a natural bridge between toric geometry and the $2d$ $(0,2)$ gauge theory.
This connection
allows us to describe the effect mass deformations have on the brane brick model
in terms of the brick matchings parameterizing the toric Calabi-Yau 4-fold
as first described in \cite{Franco:2023flw}. 
Mass deformations
have an effect on the geometry of the toric Calabi-Yau 4-fold, 
which in turn affects the shape of the corresponding toric diagram 
and the positions of the its vertices which correspond to GLSM fields associated to brick matchings.
This effect puts brick matchings into the following 3 categories as observed in \cite{Franco:2023flw}:
\begin{itemize}
    \item \textbf{Massive brick matchings} contain chiral fields that become massive under the mass deformation. Consequently, their chiral field content changes under the deformation. Before we turn on the deformation, these brick matchings correspond to an extremal vertex in the toric diagram. Under the deformation, the relative position of the corresponding vertex in the toric diagram remains unchanged with respect to the other vertices, up to an $SL(3,\mathbb{Z})$ transformation.
    \item \textbf{Moving brick matchings} correspond to extremal vertices of the toric diagram whose position relative other vertices changes with the deformation. For all the examples studied so far, the vertex associated to a moving brick matching lies at the end of the edge of the toric diagram containing multiple segments, namely an edge of the toric diagram with more than two collinear points. The effect of the deformation is to reduce the number of collinear points. Moving brick matchings do not contain any of the chiral fields that become massive under the mass deformation.
    \item \textbf{Unaffected brick matchings} refer to the remaining brick matchings that are neither massive nor moving. Their chiral field content is partially, and sometimes entirely, preserved under the mass deformation. They can correspond to both extremal and non-extremal vertices in the toric diagram.
\end{itemize}

\subsection{Triality} \label{sec:triality}

$2d$ $(0,2)$ supersymmetric gauge theories exhibit a low energy equivalence known as {\it triality} \cite{Gadde:2013lxa}. The term ``triality" follows from the fact that, in its simplest form, it relates three $2d$ SQCD-like theories in the IR. Alternatively, applying three consecutive triality transformations to the same gauge group takes the theory back to its original form.

Similarly to Seiberg duality for $4d$ $\cN=1$ gauge theories \cite{Seiberg:1994pq}, triality has a combinatorial description when applied to quiver gauge theory as we shall review in this section. A string theoretic realization of triality in terms of brane brick models was introduced in \cite{Franco:2016nwv}. This understanding was deepened in \cite{Franco:2016qxh}, where triality was connected to geometric transitions in the mirror geometry. This perspective led to the proposal of new dualities in various dimensions and to a beautiful geometric unification of them \cite{Franco:2016tcm,Franco:2017lpa}.

We refer to each of the seemingly distinct quiver theories related by triality as a {\it phase}. In the special class of theories described by brane brick models, triality manifests as different quivers with distinct $J$- and $E$-terms that share the same mesonic moduli space. Phases that can be described by brane brick models—or equivalently, by periodic quivers on $\mathbb{T}^3$—are referred to as {\it toric phases}.

The rules for the triality transformation of general quiver theories, i.e. not necessarily those associated to brane brick models or toric CY 4-folds, were fully developed in \cite{Franco:2017lpa}, where triality was shown to be a special case of graded quiver mutation.\footnote{More concretely, $m$-graded quivers with potentials exhibit order-$(m+1)$ dualities. For $m\leq 3$, this corresponds to supersymmetric gauge theories in $6-2m$ dimensions. Specifically, $m=0$, 1, 2 and 3 correspond to $6d$ $\mathcal{N}=(0,1)$, $4d$ $\mathcal{N}=1$, $2d$ $\mathcal{N}=(0,2)$ and $0d$ $\mathcal{N}=1$ field theories, respectively \cite{Franco:2017lpa,Closset:2018axq,Franco:2019bmx}. For general values of $m$, including $m>3$, the graded quivers and their mutations describe the open string sector of the topological B-model on CY $(m+2)$-folds \cite{Franco:2017lpa,Closset:2018axq}.} Below, we summarize them.

\subsubsection{Ranks}

In a quiver with several gauge nodes, triality is a local transformation, acting on the dualized node $i$ and the fields charged under it. The rank of node $i$ transforms as follows
\beal{tri_rk}
N_i' = \sum_{j} n^{\chi}_{ji} N_j - N_i \,.
\eea
Here, $N_i$ and $N_i'$ denote the rank of the dualized node before and after triality, respectively; $n^{\chi}_{ji}$ is the number of chiral arrows incoming into node $i$ from node $j$; and $N_j$ is the rank of node $j$. In short, the new rank is equal to the total number of incoming chiral fields minus the original rank.

\subsubsection{Mutation of the quiver}

Let us now explain how the quiver transforms under mutation.

\bigskip

\noindent{\bf 1. Flavors.}
We refer to the fields connected to the mutated node as {\it flavors}. They transform as follows
\beq
\begin{array}{ccc}
\textrm{Incoming chiral} \to \textrm{Outgoing chiral} & \ \ \ &  X_{ji}\to X_{ij} \\[.05 cm]
\textrm{Fermi} \to \textrm{Incoming chiral} & \ \ \ & \Lambda_{ij}\rightarrow X_{ji} \\[.05 cm]
\textrm{Outgoing chiral} \to \textrm{Fermi} & \ \ \ & X_{ij}\rightarrow \Lambda_{ij}
\end{array}
\label{transformation_flavors}
\eeq
In the table above, it is irrelevant whether a Fermi field is incoming or outgoing relative to the mutated node. This is because $2d$ $(0,2)$ theories are invariant under the conjugation of any Fermi field accompanied by the exchange of its $J$- and $E$-terms (see e.g. \cite{Franco:2015tna}). Thus, without loss of generality, we will assume that the Fermi field attached to the mutated node is oriented outward.

\bigskip

\begin{figure}[H]
\begin{center}
\resizebox{\hsize}{!}{
\includegraphics[width=3cm]{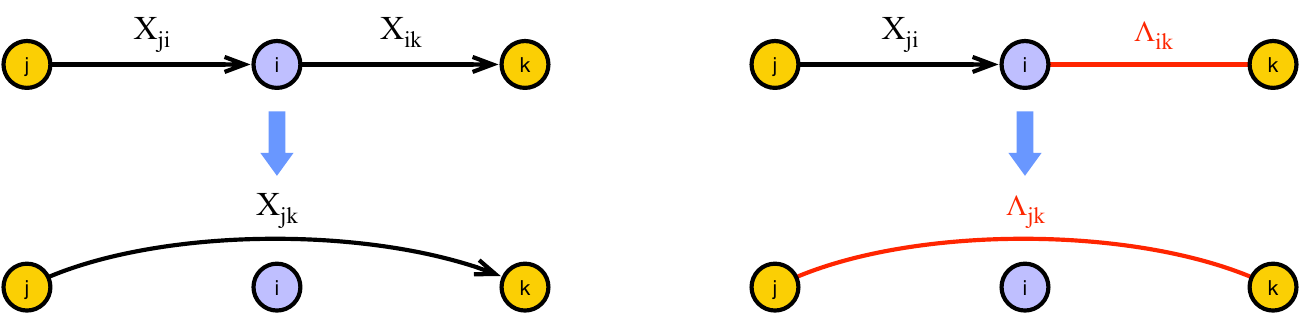}
}
\caption{Mesons resulting from the composition of an incoming chiral field with an outgoing chiral or Fermi field at node $i$. \label{fig_triality_meson}}
 \end{center}
 \end{figure}

\noindent{\bf 2. Mesons.}
We introduce mesons by composing incoming chiral fields with outgoing chiral fields and Fermi fields.
For every chiral field $X_{ji}$ 
and every outgoing chiral field $X_{ik}$,
we obtain a new chiral field $X_{jk}$, 
when we apply triality on node $i$. 
For every Fermi field $\Lambda_{ik}$ 
connected to node $i$, 
we get a new Fermi field $\Lambda_{jk}$.\footnote{As mentioned earlier, Fermi fields are not really oriented. If we use a convention in which a Fermi field terminating on node $i$ is actually $\Lambda_{ki}$, we first switch it with its conjugate and then apply this composition rule.} 
\fref{fig_triality_meson} illustrates the overall deformation of the $2d$ $(0,2)$ quiver under triality on node $i$.

\noindent{\bf 3. $J$- and $E$-Terms.}
Let us summarize the rules that control the transformation of the 
$J$- and $E$-terms under triality. 
It is convenient to describe the effect of triality on 
the $J$- and $E$-terms in terms of plaquettes. 
Furthermore, to simplify our discussion, we introduce
 a new notation involving gauge sub-indices
 where a plaquette of order $k$ goes through quiver nodes $i_1,\ldots,i_k$, 
 with $i_1$ being the node on which triality acts.

\begin{itemize}
\item[{\bf(a)}] {\bf Cubic meson-dual flavors couplings.} 
The first rule concerns new plaquettes that are in one-to-one correspondence with the mesons obtained under triality. 
For each meson, we add a cubic coupling involving the meson itself
and the dual flavors.
\fref{Quivers_cubic_couplings} illustrates how these plaquettes are formed involving
the two types of mesons in the $2d$ $(0,2)$ quiver
as shown in \fref{fig_triality_meson}.\footnote{With the orientation of the Fermis that we consider, there new couplings are contributions to $J$-terms.}
We emphasize that \fref{Quivers_cubic_couplings}
illustrates actual plaquettes as contributions to the $J$- and $E$-terms, 
not merely the transformation of the quiver itself.

\begin{figure}[ht!]
\begin{center}
\resizebox{\hsize}{!}{
\includegraphics[width=3cm]{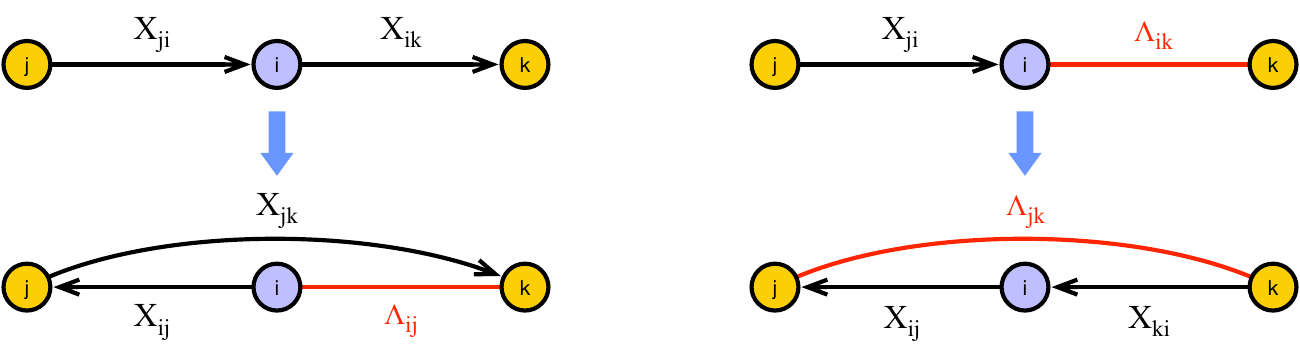}
}
\caption{Plaquettes representing cubic couplings between mesons and dual flavors.
\label{Quivers_cubic_couplings}}
 \end{center}
 \end{figure}

\end{itemize}

In addition to adding these new plaquettes to the $J$- and $E$-terms, 
the original plaquettes in the $J$- and $E$-terms are also directly affected by triality. 
Plaquettes in the original theory that do not contain the 
mutated node remain unchanged. 
For plaquettes that involve the
mutated node, there are two possibilities, depending on the type of fields in the plaquette that are connected to the mutated node. 
The first possibility is discussed in \text{(b)} , whereas \text{(c)}  and \text{(d)}  discuss the second possibility.

\begin{itemize}
\item[{\bf (b)}] In existing plaquettes, we replace every monomial of fields that compose to give a meson by the corresponding meson. 
These monomials involve an incoming chiral field and an outgoing chiral or Fermi, 
namely they are of the general form $X_{i_k i_1}X_{i_1 i_2}$ and $X_{i_k i_1}\Lambda_{i_1 i_2}$. 
\fref{triality_rule_3b} provides a graphical representation of this rule. 

\begin{figure}[h]
	\centering
	\includegraphics[width=\textwidth]{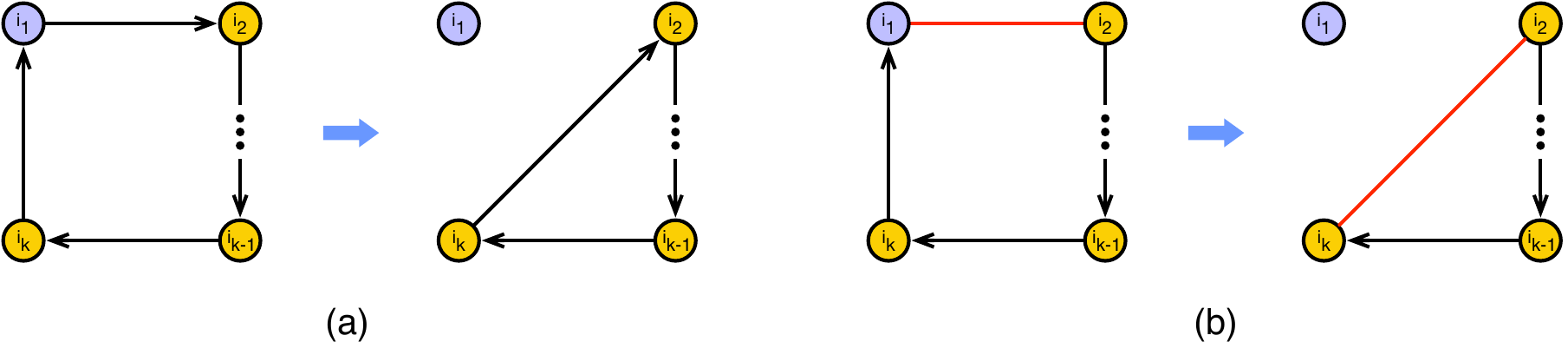}
\caption{Transformation of plaquettes containing monomials of fields giving rise to mesons. We illustrate the two possible cases: a) chiral meson and b) Fermi meson. We indicate the dualized node in purple.}
	\label{triality_rule_3b}
\end{figure}

\end{itemize}

The following two rules relate to plaquettes that 
involve the mutated node under triality, 
but do not contain any chiral fields that end on it. 
These plaquettes contain monomials of the form $\Lambda_{i_k i_1} X_{i_1 i_2}$.
\begin{itemize}
\item[{\bf (c)}] A monomial of the form $\Lambda_{i_k i_1} X_{i_1 i_2}$ in a plaquette is replaced by the monomial of the corresponding dual flavors according to \eqref{transformation_flavors}, which takes the general form $X_{i_k i_1} \Lambda_{i_1 i_2}$. \fref{triality_rule_3c} illustrates this replacement. 

\begin{figure}[h]
	\centering
	\includegraphics[height=2.6cm]{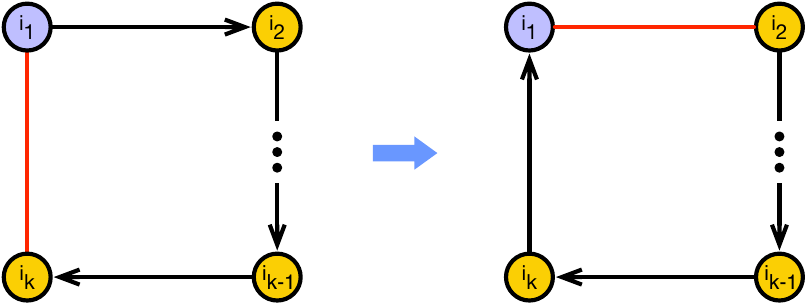}
\caption{Transformation of a plaquette that goes through the mutated node but does not contain a meson.}
	\label{triality_rule_3c}
\end{figure}

\item[{\bf (d)}] Additionally, if there is an incoming chiral field $X_{i_0 i_1}$ from a node $i_0$ to the mutated node $i_1$, 
we introduce an additional plaquette which is a copy of the original plaquette
with instances of the form $\Lambda_{i_k i_1} X_{i_1 i_2}$ replaced with the products of mesons that are obtained by introducing
$\Lambda_{i_k i_1}$ and $X_{i_1 i_2}$.\footnote{Where $\Lambda_{i_k i_1}$ needs to be conjugated as necessary to form the composition.} 
This composition of the new plaquette is illustrated in \fref{triality_rule_3d}.

\begin{figure}[h]
	\centering
	\includegraphics[height=2.6cm]{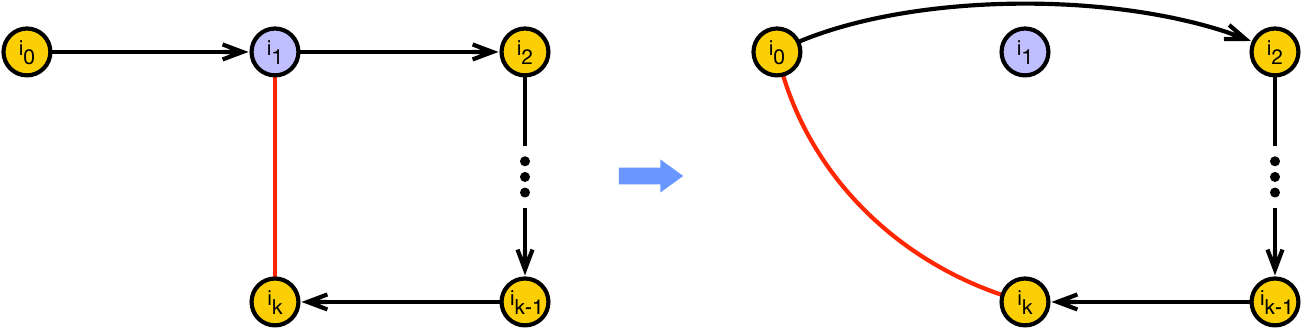}
\caption{Transformation of a plaquette in the presence of an additional chiral field incoming into the dualized node.}
	\label{triality_rule_3d}
\end{figure}

\end{itemize}
Whenever we apply \text{(c)} , we also have to apply \text{(d)}. 
\text{(c)} and \text{(d)}  
are new features of $2d$ $(0,2)$ triality 
with no analogous counterpart in $4d$ $\mathcal{N}=1$ Seiberg duality. 
As we are going to see in the following sections, 
these new rules have important implications for relevant deformations of $2d$ $(0,2)$ theories given by brane brick models.
We also note that if the above rules under triality on a quiver node 
generate mass terms in the $J$- and $E$-terms,
the corresponding massive fields are integrated out as discussed in \cite{Franco:2015tna}.
\\

\section{Divisor Volumes and Relevance of Deformations} \label{sec:vol}

A primary goal of this paper is the study of non-massive relevant deformations of $2d$ $(0,2)$ gauge theories associated to toric CY 4-folds. Such deformations modify the $J$- and $E$-terms of a theory, while leaving the quiver invariant. As explained in Section \ref{sec:deform}, it is possible to redefine fields such that the final $J$- and $E$-terms satisfy the toric condition, leading to a new brane brick model.

The theories under consideration are strongly coupled in the IR, so the dimensions of operators do not follow from naive counting of fields, i.e. from the free field dimensions. At present, we do not know of field theoretic methods to determine these dimensions in $2d$ (0,2) theories. As a result, it is in principle not known whether a general deformation is relevant or not. This stands in contrast to, for example, the case of $4d$ $\mathcal{N}=1$ theories, where they can be determined via $a$-maximization \cite{Intriligator:2003jj}.

In \cite{Franco:2023tyf}, it was found that the volume of the Sasaki-Einstein (SE) 7-manifold at the base of the CY 4-fold under consideration grows under mass deformations. 
This suggests that these volumes of SE 7-manifold $\rm Y_7^{UV}$ and $\rm Y_7^{IR}$, more broadly, satisfy
\beq
{\rm Vol}(\rm Y^{UV}_7) < {\rm Vol}(\rm Y^{IR}_7)
\label{V_Y7_UV_IR}
\eeq
and can thus be used to determine the UV $\to$ IR direction of RG flows under general deformations. 

The analogous monotonicity in the volume of SE 5-manifolds is well understood in the context of the RG flows among $4d$ $\cN=1 $ gauge theories,
since the central charge $a$ of $4d$ $\cN=1$ theories turns out to be inversely proportional to the volume of a SE $5$-manifold \cite{Gubser:1998vd,Herzog:2002ih,Benvenuti:2004dy}. 
The more detailed dictionary between $\textrm{CY}_3$ geometry and $4d$ $\cN=1$ theory enables the calculation of the scaling dimension of operators as well (see e.g. \cite{Gubser:1998fp,Herzog:2002ih,Benvenuti:2004dy}).

Inspired by such progress, we are going to extend the results of \eqref{V_Y7_UV_IR} in two directions:
\begin{itemize}
\item Investigate the change of the volume of the SE$_7$ under non-massive deformations.
\item Establish a map between divisors in the SE$_7$ base and fields in the $2d$ $(0,2)$ quiver. This, in turn, allows us to map the volumes of these divisors to the scaling dimension of fields in the quiver, from which we can determine whether a deformation of the brane brick model is relevant or not.
\end{itemize}
Besides these two directions, given that we focus on toric Calabi-Yau 4-folds characterized by 3-dimensional toric diagrams, we are going to investigate with the use of toric geometry 
how global symmetry breaks under a deformation.
\\

\subsection{From Geometry to Scaling Dimensions}

The non-compact toric CY 4-folds that we consider 
are cones over SE $7$-manifolds. They contain a basis of 5-cycles, which are in one-to-one correspondence with extremal points of the toric diagram. 
The volumes of these cycles can be calculated from the toric data via an extremization procedure (see e.g. \cite{Martelli:2006yb, Amariti:2012tj}). These volumes can then be used to determine the scaling dimension of fields. Let us review how this is done.

To every extremal vertex of the toric diagram, we associate a $4d$ vector $v_\mu=(1,\tilde{v}_\mu)$, with $\tilde{v}_\mu$ the $\mathbb{Z}^3$ position vector of the extremal vertex. Next, we consider a clockwise (when looking into the toric diagram) sequence of $n_\mu$ adjacent vectors $w_\nu$ for $ \nu=1,...,n_\mu $. To compute the volume of the 5-cycle $\Sigma_\mu$ associated to the extremal vertex, we consider
\begin{equation}\label{MSvol}
{\rm Vol}(\Sigma_\mu) = \sum_{\nu=2}^{n_\mu-1} \frac{
\langle v_\mu,w_{\nu-1},w_\nu,w_{\nu+1}\rangle \langle v_\mu,w_k,w_1,w_{n_\mu}\rangle}{\langle v_\mu,b,w_\nu,w_{\nu+1}\rangle
\langle v_\mu,b,w_{\nu-1},w_\nu \rangle \langle v_\mu,b,w_1,w_{n_\mu}\rangle} ,
\end{equation}
where the volume has been normalized by the volume of $\rm S^5$.
We also have $\langle v_1, v_2, v_3, v_4\rangle$, which is the determinant of a $4\times4$ matrix whose columns correspond to the $4$-dimensional vectors $v_1, \dots, v_4$ inside the bracket.
Lastly, 
we have the Reeb vector of the general form $b=(n,b_1, b_2, b_3)$ \cite{Martelli:2005tp,Martelli:2006yb}.
Here, given that we are considering affine toric Calabi-Yau 4-folds,
due to the Gorenstein condition the vectors $v_\mu$
corresponding to extremal vertices of the toric diagram are all co-planar allowing us to set $n=4$
in the Reeb vector $b$ \cite{Martelli:2005tp,Martelli:2006yb}.

Additionally, we consider the function
\begin{equation}\label{Zfunction}
{\rm Z_{\rm MSY}} = \sum_{\mu=1}^{N_v} \rm{Vol}(\Sigma_\mu) ,
\end{equation}
where the sum runs over the $N_v$ corners of the toric diagram. Minimizing this function determines the remaining components of the Reeb vector and the volume of the SE$_7$ base manifold. The latter is given by
\begin{equation}\label{volumeformula}
{\rm Vol}(\rm Y_7) =\frac{1}{4}{\rm Vol}(S^7) \, Z_{\rm MSY}=\frac{\pi^4}{12} Z_{\rm MSY} \,.
\end{equation}
Before translating the volume data of divisors into the scaling dimension of matter fields, 
it is worth noting that each extremal vertex with vector $v_\mu$ in the toric diagram corresponds to an extremal GLSM field $p_{\mu}$. 
The scaling dimension of such an extremal GLSM field is obtained
from the volume of corresponding 5-cycle as follows\footnote{We used the following facts: $\rm{Vol} (S^5) = \pi^3 \,, \rm{Vol} (S^7) = \frac{\pi^4}{3} \,.$}
\begin{equation}\label{scaledim}
  \Delta_\mu = \frac{\pi}{6}\frac{\rm{Vol}(\Sigma_\mu) \, \rm{Vol} (S^5) }{\rm Vol(\rm Y_7)}=2\frac{\rm Vol(\Sigma_\mu)}{\rm Z_{\rm MSY}} \,.
\end{equation}

Given the gauge-invariant deformation plaquettes $\Lambda_{ij} \cdot \Delta J_{ji}$ and $\overline{\Lambda}_{ij} \cdot \Delta E_{ij} $, we determine their relevance by summing over the contribution from chiral, Fermi and the conjugate Fermi fields,
\begin{align} \label{scaledefm}
\begin{split}
\Delta [ \Lambda_{ij} \cdot \Delta J_{ji} ] = \Delta[ \Lambda_{ij}] + \Delta [\Delta J_{ji}] \quad \textrm{for $J$-term deformation} \,, \\
\Delta [ \overline{\Lambda}_{ij} \cdot \Delta E_{ij} ] = \Delta [ \overline{\Lambda}_{ij} ] + \Delta [ \Delta E_{ij} ] \quad \textrm{for $E$-term deformation} \,.
\end{split}
\end{align}
To read each contribution, we use the extended matrix $\bar{P}_\Lambda$ defined in \eqref{def-Pfermi}. 
The scaling dimension of chiral field $X_{ij}$ reads
\beal{scalechi}
    \Delta [X_{ij}] = \sum_{\mu=1}^{N_v} \Delta_\mu \cdot ( \bar{P}_\Lambda ) _{X_{ij},\mu}   \,,
\eea
and they are summed to the scaling dimension of deformation terms $\Delta [ \Delta E] $ and $ \Delta [\Delta J] $. The contribution of Fermi $\Lambda_{ij}$ and that of the conjugate Fermi field $\overline{\Lambda}_{ij}$ read
\begin{align} \label{scalefermi}
    \Delta [ \Lambda_{ij} ] = \sum_{\mu=1}^{N_v}  \Delta_\mu \cdot ( \bar{P}_\Lambda )_{\Lambda_{ij},\mu} \quad \,, \quad 
    \Delta [\overline{\Lambda}_{ij} ] = \sum_{\mu=1}^{N_v} \Delta_\mu 
 \cdot ( \bar{P}_\Lambda)_{\overline{\Lambda}_{ij},\mu} \,.
\end{align}
Note that the sum ranges over $N_v$ GLSM fields that correspond to the extremal vertices in the toric diagram, not over all GLSM fields appearing in the forward algorithm referred to in Section \ref{sec:pmat}.
\\

\section{Examples}

\label{section_examples}

In this section, we illustrate the previous ideas with two explicit examples.

\subsection{A Relevant Non-Mass Deformation from $P_{+-}(\textrm{SPP}/\mathbb{Z}_2)$ to $P^2_{+-}(\textrm{PdP}_3)$} \label{sec:p2sppz2}

The first example we consider features a relevant deformation composed entirely of cubic plaquettes. That is, the deformation involves no masses and thus serves as a pure prototype of the new class of deformations explored in this paper. Since no masses are turned on, the quivers of the initial and deformed theories are identical.

\begin{figure}[h]
	\centering
	\includegraphics[height=5cm]{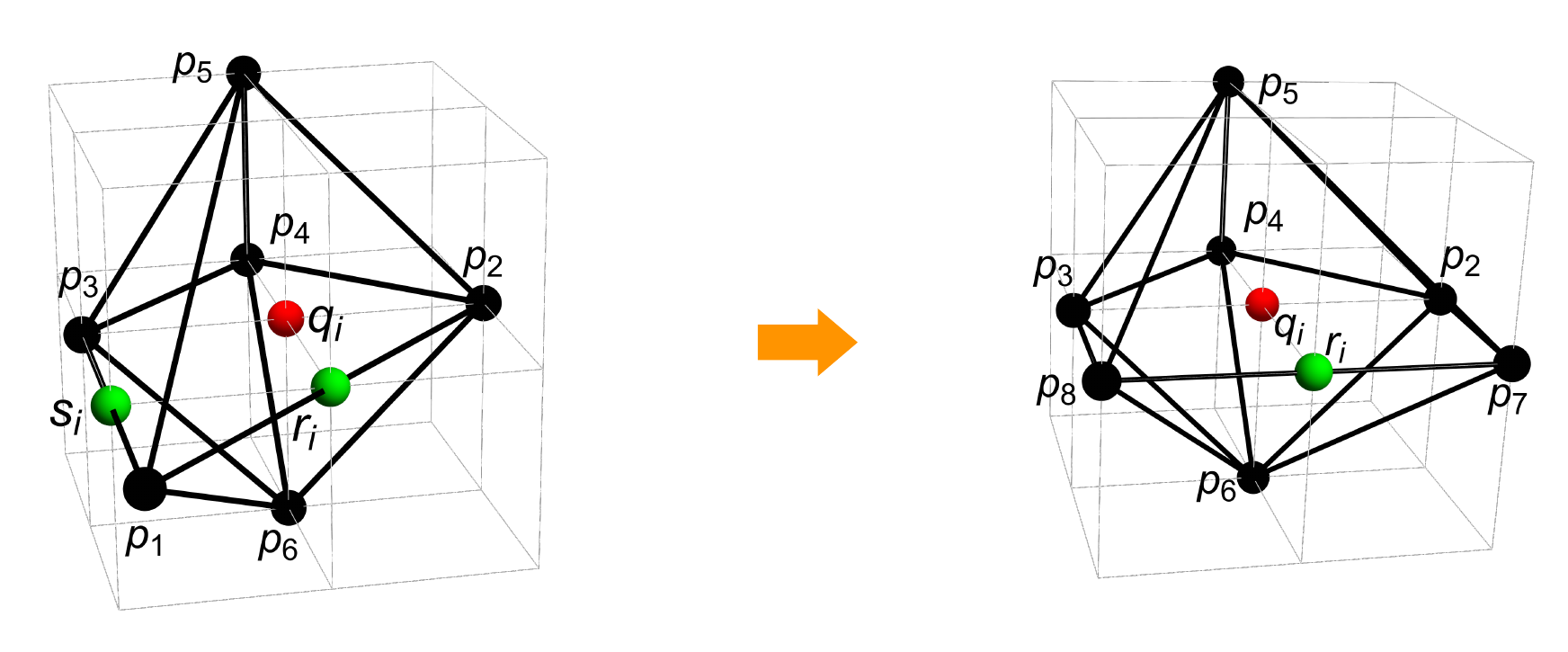}
\caption{Toric diagrams for $P_{+-}(\textrm{SPP}/\mathbb{Z}_2)$ and $P^2_{+-}(\textrm{PdP}_3)$. We have labeled points anticipating the corresponding brick matchings.}
	\label{toric_diagrams_example_1}
\end{figure}

\fref{toric_diagrams_example_1} shows the toric diagrams for the two geometries connected by the deformation, $P_{+-}(\textrm{SPP}/\mathbb{Z}_2)$ and $P^2_{+-}(\textrm{PdP}_3)$. The naming convention was first introduced in the classification of brane brick models corresponding to toric Fano 3-folds in \cite{Franco:2022gvl} and is based on a technique known as orbifold reduction \cite{Franco:2016fxm} that enables the construction of brane brick models from brane tilings associated to toric Calabi-Yau 3-folds. This method has been generalized to arbitrary toric Calabi-Yau 4-folds and is known as $3d$ printing \cite{Franco:2018qsc}.

\subsubsection*{Starting Point: a Gauge Theory for $P_{+-}(\textrm{SPP}/\mathbb{Z}_2)$}

\begin{figure}[H]
\begin{center}
\resizebox{0.6\hsize}{!}{
\includegraphics[height=5cm]{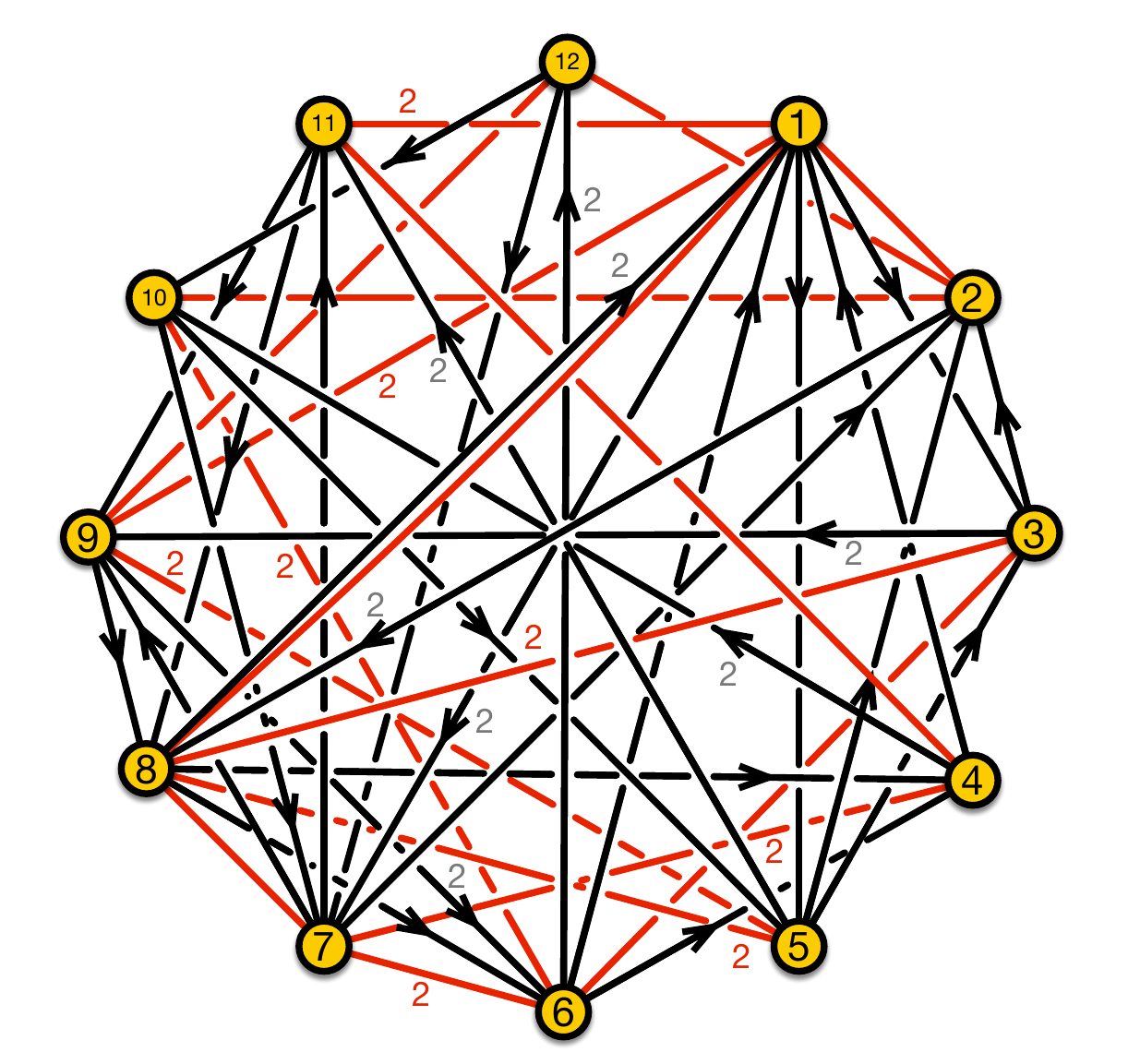}
}
\caption{Quiver diagram for a toric phase of $P_{+-}(\textrm{SPP}/\mathbb{Z}_2)$.
\label{fig_q_psppz2}}
 \end{center}
 \end{figure}

The quiver diagram for a toric phase of $P_{+-}(\textrm{SPP}/\mathbb{Z}_2)$ is shown in \fref{fig_q_psppz2}. We have constructed this theory using the $3d$ printing algorithm of \cite{Franco:2018qsc}. The analysis that we will present below will show that there is a toric phase for $P^2_{+-} (\textrm{PdP}_3)$ with the same quiver but different $J$- and $E$-terms.

The $J$- and $E$-terms for this theory are
\begin{align}
\resizebox{0.9\textwidth}{!}{$
\begin{array}{rcrclcrcl}
& &  \ \ \ \ \ \ \ \ \ \ \ \ \ \ \ \ \ & J& &&& E&  \ \ \ \ \ \ \ \ \ \ \ \ \ \  \\
 \Lambda_{21} : & \ \ \  & X_{13}\cdot X_{32} & - & X_{15}\cdot X_{52}& \ \ \ \  & P_{28}\cdot U_{81}& -& Q_{28}\cdot V_{81}\\
 \Lambda_{36} : & \ \ \  & X_{61}\cdot X_{15}\cdot X_{53} & - & X_{64}\cdot X_{41}\cdot X_{13}& \ \ \ \  & P_{39}\cdot U_{96}& -& Q_{39}\cdot V_{96}\\
 \Lambda_{87} : & \ \ \  & X_{79}\cdot X_{98} & - & X_{7,11}\cdot X_{11,8}& \ \ \ \  & V_{81}\cdot Q_{17}& -& U_{81}\cdot P_{17} \\
 \Lambda_{9,12} : & \ \ \  & X_{12,7}\cdot X_{7,11}\cdot X_{11,9} & - & X_{12,10}\cdot X_{10,7}\cdot X_{79}& \ \ \ \  &V_{96}\cdot Q_{6,12} & -& U_{96}\cdot P_{6,12}\\
 \Lambda_{18} : & \ \ \  & V_{84}\cdot X_{41} & - & V_{86}\cdot X_{61}& \ \ \ \  & P_{17}\cdot U_{72}\cdot Q_{28}& -& Q_{17}\cdot U_{72}\cdot P_{28}\\
 \Lambda_{4,11} : & \ \ \  & X_{11,8}\cdot V_{86}\cdot X_{64} & - & X_{11,9}\cdot X_{98}\cdot V_{84}& \ \ \ \  &P_{4,10}\cdot U_{10,5}\cdot Q_{5,11} & -&Q_{4,10}\cdot U_{10,5}\cdot P_{5,11} \\
 \Lambda_{2,10} : & \ \ \  & X_{10,7}\cdot U_{72}& - & U_{10,5}\cdot X_{53}\cdot X_{32}& \ \ \ \  & P_{28}\cdot V_{84}\cdot Q_{4,10}& -& Q_{28}\cdot V_{84}\cdot P_{4,10}\\
 \Lambda_{2,12} : & \ \ \  & X_{12,10}\cdot U_{10,5}\cdot X_{52} & - & X_{12,7}\cdot U_{72}& \ \ \ \  &P_{28}\cdot V_{86}\cdot Q_{6,12} & -& Q_{28}\cdot V_{86}\cdot P_{6,12}\\
 \Lambda^1_{19} : & \ \ \  & X_{98}\cdot U_{81}& - & U_{96}\cdot X_{64}\cdot X_{41}& \ \ \ \  & P_{17}\cdot X_{79}& -& X_{13}\cdot P_{39}\\
 \Lambda^2_{19} : & \ \ \  & X_{98}\cdot V_{81}& - & V_{96}\cdot X_{64}\cdot X_{41}& \ \ \ \  & X_{13}\cdot Q_{39}& -& Q_{17}\cdot X_{79}\\
 \Lambda^1_{38} : & \ \ \  & U_{81}\cdot X_{13}& - & V_{84}\cdot Q_{4,10}\cdot U_{10,5}\cdot X_{53}& \ \ \ \  & P_{39}\cdot X_{98}& -& X_{32}\cdot P_{28}\\
 \Lambda^2_{38} : & \ \ \  & V_{81}\cdot X_{13}& - & V_{84}\cdot P_{4,10}\cdot U_{10,5}\cdot X_{53}& \ \ \ \  & X_{32}\cdot Q_{28}& -& Q_{39}\cdot X_{98}\\
 \Lambda^1_{47} : & \ \ \  & U_{72}\cdot Q_{28}\cdot V_{84}& - & X_{79}\cdot U_{96}\cdot X_{64}& \ \ \ \  & P_{4,10}\cdot X_{10,7}& -& X_{41}\cdot P_{17}\\
 \Lambda^2_{47} : & \ \ \  & U_{72}\cdot P_{28}\cdot V_{84}& - & X_{79}\cdot V_{96}\cdot X_{64}& \ \ \ \  &X_{41}\cdot Q_{17}& -&  Q_{4,10}\cdot X_{10,7}\\
 \Lambda^1_{59} : & \ \ \  & U_{96}\cdot X_{61}\cdot X_{15}& - & X_{98}\cdot V_{84}\cdot Q_{4,10}\cdot U_{10,5}& \ \ \ \  & P_{5,11}\cdot X_{11,9}& -& X_{53}\cdot P_{39}\\
 \Lambda^2_{59} : & \ \ \  & V_{96}\cdot X_{61}\cdot X_{15}& - & X_{98}\cdot V_{84}\cdot P_{4,10}\cdot U_{10,5}& \ \ \ \  &X_{53}\cdot Q_{39} & -&Q_{5,11}\cdot X_{11,9} \\
\Lambda^1_{67} : & \ \ \  & X_{7,11}\cdot X_{11,9}\cdot U_{96}& - & U_{72}\cdot Q_{28}\cdot V_{86}& \ \ \ \  & P_{6,12}\cdot X_{12,7}& -& X_{61}\cdot P_{17}\\
\Lambda^2_{67} : & \ \ \  & X_{7,11}\cdot X_{11,9}\cdot V_{96}& - & U_{72}\cdot P_{28}\cdot V_{86} & \ \ \ \  & X_{61}\cdot Q_{17}& -& Q_{6,12}\cdot X_{12,7}\\
\Lambda^1_{1,11} : & \ \ \  & X_{11,9}\cdot U_{96}\cdot X_{61}& - & X_{11,8}\cdot U_{81}& \ \ \ \  & P_{17}\cdot X_{7,11} & -& X_{15}\cdot P_{5,11}\\
\Lambda^2_{1,11} : & \ \ \  & X_{11,9}\cdot V_{96}\cdot X_{61}& - & X_{11,8}\cdot V_{81}& \ \ \ \  & X_{15}\cdot Q_{5,11} & -& Q_{17}\cdot X_{7,11} \\
\Lambda^1_{6,10} : & \ \ \  & U_{10,5}\cdot X_{52}\cdot Q_{28}\cdot V_{86}& - & X_{10,7}\cdot X_{79}\cdot U_{96}& \ \ \ \  & P_{6,12}\cdot X_{12,10} & - & X_{64}\cdot P_{4,10}\\
\Lambda^2_{6,10} : & \ \ \  & U_{10,5}\cdot P_{5,11}\cdot X_{11,8}\cdot V_{86}& - & X_{10,7}\cdot X_{79}\cdot V_{96}& \ \ \ \  & X_{64}\cdot Q_{4,10} & - & Q_{6,12}\cdot X_{12,10}\\
\Lambda^1_{58} : & \ \ \  & V_{86}\cdot Q_{6,12}\cdot X_{12,10}\cdot U_{10,5}& - & U_{81}\cdot X_{15}& \ \ \ \  & P_{5,11}\cdot X_{11,8} & - & X_{52}\cdot P_{28}\\
\Lambda^2_{58} : & \ \ \  & V_{86}\cdot X_{64}\cdot P_{4,10}\cdot U_{10,5}& - & V_{81}\cdot X_{15}& \ \ \ \  & X_{52}\cdot Q_{28} & - & Q_{5,11}\cdot X_{11,8}\\
\end{array}
~.~ $}
\label{SPP(+-)}
\end{align}
From the quiver and $J$- and $E$-terms, we determine the following extended $\bar{P}$-matrix with an extra column showing the contribution of each field to the scaling dimension. The rightmost column is obtained by the procedure outlined in Section \ref{sec:vol}.
\begin{align}
\resizebox{13.5cm}{!}{$
\left(
\begin{array}{c|cccccc|cc|cc|cccccccccccccccccccc|c}
        ~ & p_1 & p_2 & p_3 & p_4 & p_5 & p_6 & s_1 & s_2 & r_1 & r_2 & q_1 & q_2 & q_3 & q_4 & q_5 & q_6 & q_7 & q_8 & q_9 & q_{10} & q_{11} & q_{12} & q_{13} & q_{14} & q_{15} & q_{16} & q_{17} & q_{18} & q_{19} & q_{20} & \Delta \\
         \hline
X_{13} & 0 & 1 & 0 & 0 & 0 & 0 & 0 & 0 & 1 & 0 & 0 & 1 & 0 & 1 & 0 & 0 & 1 & 1 & 1 & 1 & 1 & 0 & 0 & 0 & 0 & 0 & 0 & 0 & 0 & 0 & 0.391205133 \\
X_{79} & 0 & 1 & 0 & 0 & 0 & 0 & 0 & 0 & 1 & 0 & 0 & 1 & 0 & 1 & 0 & 0 & 0 & 0 & 0 & 1 & 0 & 1 & 1 & 0 & 1 & 1 & 0 & 0 & 0 & 0 & 0.391205133 \\
X_{32} & 1 & 0 & 0 & 0 & 0 & 0 & 0 & 1 & 0 & 1 & 0 & 0 & 1 & 0 & 0 & 0 & 0 & 0 & 0 & 0 & 0 & 1 & 1 & 1 & 0 & 0 & 0 & 0 & 0 & 0 & 0.443421069 \\
X_{98} & 1 & 0 & 0 & 0 & 0 & 0 & 0 & 1 & 0 & 1 & 0 & 0 & 1 & 0 & 0 & 0 & 0 & 0 & 0 & 0 & 0 & 0 & 0 & 0 & 0 & 0 & 1 & 0 & 0 & 0 & 0.443421069 \\
X_{41} & 1 & 0 & 0 & 0 & 0 & 0 & 1 & 0 & 0 & 1 & 0 & 0 & 0 & 0 & 1 & 0 & 0 & 0 & 0 & 0 & 0 & 0 & 0 & 0 & 0 & 0 & 0 & 0 & 0 & 1 & 0.443421069 \\
X_{10,7} & 1 & 0 & 0 & 0 & 0 & 0 & 1 & 0 & 0 & 1 & 0 & 0 & 0 & 0 & 1 & 0 & 0 & 0 & 1 & 0 & 1 & 0 & 0 & 1 & 0 & 0 & 0 & 0 & 0 & 0 & 0.443421069 \\
X_{53} & 0 & 0 & 1 & 0 & 0 & 0 & 1 & 0 & 0 & 0 & 0 & 0 & 0 & 1 & 0 & 0 & 0 & 0 & 0 & 1 & 1 & 0 & 0 & 0 & 0 & 0 & 0 & 0 & 0 & 0 & 0.265836439 \\
X_{11,9} & 0 & 0 & 1 & 0 & 0 & 0 & 1 & 0 & 0 & 0 & 0 & 0 & 0 & 1 & 0 & 0 & 0 & 0 & 0 & 0 & 0 & 0 & 1 & 0 & 0 & 1 & 0 & 0 & 0 & 0 & 0.265836439 \\
X_{61} & 0 & 1 & 0 & 0 & 0 & 0 & 0 & 0 & 0 & 1 & 0 & 0 & 0 & 0 & 1 & 1 & 0 & 0 & 0 & 0 & 0 & 0 & 0 & 0 & 0 & 0 & 0 & 1 & 1 & 1 & 0.391205133 \\
X_{12,7} & 0 & 1 & 0 & 0 & 0 & 0 & 0 & 0 & 0 & 1 & 0 & 0 & 0 & 0 & 1 & 1 & 0 & 1 & 1 & 0 & 1 & 0 & 0 & 1 & 0 & 0 & 0 & 0 & 0 & 1 & 0.391205133 \\
X_{15} & 1 & 0 & 0 & 0 & 0 & 0 & 0 & 1 & 1 & 0 & 0 & 1 & 0 & 0 & 0 & 0 & 1 & 1 & 1 & 0 & 0 & 0 & 0 & 0 & 0 & 0 & 0 & 0 & 0 & 0 & 0.443421069 \\
X_{7,11} & 1 & 0 & 0 & 0 & 0 & 0 & 0 & 1 & 1 & 0 & 0 & 1 & 0 & 0 & 0 & 0 & 0 & 0 & 0 & 1 & 0 & 1 & 0 & 0 & 1 & 0 & 0 & 0 & 0 & 0 & 0.443421069 \\
X_{64} & 0 & 0 & 1 & 0 & 0 & 0 & 0 & 1 & 0 & 0 & 0 & 0 & 0 & 0 & 0 & 1 & 0 & 0 & 0 & 0 & 0 & 0 & 0 & 0 & 0 & 0 & 0 & 1 & 1 & 0 & 0.265836439 \\
X_{12,10} & 0 & 0 & 1 & 0 & 0 & 0 & 0 & 1 & 0 & 0 & 0 & 0 & 0 & 0 & 0 & 1 & 0 & 1 & 0 & 0 & 0 & 0 & 0 & 0 & 0 & 0 & 0 & 0 & 0 & 1 & 0.265836439 \\
X_{52} & 0 & 1 & 0 & 0 & 0 & 0 & 0 & 0 & 0 & 1 & 0 & 0 & 1 & 1 & 0 & 0 & 0 & 0 & 0 & 1 & 1 & 1 & 1 & 1 & 0 & 0 & 0 & 0 & 0 & 0 & 0.391205133 \\
X_{11,8} & 0 & 1 & 0 & 0 & 0 & 0 & 0 & 0 & 0 & 1 & 0 & 0 & 1 & 1 & 0 & 0 & 0 & 0 & 0 & 0 & 0 & 0 & 1 & 0 & 0 & 1 & 1 & 0 & 0 & 0 & 0.391205133 \\
U_{81} & 0 & 0 & 1 & 1 & 0 & 1 & 1 & 0 & 0 & 0 & 1 & 0 & 0 & 0 & 1 & 1 & 0 & 0 & 0 & 0 & 0 & 0 & 0 & 0 & 0 & 0 & 0 & 1 & 1 & 1 & 0.801980737 \\
U_{96} & 0 & 0 & 0 & 1 & 0 & 1 & 0 & 0 & 0 & 0 & 1 & 0 & 1 & 0 & 0 & 0 & 0 & 0 & 0 & 0 & 0 & 0 & 0 & 0 & 0 & 0 & 1 & 0 & 0 & 0 & 0.536144298 \\
U_{72} & 0 & 0 & 1 & 1 & 0 & 0 & 0 & 1 & 0 & 0 & 0 & 1 & 1 & 1 & 0 & 0 & 0 & 0 & 0 & 1 & 0 & 1 & 1 & 0 & 0 & 0 & 0 & 1 & 0 & 0 & 0.438587676 \\
U_{10,5} & 0 & 0 & 0 & 1 & 0 & 0 & 0 & 0 & 0 & 0 & 0 & 1 & 0 & 0 & 1 & 0 & 0 & 0 & 1 & 0 & 0 & 0 & 0 & 0 & 0 & 0 & 0 & 0 & 0 & 0 & 0.172751237 \\
V_{81} & 0 & 0 & 1 & 1 & 1 & 0 & 1 & 0 & 0 & 0 & 1 & 0 & 0 & 0 & 1 & 1 & 0 & 0 & 0 & 0 & 0 & 0 & 0 & 0 & 0 & 0 & 0 & 1 & 1 & 1 & 0.801980737 \\
V_{96} & 0 & 0 & 0 & 1 & 1 & 0 & 0 & 0 & 0 & 0 & 1 & 0 & 1 & 0 & 0 & 0 & 0 & 0 & 0 & 0 & 0 & 0 & 0 & 0 & 0 & 0 & 1 & 0 & 0 & 0 & 0.536144298 \\
V_{84} & 0 & 1 & 0 & 0 & 0 & 0 & 0 & 0 & 1 & 0 & 1 & 0 & 0 & 0 & 0 & 1 & 0 & 0 & 0 & 0 & 0 & 0 & 0 & 0 & 0 & 0 & 0 & 1 & 1 & 0 & 0.391205133 \\
V_{86} & 1 & 0 & 0 & 0 & 0 & 0 & 1 & 0 & 1 & 0 & 1 & 0 & 0 & 0 & 0 & 0 & 0 & 0 & 0 & 0 & 0 & 0 & 0 & 0 & 0 & 0 & 0 & 0 & 0 & 0 & 0.443421069 \\
P_{17} & 0 & 0 & 0 & 0 & 1 & 0 & 0 & 0 & 0 & 0 & 0 & 0 & 0 & 0 & 0 & 0 & 1 & 1 & 1 & 0 & 1 & 0 & 0 & 1 & 0 & 0 & 0 & 0 & 0 & 0 & 0.363393061 \\
P_{28} & 0 & 0 & 0 & 0 & 1 & 0 & 0 & 0 & 0 & 0 & 0 & 0 & 0 & 0 & 0 & 0 & 0 & 0 & 0 & 0 & 0 & 0 & 0 & 0 & 1 & 1 & 1 & 0 & 0 & 0 & 0.363393061 \\
P_{39} & 0 & 0 & 0 & 0 & 1 & 0 & 0 & 0 & 0 & 0 & 0 & 0 & 0 & 0 & 0 & 0 & 0 & 0 & 0 & 0 & 0 & 1 & 1 & 1 & 1 & 1 & 0 & 0 & 0 & 0 & 0.363393061 \\
P_{4,10} & 0 & 0 & 0 & 0 & 1 & 0 & 0 & 0 & 0 & 0 & 0 & 0 & 0 & 0 & 0 & 0 & 1 & 1 & 0 & 0 & 0 & 0 & 0 & 0 & 0 & 0 & 0 & 0 & 1 & 1 & 0.363393061 \\
P_{5,11} & 0 & 0 & 0 & 0 & 1 & 0 & 0 & 0 & 0 & 0 & 0 & 0 & 0 & 0 & 0 & 0 & 0 & 0 & 0 & 1 & 1 & 1 & 0 & 1 & 1 & 0 & 0 & 0 & 0 & 0 & 0.363393061 \\
P_{6,12} & 0 & 0 & 0 & 0 & 1 & 0 & 0 & 0 & 0 & 0 & 0 & 0 & 0 & 0 & 0 & 0 & 1 & 0 & 0 & 0 & 0 & 0 & 0 & 0 & 0 & 0 & 0 & 1 & 1 & 0 & 0.363393061 \\
Q_{17} & 0 & 0 & 0 & 0 & 0 & 1 & 0 & 0 & 0 & 0 & 0 & 0 & 0 & 0 & 0 & 0 & 1 & 1 & 1 & 0 & 1 & 0 & 0 & 1 & 0 & 0 & 0 & 0 & 0 & 0 & 0.363393061 \\
Q_{28} & 0 & 0 & 0 & 0 & 0 & 1 & 0 & 0 & 0 & 0 & 0 & 0 & 0 & 0 & 0 & 0 & 0 & 0 & 0 & 0 & 0 & 0 & 0 & 0 & 1 & 1 & 1 & 0 & 0 & 0 & 0.363393061 \\
Q_{39} & 0 & 0 & 0 & 0 & 0 & 1 & 0 & 0 & 0 & 0 & 0 & 0 & 0 & 0 & 0 & 0 & 0 & 0 & 0 & 0 & 0 & 1 & 1 & 1 & 1 & 1 & 0 & 0 & 0 & 0 & 0.363393061 \\
Q_{4,10} & 0 & 0 & 0 & 0 & 0 & 1 & 0 & 0 & 0 & 0 & 0 & 0 & 0 & 0 & 0 & 0 & 1 & 1 & 0 & 0 & 0 & 0 & 0 & 0 & 0 & 0 & 0 & 0 & 1 & 1 & 0.363393061 \\
Q_{5,11} & 0 & 0 & 0 & 0 & 0 & 1 & 0 & 0 & 0 & 0 & 0 & 0 & 0 & 0 & 0 & 0 & 0 & 0 & 0 & 1 & 1 & 1 & 0 & 1 & 1 & 0 & 0 & 0 & 0 & 0 & 0.363393061 \\
Q_{6,12} & 0 & 0 & 0 & 0 & 0 & 1 & 0 & 0 & 0 & 0 & 0 & 0 & 0 & 0 & 0 & 0 & 1 & 0 & 0 & 0 & 0 & 0 & 0 & 0 & 0 & 0 & 0 & 1 & 1 & 0 & 0.363393061 \\
\hline
\Lambda_{21} & 0 & 0 & 1 & 1 & 1 & 1 & 1 & 0 & 0 & 0 & 1 & 0 & 0 & 0 & 1 & 1 & 1 & 0 & 0 & 0 & 0 & 0 & 0 & 0 & 1 & 1 & 0 & 1 & 1 & 1 & 1.165373797 \\
\Lambda_{36} & 0 & 0 & 0 & 1 & 1 & 1 & 0 & 0 & 0 & 0 & 1 & 0 & 1 & 0 & 0 & 0 & 0 & 0 & 0 & 0 & 0 & 1 & 1 & 1 & 1 & 1 & 1 & 0 & 0 & 0 & 0.899537359 \\
\Lambda_{87} & 0 & 0 & 1 & 1 & 1 & 1 & 1 & 0 & 0 & 0 & 1 & 0 & 0 & 0 & 1 & 1 & 1 & 1 & 1 & 0 & 1 & 0 & 0 & 1 & 0 & 0 & 0 & 1 & 1 & 1 & 1.165373797 \\
\Lambda_{9,12} & 0 & 0 & 0 & 1 & 1 & 1 & 0 & 0 & 0 & 0 & 1 & 0 & 1 & 0 & 0 & 0 & 1 & 0 & 0 & 0 & 0 & 0 & 0 & 0 & 0 & 0 & 1 & 1 & 1 & 0 & 0.899537359 \\
\Lambda_{18} & 0 & 0 & 1 & 1 & 1 & 1 & 0 & 1 & 0 & 0 & 0 & 1 & 1 & 1 & 0 & 0 & 1 & 1 & 1 & 1 & 1 & 1 & 1 & 1 & 1 & 1 & 1 & 1 & 0 & 0 & 1.165373797 \\
\Lambda_{4,11} & 0 & 0 & 0 & 1 & 1 & 1 & 0 & 0 & 0 & 0 & 0 & 1 & 0 & 0 & 1 & 0 & 1 & 1 & 1 & 0 & 1 & 1 & 0 & 1 & 1 & 0 & 0 & 0 & 1 & 1 & 0.899537359 \\
\Lambda_{2,10} & 0 & 1 & 0 & 0 & 1 & 1 & 0 & 0 & 1 & 0 & 1 & 0 & 0 & 0 & 0 & 1 & 1 & 1 & 0 & 0 & 0 & 0 & 0 & 0 & 1 & 1 & 1 & 1 & 1 & 1 & 1.117991255 \\
\Lambda_{2,12} & 1 & 0 & 0 & 0 & 1 & 1 & 1 & 0 & 1 & 0 & 1 & 0 & 0 & 0 & 0 & 0 & 1 & 0 & 0 & 0 & 0 & 0 & 0 & 0 & 1 & 1 & 1 & 1 & 1 & 0 & 1.170207191 \\
\Lambda^1_{19} & 0 & 1 & 0 & 0 & 1 & 0 & 0 & 0 & 1 & 0 & 0 & 1 & 0 & 1 & 0 & 0 & 1 & 1 & 1 & 1 & 1 & 1 & 1 & 1 & 1 & 1 & 0 & 0 & 0 & 0 & 0.754598194 \\
\Lambda^2_{19} & 0 & 1 & 0 & 0 & 0 & 1 & 0 & 0 & 1 & 0 & 0 & 1 & 0 & 1 & 0 & 0 & 1 & 1 & 1 & 1 & 1 & 1 & 1 & 1 & 1 & 1 & 0 & 0 & 0 & 0 & 0.754598194 \\
\Lambda^1_{38} & 1 & 0 & 0 & 0 & 1 & 0 & 0 & 1 & 0 & 1 & 0 & 0 & 1 & 0 & 0 & 0 & 0 & 0 & 0 & 0 & 0 & 1 & 1 & 1 & 1 & 1 & 1 & 0 & 0 & 0 & 0.80681413 \\
\Lambda^2_{38} & 1 & 0 & 0 & 0 & 0 & 1 & 0 & 1 & 0 & 1 & 0 & 0 & 1 & 0 & 0 & 0 & 0 & 0 & 0 & 0 & 0 & 1 & 1 & 1 & 1 & 1 & 1 & 0 & 0 & 0 & 0.80681413 \\
\Lambda^1_{47} & 1 & 0 & 0 & 0 & 1 & 0 & 1 & 0 & 0 & 1 & 0 & 0 & 0 & 0 & 1 & 0 & 1 & 1 & 1 & 0 & 1 & 0 & 0 & 1 & 0 & 0 & 0 & 0 & 1 & 1 & 0.80681413 \\
\Lambda^2_{47} & 1 & 0 & 0 & 0 & 0 & 1 & 1 & 0 & 0 & 1 & 0 & 0 & 0 & 0 & 1 & 0 & 1 & 1 & 1 & 0 & 1 & 0 & 0 & 1 & 0 & 0 & 0 & 0 & 1 & 1 & 0.80681413 \\
\Lambda^1_{59} & 0 & 0 & 1 & 0 & 1 & 0 & 1 & 0 & 0 & 0 & 0 & 0 & 0 & 1 & 0 & 0 & 0 & 0 & 0 & 1 & 1 & 1 & 1 & 1 & 1 & 1 & 0 & 0 & 0 & 0 & 0.629229499 \\
\Lambda^2_{59} & 0 & 0 & 1 & 0 & 0 & 1 & 1 & 0 & 0 & 0 & 0 & 0 & 0 & 1 & 0 & 0 & 0 & 0 & 0 & 1 & 1 & 1 & 1 & 1 & 1 & 1 & 0 & 0 & 0 & 0 & 0.629229499 \\
\Lambda^1_{67} & 0 & 1 & 0 & 0 & 1 & 0 & 0 & 0 & 0 & 1 & 0 & 0 & 0 & 0 & 1 & 1 & 1 & 1 & 1 & 0 & 1 & 0 & 0 & 1 & 0 & 0 & 0 & 1 & 1 & 1 & 0.754598194 \\
\Lambda^2_{67} & 0 & 1 & 0 & 0 & 0 & 1 & 0 & 0 & 0 & 1 & 0 & 0 & 0 & 0 & 1 & 1 & 1 & 1 & 1 & 0 & 1 & 0 & 0 & 1 & 0 & 0 & 0 & 1 & 1 & 1 & 0.754598194 \\
\Lambda^1_{1,11} & 1 & 0 & 0 & 0 & 1 & 0 & 0 & 1 & 1 & 0 & 0 & 1 & 0 & 0 & 0 & 0 & 1 & 1 & 1 & 1 & 1 & 1 & 0 & 1 & 1 & 0 & 0 & 0 & 0 & 0 & 0.80681413 \\
\Lambda^2_{1,11} & 1 & 0 & 0 & 0 & 0 & 1 & 0 & 1 & 1 & 0 & 0 & 1 & 0 & 0 & 0 & 0 & 1 & 1 & 1 & 1 & 1 & 1 & 0 & 1 & 1 & 0 & 0 & 0 & 0 & 0 & 0.80681413\\
\Lambda^1_{6,10} & 0 & 0 & 1 & 0 & 1 & 0 & 0 & 1 & 0 & 0 & 0 & 0 & 0 & 0 & 0 & 1 & 1 & 1 & 0 & 0 & 0 & 0 & 0 & 0 & 0 & 0 & 0 & 1 & 1 & 1 & 0.629229499 \\
\Lambda^2_{6,10} & 0 & 0 & 1 & 0 & 0 & 1 & 0 & 1 & 0 & 0 & 0 & 0 & 0 & 0 & 0 & 1 & 1 & 1 & 0 & 0 & 0 & 0 & 0 & 0 & 0 & 0 & 0 & 1 & 1 & 1 & 0.629229499 \\
\Lambda^1_{58} & 0 & 1 & 0 & 0 & 1 & 0 & 0 & 0 & 0 & 1 & 0 & 0 & 1 & 1 & 0 & 0 & 0 & 0 & 0 & 1 & 1 & 1 & 1 & 1 & 1 & 1 & 1 & 0 & 0 & 0 & 0.754598194 \\
\Lambda^2_{58} & 0 & 1 & 0 & 0 & 0 & 1 & 0 & 0 & 0 & 1 & 0 & 0 & 1 & 1 & 0 & 0 & 0 & 0 & 0 & 1 & 1 & 1 & 1 & 1 & 1 & 1 & 1 & 0 & 0 & 0 & 0.754598194 \\
   \end{array}
\right)~.~$}
\label{P_matrix_SPP(+-)}
\end{align}
In \fref{toric_diagrams_example_1}, we have indicated the point in the toric diagram associated with each of these brick matchings.  The complete forward algorithm introduced in \cite{Franco:2015tna} confirms that the gauge theory has the desired geometry as its mesonic moduli space.

The volume of the base SE$_7$ manifold of $P_{+-} (\textrm{SPP} / \IZ_2 ) $ geometry is
\begin{align}
   \textrm{Vol} (P_{+-}(\textrm{SPP}/\IZ_2))=  3.75366 \, .
\label{Vol_SPP_Z2}
\end{align}
Note that here and throughout the paper, we will often refer to the CY$_4$ and its SE$_7$ base using 
the same name. Their distinction would be clear in the context.
\begin{table}[ht!]
\centering
\begin{tabular}{|c|c|l|}
\hline
\; &  $U(1)_R$ & fugacity \\
\hline
$p_1$ & $ r_3  $ &  $t_1 = \overline{t}_3 $ \\
$p_2$ & $ r_1 + r_3  $ &  $t_2 = \overline{t}_1 \overline{t}_3 $ \\
$p_3$ & $ r_1 $ &  $t_3 = \overline{t}_1  $ \\
$p_4$ & $ r_4  $ &  $t_4 = \overline{t}_4  $ \\
$p_5$ & $ r_2  $ &  $t_5 = \overline{t}_2  $ \\
$p_6$ & $ r_2 $ &  $t_6 = \overline{t}_2  $ \\
\hline
\end{tabular}
\caption{Charges under the $U(1)_R$ symmetry of the $P_{+-} ( \textrm{SPP} / \IZ ) $ model of the extremal GLSM fields $p_a$.
Here, $U(1)_R$ charges $r_1 \,, r_2 \,, r_3 $ and $r_4 $ are chosen such that
the $J$- and $E$-terms coupled to Fermi fields have an overall $U(1)_R$ charge of $2$
with $2r_1 + 2r_2 + 2r_3 + r_4 = 2 \,.$
 \label{tab_10}}
\end{table}

The $U(1)_R$ charges are summarized in \tref{tab_10}. The Hilbert series refined only under the $U(1)_R$ symmetry using the fugacities in \tref{tab_10} takes the following form,
\beal{HS10}
g(\bar{t}_1, \dots, \bar{t}_4 ; P_{+-} (\textrm{SPP}/\IZ_2) ) = 
\frac{P(\bar{t}_1, \dots, \bar{t}_4)}{
(1- \bar{t}_1^2 \bar{t}_2 \bar{t}_3^4)^3 (1- \bar{t}_1^2 \bar{t}_2^2 \bar{t}_3^2 \bar{t}_4)^2 (1- \bar{t}_1^2 \bar{t}_2^3 \bar{t}_4^2)^2
}~,~
\eea
where
the numerator is given by,
\beal{HS10b}
&&
P(\bar{t}_1, \dots, \bar{t}_4) =
1 + 3 \bar{t}_1^2 \bar{t}_2 \bar{t}_3^4 + 7 \bar{t}_1^2 \bar{t}_2^2 \bar{t}_3^2 \bar{t}_4 - 13 \bar{t}_1^4 \bar{t}_2^3 \bar{t}_3^6 \bar{t}_4 +  2 \bar{t}_1^6 \bar{t}_2^4 \bar{t}_3^{10} \bar{t}_4 + 2 \bar{t}_1^2 \bar{t}_2^3 \bar{t}_4^2 - 10 \bar{t}_1^4 \bar{t}_2^4 \bar{t}_3^4 \bar{t}_4^2 
\nn\\
&&
\hspace{0.5cm}
-  4 \bar{t}_1^6 \bar{t}_2^5 \bar{t}_3^8 \bar{t}_4^2 + 4 \bar{t}_1^8 \bar{t}_2^6 \bar{t}_3^{12} \bar{t}_4^2 -  4 \bar{t}_1^4 \bar{t}_2^5 \bar{t}_3^2 \bar{t}_4^3 + 4 \bar{t}_1^6 \bar{t}_2^6 \bar{t}_3^6 \bar{t}_4^3 +  10 \bar{t}_1^8 \bar{t}_2^7 \bar{t}_3^{10} \bar{t}_4^3 - 2 \bar{t}_1^{10} \bar{t}_2^8 \bar{t}_3^{14} \bar{t}_4^3 
\nn\\
&&
\hspace{0.5cm}
-  2 \bar{t}_1^6 \bar{t}_2^7 \bar{t}_3^4 \bar{t}_4^4  + 13 \bar{t}_1^8 \bar{t}_2^8 \bar{t}_3^8 \bar{t}_4^4 -  7 \bar{t}_1^{10} \bar{t}_2^9 \bar{t}_3^{12} \bar{t}_4^4 - 3 \bar{t}_1^{10} \bar{t}_2^{10} \bar{t}_3^{10} \bar{t}_4^5 -  \bar{t}_1^{12} \bar{t}_2^{11} \bar{t}_3^{14} \bar{t}_4^5
~.~
\eea

\subsubsection*{Relevant Non-Mass Deformation to $P_{+-}^2 (\textrm{PdP}_3 )$}

Let us now consider the following deformation of the $J$-terms, where the new contributions are indicated in blue,
\begin{align}
\resizebox{\textwidth}{!}{$
\begin{array}{rcrclcrcl}
& &  \ \ \ \ \ \ \ \ \ \ \ \ \ \ \ \ \ & J& + \, \textcolor{blue}{\Delta J} &&& E&  \ \ \ \ \ \ \ \ \ \ \ \ \ \  \\
\Lambda_{36} : & \ \ \  & X_{61}\cdot X_{15}\cdot X_{53} & - & X_{64}\cdot X_{41}\cdot X_{13}-\textcolor{blue}{\mu \, X_{61}\cdot X_{13}}& \ \ \ \  & P_{39}\cdot U_{96}& -& Q_{39}\cdot V_{96}\\
\Lambda_{9,12} : & \ \ \  & X_{12,7}\cdot X_{7,11}\cdot X_{11,9} & - & X_{12,10}\cdot X_{10,7}\cdot X_{79}-\textcolor{blue}{\mu \, X_{12,7}\cdot X_{79}}& \ \ \ \  &V_{96}\cdot Q_{6,12} & -& U_{96}\cdot P_{6,12}\\
\Lambda_{4,11} : & \ \ \  & X_{11,8}\cdot V_{86}\cdot X_{64} & - & X_{11,9}\cdot X_{98}\cdot V_{84}+\textcolor{blue}{\mu \, X_{11,8}\cdot V_{84}}& \ \ \ \  & Q_{4,10}\cdot U_{10,5}\cdot P_{5,11}& -& P_{4,10}\cdot U_{10,5}\cdot Q_{5,11}\\
 \Lambda_{2,10} : & \ \ \  & X_{10,7}\cdot U_{72}& - & U_{10,5}\cdot X_{53}\cdot X_{32}+\textcolor{blue}{\mu \, U_{10,5}\cdot X_{52}}& \ \ \ \  & P_{28}\cdot V_{84}\cdot Q_{4,10}& -& Q_{28}\cdot V_{84}\cdot P_{4,10}\\
 \Lambda^1_{19} : & \ \ \  & X_{98}\cdot U_{81}& - & U_{96}\cdot X_{64}\cdot X_{41}-\textcolor{blue}{\mu \, U_{96}\cdot X_{61}}& \ \ \ \  & P_{17}\cdot X_{79}& -& X_{13}\cdot P_{39}\\
 \Lambda^2_{19} : & \ \ \  & X_{98}\cdot V_{81}& - & V_{96}\cdot X_{64}\cdot X_{41}-\textcolor{blue}{\mu \, V_{96}\cdot X_{61}}& \ \ \ \  & X_{13}\cdot Q_{39}& -& Q_{17}\cdot X_{79}\\
\Lambda^1_{67} : & \ \ \  & X_{7,11}\cdot X_{11,9}\cdot U_{96}& - & U_{72}\cdot Q_{28}\cdot V_{86}-\textcolor{blue}{\mu \, X_{79}\cdot U_{96}} & \ \ \ \  & P_{6,12}\cdot X_{12,7}& -& X_{61}\cdot P_{17}\\
\Lambda^2_{67} : & \ \ \  & X_{7,11}\cdot X_{11,9}\cdot V_{96}& - & U_{72}\cdot P_{28}\cdot V_{86}-\textcolor{blue}{\mu \, X_{79}\cdot V_{96}} & \ \ \ \  & X_{61}\cdot Q_{17}& -& Q_{6,12}\cdot X_{12,7}\\
\Lambda^1_{58} : & \ \ \  & V_{86}\cdot Q_{6,12}\cdot X_{12,10}\cdot U_{10,5}& - & U_{81}\cdot X_{15}+\textcolor{blue}{\mu \, V_{84}\cdot Q_{4,10}\cdot U_{10,5}}& \ \ \ \  & P_{5,11}\cdot X_{11,8} & - & X_{52}\cdot P_{28}\\
\Lambda^2_{58} : & \ \ \  & V_{86}\cdot X_{64}\cdot P_{4,10}\cdot U_{10,5}& - & V_{81}\cdot X_{15}+\textcolor{blue}{\mu \, V_{84}\cdot P_{4,10}\cdot U_{10,5}}& \ \ \ \  & X_{52}\cdot Q_{28} & - & Q_{5,11}\cdot X_{11,8}\\
\end{array}
~.~ $}
\label{deformationSPP(+-)}
\end{align}
Here, we only list the Fermi fields whose interactions are deformed. Since the deformation involves only cubic and quartic plaquettes, no fields become massive. Therefore, the quiver of the theory remains the same as the one in \fref{fig_q_psppz2}. As sketched in Section \ref{sec:vol}, we use the extended $\bar{P}$-matrix in \eqref{P_matrix_SPP(+-)} to determine that all the deformation plaquettes have the same extremal GLSM field content; $p_1 \cdot p_2 \cdot p_3^2 \cdot p_5$. 
Using the scaling dimension rule in \eqref{scaledefm}, we find that these plaquettes have the following scaling dimension $\Delta [ \Lambda \cdot \Delta J ] \simeq 1.68195 < 2$, which implies that the deformation in \eqref{deformationSPP(+-)} is relevant.

Next, we introduce the following change of field variables,
\begin{align}
  X_{52}\to & -\frac{1}{\mu}X_{52} + \frac{1}{\mu}X_{53}\cdot X_{32} \,, \nonumber \\
  X_{11,8}\to & -\frac{1}{\mu}X_{11,8} + \frac{1}{\mu}X_{11,9}\cdot X_{98} \,, \nonumber \\
  X_{13}\to & -\frac{1}{\mu}X_{13} + \frac{1}{\mu}X_{15}\cdot X_{53} \,, \nonumber \\
  X_{79}\to & -\frac{1}{\mu}X_{79} + \frac{1}{\mu}X_{7,11}\cdot X_{11,9} \,, \nonumber \\
  X_{61}\to & \frac{1}{\mu}X_{61} - \frac{1}{\mu}X_{64}\cdot X_{41} \,, \nonumber \\
  X_{12,7}\to & \frac{1}{\mu}X_{12,7} - \frac{1}{\mu}X_{12,10}\cdot X_{10,7} \,, \nonumber \\
  V_{84}\to & \frac{1}{\mu}V_{84} - \frac{1}{\mu}V_{86}\cdot X_{64} \,, \nonumber \\
  \Lambda_{2,10}\to & \frac{1}{\mu}\Lambda_{2,10} + \frac{1}{\mu}\Lambda_{2,12}\cdot X_{12,10} - \frac{1}{\mu}Q_{28}\cdot V_{86}\cdot\Lambda^1_{6,10}- \frac{1}{\mu}P_{28}\cdot V_{86}\cdot\Lambda^2_{6,10} \,, \nonumber \\
  \Lambda^1_{58}\to & -\frac{1}{\mu}\Lambda^{1}_{58} + \frac{1}{\mu}\Lambda^1_{59}\cdot X_{98} + \frac{1}{\mu}X_{53}\cdot\Lambda^1_{38} \,, \nonumber \\
  \Lambda^2_{58}\to & -\frac{1}{\mu}\Lambda^{2}_{58} + \frac{1}{\mu}\Lambda^2_{59}\cdot X_{98} + \frac{1}{\mu}X_{53}\cdot\Lambda^2_{38} \,, \nonumber \\
  \Lambda^1_{19}\to & -\frac{1}{\mu}\Lambda^{1}_{19} + \frac{1}{\mu}\Lambda^1_{1,11}\cdot X_{11,9} + \frac{1}{\mu}X_{15}\cdot\Lambda^1_{59} \,, \nonumber \\
  \Lambda^2_{19}\to & -\frac{1}{\mu}\Lambda^{2}_{19} + \frac{1}{\mu}\Lambda^2_{1,11}\cdot X_{11,9} + \frac{1}{\mu}X_{15}\cdot\Lambda^2_{59} \,, \nonumber \\
  \Lambda^1_{67}\to & \frac{1}{\mu}\Lambda^{1}_{67} - \frac{1}{\mu}\Lambda^1_{6,10}\cdot X_{10,7} - \frac{1}{\mu}X_{64}\cdot\Lambda^1_{47} \,, \nonumber \\
  \Lambda^2_{67}\to & \frac{1}{\mu}\Lambda^{2}_{67} - \frac{1}{\mu}\Lambda^2_{6,10}\cdot X_{10,7} - \frac{1}{\mu}X_{64}\cdot\Lambda^2_{47} \,. 
\label{dP3redefs}
\end{align}

Some of the $J$-terms remain non-toric after the change of variables. Nevertheless, their binomial property is restored by applying the vanishing $E$-term relations of other Fermi fields. Note that the quiver diagram remains the same as in \fref{fig_q_psppz2}, and the $J$- and $E$-terms become
\begin{align}
\resizebox{\textwidth}{!}{$
\begin{array}{rcrclcrcl}
& &  \ \ \ \ \ \ \ \ \ \ \ \ \ \ \ \ \ & J& &&& E&  \ \ \ \ \ \ \ \ \ \ \ \ \ \  \\
 \Lambda_{21} : & \ \ \  & X_{15}\cdot X'_{52} & - & X'_{13}\cdot X_{32}& \ \ \ \  & P_{28}\cdot U_{81}& -& Q_{28}\cdot V_{81}\\
 \Lambda_{36} : & \ \ \  & X'_{61}\cdot X'_{13} & - & X_{64}\cdot X_{41}\cdot X_{15}\cdot X_{53}& \ \ \ \  & P_{39}\cdot U_{96}& -& Q_{39}\cdot V_{96}\\
 \Lambda_{87} : & \ \ \  & X_{7,11}\cdot X'_{11,8} & - & X'_{79}\cdot X_{98}& \ \ \ \  & V_{81}\cdot Q_{17} & -&  U_{81}\cdot P_{17}\\
 \Lambda_{9,12} : & \ \ \  & X'_{12,7}\cdot X'_{79} & - & X_{12,10}\cdot X_{10,7}\cdot X_{7,11}\cdot X_{11,9}& \ \ \ \  & V_{96}\cdot Q_{6,12}& -& U_{96}\cdot P_{6,12}\\
 \Lambda_{18} : & \ \ \  & V'_{84}\cdot X_{41} & - & V_{86}\cdot X'_{61}& \ \ \ \  & P_{17}\cdot U_{72}\cdot Q_{28} & -& Q_{17}\cdot U_{72}\cdot P_{28} \\
 \Lambda_{4,11} : & \ \ \  & X_{11,9}\cdot X_{98}\cdot V_{86}\cdot X_{64}  & - & X'_{11,8}\cdot V'_{84}& \ \ \ \  & P_{4,10}\cdot U_{10,5}\cdot Q_{5,11} & -& Q_{4,10}\cdot U_{10,5}\cdot P_{5,11} \\
 \Lambda'_{2,10} : & \ \ \  & X_{10,7}\cdot U_{72}& - & U_{10,5}\cdot X'_{52}& \ \ \ \  & P_{28}\cdot V'_{84}\cdot Q_{4,10}& -& Q_{28}\cdot V'_{84}\cdot P_{4,10}\\
 \Lambda_{2,12} : & \ \ \  & X_{12,10}\cdot U_{10,5}\cdot X_{53}\cdot X_{32} & - & X'_{12,7}\cdot U_{72}& \ \ \ \  & P_{28}\cdot V_{86}\cdot Q_{6,12} & -& Q_{28}\cdot V_{86}\cdot P_{6,12} \\
  \Lambda^{1'}_{19} : & \ \ \  & U_{96}\cdot X'_{61}& - & X_{98}\cdot U_{81}& \ \ \ \  & P_{17}\cdot X'_{79}& -& X'_{13}\cdot P_{39}\\
 \Lambda^{2'}_{19} : & \ \ \  & V_{96}\cdot X'_{61}& - & X_{98}\cdot V_{81}& \ \ \ \  & X'_{13}\cdot Q_{39}& -& Q_{17}\cdot X'_{79}\\
 \Lambda^1_{38} : & \ \ \  & V_{86}\cdot Q_{6,12}\cdot X_{12,10}\cdot U_{10,5}\cdot X_{53}& - & U_{81}\cdot X'_{13}& \ \ \ \  & P_{39}\cdot X_{98}& -& X_{32}\cdot P_{28}\\
 \Lambda^2_{38} : & \ \ \  & V_{86}\cdot X_{64}\cdot P_{4,10}\cdot U_{10,5}\cdot X_{53}& - & V_{81}\cdot X'_{13}& \ \ \ \  & X_{32}\cdot Q_{28}& -& Q_{39}\cdot X_{98}\\
 \Lambda^1_{47} : & \ \ \  & U_{72}\cdot Q_{28}\cdot V'_{84}& - & X_{7,11}\cdot X_{11,9}\cdot U_{96}\cdot X_{64}& \ \ \ \  & P_{4,10}\cdot X_{10,7}& -& X_{41}\cdot P_{17}\\
 \Lambda^2_{47} : & \ \ \  & U_{72}\cdot P_{28}\cdot V'_{84}& - & X_{7,11}\cdot X_{11,9}\cdot V_{96}\cdot X_{64}& \ \ \ \  & X_{41}\cdot Q_{17}& -&  Q_{4,10}\cdot X_{10,7}\\
  \Lambda^1_{59} : & \ \ \  & X_{98}\cdot V_{86}\cdot Q_{6,12}\cdot X_{12,10}\cdot U_{10,5}& - & U_{96}\cdot X_{64}\cdot X_{41}\cdot X_{15}& \ \ \ \  & P_{5,11}\cdot X_{11,9}& -& X_{53}\cdot P_{39}\\
 \Lambda^2_{59} : & \ \ \  & X_{98}\cdot V_{86}\cdot X_{64}\cdot P_{4,10}\cdot U_{10,5}& - & V_{96}\cdot X_{64}\cdot X_{41}\cdot X_{15}& \ \ \ \  &X_{53}\cdot Q_{39} & -&Q_{5,11}\cdot X_{11,9} \\
 \Lambda^{1'}_{67} : & \ \ \  & X'_{79}\cdot U_{96}& - & U_{72}\cdot Q_{28}\cdot V_{86} & \ \ \ \  & P_{6,12}\cdot X'_{12,7} & -& X'_{61}\cdot P_{17}\\
\Lambda^{2'}_{67} : & \ \ \  & X'_{79}\cdot V_{96}& - & U_{72}\cdot P_{28}\cdot V_{86} & \ \ \ \  & X'_{61}\cdot Q_{17} & - & Q_{6,12}\cdot X'_{12,7}\\
\Lambda^1_{1,11} : & \ \ \  & X'_{11,8}\cdot U_{81}& - & X_{11,9}\cdot U_{96}\cdot X_{64}\cdot X_{41}& \ \ \ \  & P_{17}\cdot X_{7,11} & -& X_{15}\cdot P_{5,11}\\
\Lambda^2_{1,11} : & \ \ \  & X'_{11,8}\cdot V_{81}& - & X_{11,9}\cdot V_{96}\cdot X_{64}\cdot X_{41}& \ \ \ \  & X_{15}\cdot Q_{5,11} & -& Q_{17}\cdot X_{7,11} \\
\Lambda^1_{6,10} : & \ \ \  & U_{10,5}\cdot X_{53}\cdot X_{32}\cdot Q_{28}\cdot V_{86}& - & X_{10,7}\cdot X_{7,11}\cdot X_{11,9}\cdot U_{96}& \ \ \ \  & P_{6,12}\cdot X_{12,10} & - & X_{64}\cdot P_{4,10}\\
\Lambda^2_{6,10} : & \ \ \  & U_{10,5}\cdot P_{5,11}\cdot X_{11,9}\cdot X_{98}\cdot V_{86}& - & X_{10,7}\cdot X_{7,11}\cdot X_{11,9}\cdot V_{96}& \ \ \ \  & X_{64}\cdot Q_{4,10} & - & Q_{6,12}\cdot X_{12,10}\\
\Lambda^{1'}_{58} : & \ \ \  & U_{81}\cdot X_{15}& - & V'_{84}\cdot Q_{4,10}\cdot U_{10,5}& \ \ \ \  & P_{5,11}\cdot X'_{11,8} & - & X'_{52}\cdot P_{28}\\
\Lambda^{2'}_{58} : & \ \ \  & V_{81}\cdot X_{15}& - & V'_{84}\cdot P_{4,10}\cdot U_{10,5}& \ \ \ \  & X'_{52}\cdot Q_{28} & - & Q_{5,11}\cdot X'_{11,8}\\
\end{array}
\,. $} 
\label{PdP3ir(+-)}
\end{align}
The extended $\bar{P}$-matrix is as follows, with an extra column for the scaling dimension of chiral fields,
\begin{align}
\resizebox{13.2cm}{!}{$
\left(
\begin{array}{c|ccccccc|cc|cccccccccccccccccccc|c}
        \; & p_2 & p_3 & p_4 & p_5 & p_6 & p_7 & p_8 & r_1 & r_2 & q_1 & q_2 & q_3 & q_4 & q_5 & q_6 & q_7 & q_8 & q_9 & q_{10} & q_{11} & q_{12} & q_{13} & q_{14} & q_{15} & q_{16} & q_{17} & q_{18} & q_{19} & q_{20} & \Delta \\
         \hline
        X_{13} & 1 & 0 & 0 & 0 & 0 & 1 & 0 & 1 & 0 & 0 & 1 & 0 & 1 & 0 & 0 & 1 & 1 & 1 & 1 & 1 & 0 & 0 & 0 & 0 & 0 & 0 & 0 & 0 & 0 & 0.535935531 \\
        X_{79} & 1 & 0 & 0 & 0 & 0 & 1 & 0 & 1 & 0 & 0 & 1 & 0 & 1 & 0 & 0 & 0 & 0 & 0 & 1 & 0 & 1 & 1 & 0 & 1 & 1 & 0 & 0 & 0 & 0 & 0.535935531 \\
        X_{32} & 0 & 0 & 0 & 0 & 0 & 0 & 1 & 0 & 1 & 0 & 0 & 1 & 0 & 0 & 0 & 0 & 0 & 0 & 0 & 0 & 1 & 1 & 1 & 0 & 0 & 0 & 0 & 0 & 0 & 0.36448494 \\
        X_{98} & 0 & 0 & 0 & 0 & 0 & 0 & 1 & 0 & 1 & 0 & 0 & 1 & 0 & 0 & 0 & 0 & 0 & 0 & 0 & 0 & 0 & 0 & 0 & 0 & 0 & 1 & 0 & 0 & 0 & 0.36448494 \\
        X_{41} & 0 & 0 & 0 & 0 & 0 & 1 & 0 & 0 & 1 & 0 & 0 & 0 & 0 & 1 & 0 & 0 & 0 & 0 & 0 & 0 & 0 & 0 & 0 & 0 & 0 & 0 & 0 & 0 & 1 & 0.364484937 \\
        X_{10,7} & 0 & 0 & 0 & 0 & 0 & 1 & 0 & 0 & 1 & 0 & 0 & 0 & 0 & 1 & 0 & 0 & 0 & 1 & 0 & 1 & 0 & 0 & 1 & 0 & 0 & 0 & 0 & 0 & 0 & 0.364484937 \\
        X_{53} & 0 & 1 & 0 & 0 & 0 & 0 & 0 & 0 & 0 & 0 & 0 & 0 & 1 & 0 & 0 & 0 & 0 & 0 & 1 & 1 & 0 & 0 & 0 & 0 & 0 & 0 & 0 & 0 & 0 & 0.200199517 \\
        X_{11,9} & 0 & 1 & 0 & 0 & 0 & 0 & 0 & 0 & 0 & 0 & 0 & 0 & 1 & 0 & 0 & 0 & 0 & 0 & 0 & 0 & 0 & 1 & 0 & 0 & 1 & 0 & 0 & 0 & 0 & 0.200199517 \\
        X_{61} & 0 & 1 & 0 & 0 & 0 & 0 & 1 & 0 & 1 & 0 & 0 & 0 & 0 & 1 & 1 & 0 & 0 & 0 & 0 & 0 & 0 & 0 & 0 & 0 & 0 & 0 & 1 & 1 & 1 & 0.564684457 \\
        X_{12,7} & 0 & 1 & 0 & 0 & 0 & 0 & 1 & 0 & 1 & 0 & 0 & 0 & 0 & 1 & 1 & 0 & 1 & 1 & 0 & 1 & 0 & 0 & 1 & 0 & 0 & 0 & 0 & 0 & 1 & 0.564684457 \\
        X_{15} & 0 & 0 & 0 & 0 & 0 & 0 & 1 & 1 & 0 & 0 & 1 & 0 & 0 & 0 & 0 & 1 & 1 & 1 & 0 & 0 & 0 & 0 & 0 & 0 & 0 & 0 & 0 & 0 & 0 & 0.36448494 \\
        X_{7,11} & 0 & 0 & 0 & 0 & 0 & 0 & 1 & 1 & 0 & 0 & 1 & 0 & 0 & 0 & 0 & 0 & 0 & 0 & 1 & 0 & 1 & 0 & 0 & 1 & 0 & 0 & 0 & 0 & 0 & 0.36448494 \\
        X_{64} & 1 & 0 & 0 & 0 & 0 & 0 & 0 & 0 & 0 & 0 & 0 & 0 & 0 & 0 & 1 & 0 & 0 & 0 & 0 & 0 & 0 & 0 & 0 & 0 & 0 & 0 & 1 & 1 & 0 & 0.171450594 \\
        X_{12,10} & 1 & 0 & 0 & 0 & 0 & 0 & 0 & 0 & 0 & 0 & 0 & 0 & 0 & 0 & 1 & 0 & 1 & 0 & 0 & 0 & 0 & 0 & 0 & 0 & 0 & 0 & 0 & 0 & 1 & 0.171450594 \\
        X_{52} & 1 & 0 & 0 & 0 & 0 & 1 & 0 & 0 & 1 & 0 & 0 & 1 & 1 & 0 & 0 & 0 & 0 & 0 & 1 & 1 & 1 & 1 & 1 & 0 & 0 & 0 & 0 & 0 & 0 & 0.535935531 \\
        X_{11,8} & 1 & 0 & 0 & 0 & 0 & 1 & 0 & 0 & 1 & 0 & 0 & 1 & 1 & 0 & 0 & 0 & 0 & 0 & 0 & 0 & 0 & 1 & 0 & 0 & 1 & 1 & 0 & 0 & 0 & 0.535935531 \\
        U_{81} & 0 & 1 & 1 & 0 & 1 & 0 & 0 & 0 & 0 & 1 & 0 & 0 & 0 & 1 & 1 & 0 & 0 & 0 & 0 & 0 & 0 & 0 & 0 & 0 & 0 & 0 & 1 & 1 & 1 & 0.735614816 \\
        U_{96} & 0 & 0 & 1 & 0 & 1 & 0 & 0 & 0 & 0 & 1 & 0 & 1 & 0 & 0 & 0 & 0 & 0 & 0 & 0 & 0 & 0 & 0 & 0 & 0 & 0 & 1 & 0 & 0 & 0 & 0.535415299 \\
        U_{72} & 1 & 0 & 1 & 0 & 0 & 0 & 0 & 0 & 0 & 0 & 1 & 1 & 1 & 0 & 0 & 0 & 0 & 0 & 1 & 0 & 1 & 1 & 0 & 0 & 0 & 0 & 1 & 0 & 0 & 0.342901188 \\
        U_{10,5} & 0 & 0 & 1 & 0 & 0 & 0 & 0 & 0 & 0 & 0 & 1 & 0 & 0 & 1 & 0 & 0 & 0 & 1 & 0 & 0 & 0 & 0 & 0 & 0 & 0 & 0 & 0 & 0 & 0 & 0.171450594 \\
        V_{81} & 0 & 1 & 1 & 1 & 0 & 0 & 0 & 0 & 0 & 1 & 0 & 0 & 0 & 1 & 1 & 0 & 0 & 0 & 0 & 0 & 0 & 0 & 0 & 0 & 0 & 0 & 1 & 1 & 1 & 0.735614816 \\
        V_{96} & 0 & 0 & 1 & 1 & 0 & 0 & 0 & 0 & 0 & 1 & 0 & 1 & 0 & 0 & 0 & 0 & 0 & 0 & 0 & 0 & 0 & 0 & 0 & 0 & 0 & 1 & 0 & 0 & 0 & 0.535415299 \\
        V_{84} & 0 & 1 & 0 & 0 & 0 & 0 & 1 & 1 & 0 & 1 & 0 & 0 & 0 & 0 & 1 & 0 & 0 & 0 & 0 & 0 & 0 & 0 & 0 & 0 & 0 & 0 & 1 & 1 & 0 & 0.564684457 \\
        V_{86} & 0 & 0 & 0 & 0 & 0 & 1 & 0 & 1 & 0 & 1 & 0 & 0 & 0 & 0 & 0 & 0 & 0 & 0 & 0 & 0 & 0 & 0 & 0 & 0 & 0 & 0 & 0 & 0 & 0 & 0.364484937 \\
        P_{17} & 0 & 0 & 0 & 1 & 0 & 0 & 0 & 0 & 0 & 0 & 0 & 0 & 0 & 0 & 0 & 1 & 1 & 1 & 0 & 1 & 0 & 0 & 1 & 0 & 0 & 0 & 0 & 0 & 0 & 0.363964705 \\
        P_{28} & 0 & 0 & 0 & 1 & 0 & 0 & 0 & 0 & 0 & 0 & 0 & 0 & 0 & 0 & 0 & 0 & 0 & 0 & 0 & 0 & 0 & 0 & 0 & 1 & 1 & 1 & 0 & 0 & 0 & 0.363964705 \\
        P_{39} & 0 & 0 & 0 & 1 & 0 & 0 & 0 & 0 & 0 & 0 & 0 & 0 & 0 & 0 & 0 & 0 & 0 & 0 & 0 & 0 & 1 & 1 & 1 & 1 & 1 & 0 & 0 & 0 & 0 & 0.363964705 \\
        P_{4,10} & 0 & 0 & 0 & 1 & 0 & 0 & 0 & 0 & 0 & 0 & 0 & 0 & 0 & 0 & 0 & 1 & 1 & 0 & 0 & 0 & 0 & 0 & 0 & 0 & 0 & 0 & 0 & 1 & 1 & 0.363964705 \\
        P_{5,11} & 0 & 0 & 0 & 1 & 0 & 0 & 0 & 0 & 0 & 0 & 0 & 0 & 0 & 0 & 0 & 0 & 0 & 0 & 1 & 1 & 1 & 0 & 1 & 1 & 0 & 0 & 0 & 0 & 0 & 0.363964705 \\
        P_{6,12} & 0 & 0 & 0 & 1 & 0 & 0 & 0 & 0 & 0 & 0 & 0 & 0 & 0 & 0 & 0 & 1 & 0 & 0 & 0 & 0 & 0 & 0 & 0 & 0 & 0 & 0 & 1 & 1 & 0 & 0.363964705 \\
        Q_{17} & 0 & 0 & 0 & 0 & 1 & 0 & 0 & 0 & 0 & 0 & 0 & 0 & 0 & 0 & 0 & 1 & 1 & 1 & 0 & 1 & 0 & 0 & 1 & 0 & 0 & 0 & 0 & 0 & 0 & 0.363964705 \\
        Q_{28} & 0 & 0 & 0 & 0 & 1 & 0 & 0 & 0 & 0 & 0 & 0 & 0 & 0 & 0 & 0 & 0 & 0 & 0 & 0 & 0 & 0 & 0 & 0 & 1 & 1 & 1 & 0 & 0 & 0 & 0.363964705 \\
        Q_{39} & 0 & 0 & 0 & 0 & 1 & 0 & 0 & 0 & 0 & 0 & 0 & 0 & 0 & 0 & 0 & 0 & 0 & 0 & 0 & 0 & 1 & 1 & 1 & 1 & 1 & 0 & 0 & 0 & 0 & 0.363964705 \\
        Q_{4,10} & 0 & 0 & 0 & 0 & 1 & 0 & 0 & 0 & 0 & 0 & 0 & 0 & 0 & 0 & 0 & 1 & 1 & 0 & 0 & 0 & 0 & 0 & 0 & 0 & 0 & 0 & 0 & 1 & 1 & 0.363964705 \\
        Q_{5,11} & 0 & 0 & 0 & 0 & 1 & 0 & 0 & 0 & 0 & 0 & 0 & 0 & 0 & 0 & 0 & 0 & 0 & 0 & 1 & 1 & 1 & 0 & 1 & 1 & 0 & 0 & 0 & 0 & 0 & 0.363964705 \\
        Q_{6,12} & 0 & 0 & 0 & 0 & 1 & 0 & 0 & 0 & 0 & 0 & 0 & 0 & 0 & 0 & 0 & 1 & 0 & 0 & 0 & 0 & 0 & 0 & 0 & 0 & 0 & 0 & 1 & 1 & 0 & 0.363964705 \\ \hline
        \Lambda_{21} & 0 & 1 & 1 & 1 & 1 & 0 & 0 & 0 & 0 & 1 & 0 & 0 & 0 & 1 & 1 & 1 & 0 & 0 & 0 & 0 & 0 & 0 & 0 & 1 & 1 & 0 & 1 & 1 & 1 & 1.099579522 \\
        \Lambda_{36} & 0 & 0 & 1 & 1 & 1 & 0 & 0 & 0 & 0 & 1 & 0 & 1 & 0 & 0 & 0 & 0 & 0 & 0 & 0 & 0 & 1 & 1 & 1 & 1 & 1 & 1 & 0 & 0 & 0 & 0.899380005 \\
        \Lambda_{87} & 0 & 1 & 1 & 1 & 1 & 0 & 0 & 0 & 0 & 1 & 0 & 0 & 0 & 1 & 1 & 1 & 1 & 1 & 0 & 1 & 0 & 0 & 1 & 0 & 0 & 0 & 1 & 1 & 1 & 1.099579522 \\
        \Lambda_{9,12} & 0 & 0 & 1 & 1 & 1 & 0 & 0 & 0 & 0 & 1 & 0 & 1 & 0 & 0 & 0 & 1 & 0 & 0 & 0 & 0 & 0 & 0 & 0 & 0 & 0 & 1 & 1 & 1 & 0 & 0.899380005\\
        \Lambda_{18} & 0 & 0 & 1 & 1 & 1 & 1 & 0 & 0 & 0 & 0 & 1 & 1 & 1 & 0 & 0 & 1 & 1 & 1 & 1 & 1 & 1 & 1 & 1 & 1 & 1 & 1 & 1 & 0 & 0 & 1.263864942 \\
        \Lambda_{4,11} & 0 & 0 & 1 & 1 & 1 & 0 & 0 & 0 & 0 & 0 & 1 & 0 & 0 & 1 & 0 & 1 & 1 & 1 & 0 & 1 & 1 & 0 & 1 & 1 & 0 & 0 & 0 & 1 & 1 & 0.899380005\\
        \Lambda_{2,10} & 0 & 1 & 0 & 1 & 1 & 0 & 1 & 1 & 0 & 1 & 0 & 0 & 0 & 0 & 1 & 1 & 1 & 0 & 0 & 0 & 0 & 0 & 0 & 1 & 1 & 1 & 1 & 1 & 1 & 1.292613868\\
        \Lambda_{2,12} & 1 & 0 & 0 & 1 & 1 & 0 & 0 & 1 & 0 & 1 & 0 & 0 & 0 & 0 & 0 & 1 & 0 & 0 & 0 & 0 & 0 & 0 & 0 & 1 & 1 & 1 & 1 & 1 & 0 & 0.899380005\\
        \Lambda^1_{19} & 1 & 0 & 0 & 1 & 0 & 1 & 0 & 1 & 0 & 0 & 1 & 0 & 1 & 0 & 0 & 1 & 1 & 1 & 1 & 1 & 1 & 1 & 1 & 1 & 1 & 0 & 0 & 0 & 0 & 0.899900237\\
        \Lambda^2_{19} & 1 & 0 & 0 & 0 & 1 & 1 & 0 & 1 & 0 & 0 & 1 & 0 & 1 & 0 & 0 & 1 & 1 & 1 & 1 & 1 & 1 & 1 & 1 & 1 & 1 & 0 & 0 & 0 & 0 & 0.899900237\\
        \Lambda^1_{38} & 0 & 0 & 0 & 1 & 0 & 0 & 1 & 0 & 1 & 0 & 0 & 1 & 0 & 0 & 0 & 0 & 0 & 0 & 0 & 0 & 1 & 1 & 1 & 1 & 1 & 1 & 0 & 0 & 0 & 0.728449645\\
        \Lambda^2_{38} & 0 & 0 & 0 & 0 & 1 & 0 & 1 & 0 & 1 & 0 & 0 & 1 & 0 & 0 & 0 & 0 & 0 & 0 & 0 & 0 & 1 & 1 & 1 & 1 & 1 & 1 & 0 & 0 & 0 & 0.728449645\\
        \Lambda^1_{47} & 0 & 0 & 0 & 1 & 0 & 1 & 0 & 0 & 1 & 0 & 0 & 0 & 0 & 1 & 0 & 1 & 1 & 1 & 0 & 1 & 0 & 0 & 1 & 0 & 0 & 0 & 0 & 1 & 1 & 0.728449643\\
        \Lambda^2_{47} & 0 & 0 & 0 & 0 & 1 & 1 & 0 & 0 & 1 & 0 & 0 & 0 & 0 & 1 & 0 & 1 & 1 & 1 & 0 & 1 & 0 & 0 & 1 & 0 & 0 & 0 & 0 & 1 & 1 & 0.728449643\\
        \Lambda^1_{59} & 0 & 1 & 0 & 1 & 0 & 0 & 0 & 0 & 0 & 0 & 0 & 0 & 1 & 0 & 0 & 0 & 0 & 0 & 1 & 1 & 1 & 1 & 1 & 1 & 1 & 0 & 0 & 0 & 0 & 0.564164222\\
        \Lambda^2_{59} & 0 & 1 & 0 & 0 & 1 & 0 & 0 & 0 & 0 & 0 & 0 & 0 & 1 & 0 & 0 & 0 & 0 & 0 & 1 & 1 & 1 & 1 & 1 & 1 & 1 & 0 & 0 & 0 & 0 & 0.564164222\\
        \Lambda^1_{67} & 0 & 1 & 0 & 1 & 0 & 0 & 1 & 0 & 1 & 0 & 0 & 0 & 0 & 1 & 1 & 1 & 1 & 1 & 0 & 1 & 0 & 0 & 1 & 0 & 0 & 0 & 1 & 1 & 1 & 0.928649162\\
        \Lambda^2_{67} & 0 & 1 & 0 & 0 & 1 & 0 & 1 & 0 & 1 & 0 & 0 & 0 & 0 & 1 & 1 & 1 & 1 & 1 & 0 & 1 & 0 & 0 & 1 & 0 & 0 & 0 & 1 & 1 & 1 & 0.928649162\\
        \Lambda^1_{1,11} & 0 & 0 & 0 & 1 & 0 & 0 & 1 & 1 & 0 & 0 & 1 & 0 & 0 & 0 & 0 & 1 & 1 & 1 & 1 & 1 & 1 & 0 & 1 & 1 & 0 & 0 & 0 & 0 & 0 & 0.728449645 \\
        \Lambda^2_{1,11} & 0 & 0 & 0 & 0 & 1 & 0 & 1 & 1 & 0 & 0 & 1 & 0 & 0 & 0 & 0 & 1 & 1 & 1 & 1 & 1 & 1 & 0 & 1 & 1 & 0 & 0 & 0 & 0 & 0 & 0.728449645 \\
        \Lambda^1_{6,10} & 1 & 0 & 0 & 1 & 0 & 0 & 0 & 0 & 0 & 0 & 0 & 0 & 0 & 0 & 1 & 1 & 1 & 0 & 0 & 0 & 0 & 0 & 0 & 0 & 0 & 0 & 1 & 1 & 1 & 0.535415299 \\
        \Lambda^2_{6,10} & 1 & 0 & 0 & 0 & 1 & 0 & 0 & 0 & 0 & 0 & 0 & 0 & 0 & 0 & 1 & 1 & 1 & 0 & 0 & 0 & 0 & 0 & 0 & 0 & 0 & 0 & 1 & 1 & 1 & 0.535415299 \\
        \Lambda^1_{58} & 1 & 0 & 0 & 1 & 0 & 1 & 0 & 0 & 1 & 0 & 0 & 1 & 1 & 0 & 0 & 0 & 0 & 0 & 1 & 1 & 1 & 1 & 1 & 1 & 1 & 1 & 0 & 0 & 0 & 0.899900237 \\
        \Lambda^2_{58} & 1 & 0 & 0 & 0 & 1 & 1 & 0 & 0 & 1 & 0 & 0 & 1 & 1 & 0 & 0 & 0 & 0 & 0 & 1 & 1 & 1 & 1 & 1 & 1 & 1 & 1 & 0 & 0 & 0 & 0.899900237 \\
    \end{array}
\right)~.~$}
\end{align}
The vertices of the toric diagram in \fref{toric_diagrams_example_1} are labelled with the corresponding brick matchings in the extended $\bar{P}$-matrix. The complete forward algorithm independently confirms that the new theory has $P^2_{+-}(\textrm{PdP}_3)$ as its mesonic moduli space.

The volume of the SE$_7$ base manifold of this model is
\begin{align}
   {\rm Vol} ( P^2_{+-} (\textrm{PdP}_3))=  3.99667 \,.
\end{align}
Comparing to \eqref{Vol_SPP_Z2}, we see that
\begin{align}
\frac{ {\rm Vol } (P^2_{+-} (\textrm{PdP}_3)) }{ {\rm Vol} (P_{+-}(\textrm{SPP}/\IZ_2)) } = \frac{3.99667}{3.75355} \simeq 1.06 > 1  \, ,
\end{align}
which is consistent with the observed increase in the base volume from the UV to the IR, as summarized in \eref{V_Y7_UV_IR}. This provides further geometric evidence that the deformation under consideration is relevant.

\begin{table}[ht!]
\centering
\begin{tabular}{|c|c|l|}
\hline
\; &  $U(1)_R$ & fugacity \\
\hline
$p_2$ & $ r_1 $ &  $t_2 = \overline{t}_1 $ \\
$p_3$ & $ r_1 $ &  $t_3 = \overline{t}_1 $ \\
$p_4$ & $ r_4 $ &  $t_4 = \overline{t}_4 $ \\
$p_5$ & $ r_2 $ &  $t_5 = \overline{t}_2 $ \\
$p_6$ & $ r_2 $ &  $t_6 = \overline{t}_2 $ \\
$p_7$ & $ r_3 $ &  $t_7 = \overline{t}_3 $ \\
$p_8$ & $ r_3 $ &  $t_8 = \overline{t}_3 $ \\
\hline
\end{tabular}
\caption{Charges under the $U(1)_R$ symmetry of the $P^2_{+-} (\textrm{PdP}_3 ) $ model of the extremal GLSM fields $p_a$.
Here, $U(1)_R$ charges $r_1\,, r_2 \,, r_3$ and $r_4$ are chosen such that
the $J$- and $E$-terms coupled to Fermi fields have an overall $U(1)_R$ charge of $2$
with $ 2r_1 + 2r_2 + 2r_3 + r_4 = 2 \,.$
 \label{tab_11}}
\end{table}

The Hilbert series of the mesonic moduli space
refined only under the $U(1)_R$ symmetry with fugacities summarized in \tref{tab_11}
takes the following form,
\beal{HS11}
g(\bar{t}_1, \dots, \bar{t}_4 ; P^2_{+-} (\textrm{PdP}_3) ) = 
\frac{P(\bar{t}_1, \dots, \bar{t}_4)}{
(1- \bar{t}_1^2 \bar{t}_2 \bar{t}_3^4)^3 (1- \bar{t}_1^2 \bar{t}_2^2 \bar{t}_3^2 \bar{t}_4)^2 (1- \bar{t}_1^2 \bar{t}_2^3 \bar{t}_4^2)^2
}~,~
\eea
where the numerator is given by, 
\beal{HS11b}
&&
P(\bar{t}_1, \dots, \bar{t}_4) =
1 + 3 \bar{t}_1^2 \bar{t}_2 \bar{t}_3^4 + 7 \bar{t}_1^2 \bar{t}_2^2 \bar{t}_3^2 \bar{t}_4 - 13 \bar{t}_1^4 \bar{t}_2^3 \bar{t}_3^6 \bar{t}_4 +  2 \bar{t}_1^6 \bar{t}_2^4 \bar{t}_3^{10} \bar{t}_4 + 2 \bar{t}_1^2 \bar{t}_2^3 \bar{t}_4^2 - 10 \bar{t}_1^4 \bar{t}_2^4 \bar{t}_3^4 \bar{t}_4^2 
\nn\\
&&
\hspace{0.5cm}
-  4 \bar{t}_1^6 \bar{t}_2^5 \bar{t}_3^8 \bar{t}_4^2 + 4 \bar{t}_1^8 \bar{t}_2^6 \bar{t}_3^{12} \bar{t}_4^2 -  4 \bar{t}_1^4 \bar{t}_2^5 \bar{t}_3^2 \bar{t}_4^3 + 4 \bar{t}_1^6 \bar{t}_2^6 \bar{t}_3^6 \bar{t}_4^3 +  10 \bar{t}_1^8 \bar{t}_2^7 \bar{t}_3^{10} \bar{t}_4^3 - 2 \bar{t}_1^{10} \bar{t}_2^8 \bar{t}_3^{14} \bar{t}_4^3 
\nn\\
&&
\hspace{0.5cm}
-  2 \bar{t}_1^6 \bar{t}_2^7 \bar{t}_3^4 \bar{t}_4^4  + 13 \bar{t}_1^8 \bar{t}_2^8 \bar{t}_3^8 \bar{t}_4^4 -  7 \bar{t}_1^{10} \bar{t}_2^9 \bar{t}_3^{12} \bar{t}_4^4 - 3 \bar{t}_1^{10} \bar{t}_2^{10} \bar{t}_3^{10} \bar{t}_4^5 -  \bar{t}_1^{12} \bar{t}_2^{11} \bar{t}_3^{14} \bar{t}_4^5
~.~
\eea
Note that the Hilbert series in \eqref{HS11} is identical to \eqref{HS10}.
This is in line with the observations made in \cite{Ghim:2024asj, Ghim:2025zhs} that brane brick models whose 
associated toric Calabi-Yau 4-folds are related by a birational transformation
share the same 
Hilbert series when it is refined only under the $U(1)_R$ symmetry. 
\\

\subsection{A Mass Deformation from $\textrm{SPP} \times \IC $ to $ D_3 $ \label{sec:sppc}}

\begin{figure}[h]
	\centering
	\includegraphics[height=4.5cm]{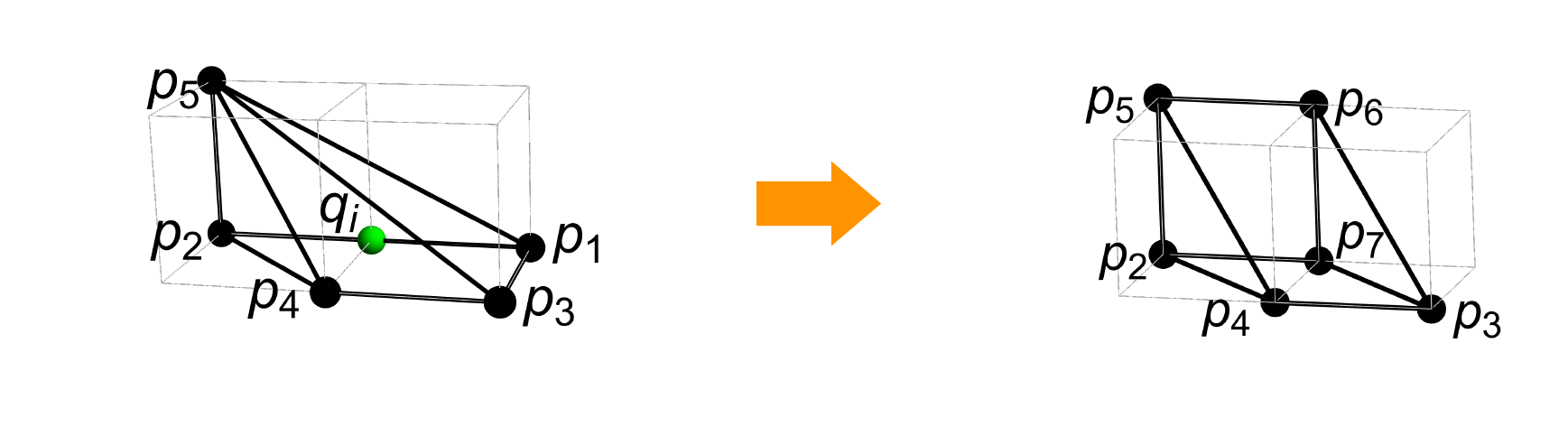}
\caption{Toric diagrams for $\textrm{SPP} \times \IC $ and $ D_3 $. We have labeled points anticipating the corresponding brick matchings.}
	\label{toric_diagrams_example_2}
\end{figure}

Let us now consider a deformation from $\textrm{SPP} \times \IC $ to $D_3$. While the original theory $\textrm{SPP} \times \IC$ has $(2,2)$ supersymmetry, the deformation halves the number of supercharges, preserving $(0,2)$ supersymmetry for the $D_3$ theory. \fref{toric_diagrams_example_2} shows the corresponding toric diagrams. Once again, we observe that the effect of the deformation is to reduce the collinearity of points on edges. 

\subsubsection*{Starting point: a brane brick model for $\textrm{SPP} \times \IC $ }

\begin{figure}[H]
\begin{center}
\resizebox{0.4\hsize}{!}{
\includegraphics[height=6cm]{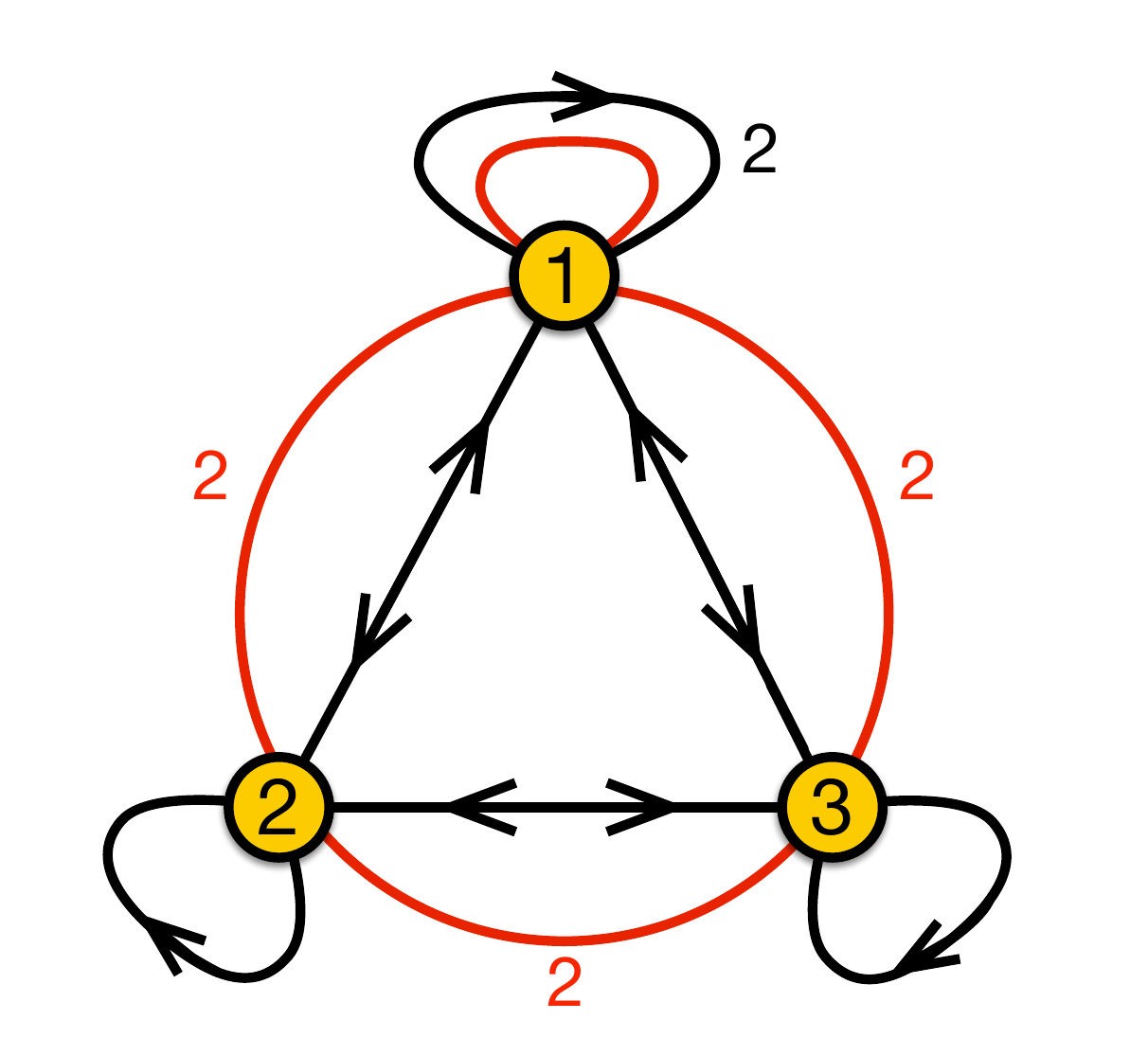}
}
\caption{
Quiver diagram for $\textrm{SPP} \times \IC $ .
\label{fig_q_sppxc}}
 \end{center}
 \end{figure}

Since this CY$_4$ is of the form CY$_3\times \mathbb{C}$, the corresponding gauge theory can be obtained by dimensional reduction from the $4d$ $\mathcal{N}=1$ theory corresponding to SPP \cite{Morrison:1998cs}. The resulting $2d$ (2,2) theory was first presented in \cite{Franco:2015tna} and its quiver diagram is shown in \fref{fig_q_sppxc}.

The $J$- and $E$-terms take the following form,
 \begin{align}
\begin{array}{rcrclcrcl}
& &  \ \ \ \ \ \ \ \ \ \ \ \ \ \ \ \ \ & J& &&& E&  \ \ \ \ \ \ \ \ \ \ \ \ \ \  \\
 \Lambda_{11} : & \ \ \  & X_{13} \cdot X_{31} & - & X_{12} \cdot X_{21} \ & \ \ \ \  & \Phi_{11} \cdot X_{11}& -& X_{11}\cdot \Phi_{11}\\
 \Lambda_{21} : & \ \ \  & X_{12} \cdot X_{23}\cdot X_{32} & - & X_{11} \cdot X_{12} & \ \ \ \  & \Phi_{22} \cdot X_{21}& -& X_{21}\cdot \Phi_{11}\\
 \Lambda_{12} : & \ \ \  & X_{23} \cdot X_{32}\cdot X_{21} & - & X_{21} \cdot X_{11} & \ \ \ \  & \Phi_{11} \cdot X_{12}& -& X_{12}\cdot \Phi_{22}\\
 \Lambda_{31} : & \ \ \  & X_{11} \cdot X_{13}  & - & X_{13} \cdot X_{32}\cdot X_{23} & \ \ \ \  & \Phi_{33} \cdot X_{31}& -& X_{31}\cdot \Phi_{11}\\
 \Lambda_{13} : & \ \ \  & X_{31} \cdot X_{11}  & - & X_{32} \cdot X_{23}\cdot X_{31} & \ \ \ \  & \Phi_{11} \cdot X_{13}& -& X_{13}\cdot \Phi_{33}\\
 \Lambda_{32} : & \ \ \  & X_{21} \cdot X_{12}\cdot X_{23} & - & X_{23}\cdot X_{31} \cdot X_{13} & \ \ \ \  & \Phi_{33} \cdot X_{32}& -& X_{32}\cdot \Phi_{22}\\
  \Lambda_{23} : & \ \ \  & X_{32} \cdot X_{21}\cdot X_{12} & - & X_{31} \cdot X_{13}\cdot X_{32} & \ \ \ \  & \Phi_{22} \cdot X_{23}& -& X_{23}\cdot \Phi_{33}\\
\end{array} \,.
\label{SPP_JE}
\end{align}

From the above $J$- and $E$-terms, we obtain combinatorially the following extended $\bar{P}$-matrix,
{\small
\begin{align}
\left(\begin{array}{c|ccccc|cc|c}
         \; & p_1 & p_2 & p_3 & p_4 & p_5 & q_1 & q_2 & \Delta \\
        \hline
       \Phi_{11}  & 0 & 0 & 0 & 0 & 1 & 0 & 0 & 0.5 \\
        \Phi_{22} & 0 & 0 & 0 & 0 & 1 & 0 & 0 & 0.5 \\
        \Phi_{33} & 0 & 0 & 0 & 0 & 1 & 0 & 0 & 0.5 \\
        X_{11}    & 0 & 0 & 1 & 1 & 0 & 0 & 0 & 0.633975 \\
        X_{13}    & 0 & 1 & 0 & 0 & 0 & 0 & 1 & 0.433013 \\
        X_{31}    & 1 & 0 & 0 & 0 & 0 & 1 & 0 & 0.433013 \\
        X_{23}    & 0 & 0 & 1 & 0 & 0 & 0 & 0 & 0.316987 \\
        X_{32}    & 0 & 0 & 0 & 1 & 0 & 0 & 0 & 0.316987 \\
        X_{12}    & 1 & 0 & 0 & 0 & 0 & 0 & 1 & 0.433013 \\
        X_{21}    & 0 & 1 & 0 & 0 & 0 & 1 & 0 & 0.433013 \\
        \hline
        \Lambda_{11} & 0 & 0 & 1 & 1 & 1 & 0 & 0 & 1.13397 \\
        \Lambda_{21} & 0 & 1 & 0 & 0 & 1 & 1 & 0 & 0.933013 \\
        \Lambda_{12} & 1 & 0 & 0 & 0 & 1 & 0 & 1 & 0.933013 \\
        \Lambda_{31} & 1 & 0 & 0 & 0 & 1 & 1 & 0 & 0.9330130 \\
        \Lambda_{13} & 0 & 1 & 0 & 0 & 1 & 0 & 1 & 0.933013 \\
        \Lambda_{32} & 0 & 0 & 0 & 1 & 1 & 0 & 0 & 0.816987 \\
        \Lambda_{23} & 0 & 0 & 1 & 0 & 1 & 0 & 0 & 0.816987 \\
      \end{array}
      \right)~.~
\label{p_sppxc}
\end{align}
}
In addition, the complete forward algorithm separately confirms that the theory has $\textrm{SPP} \times \IC$ as its mesonic moduli space.
The vertices in the toric diagram on the left-hand side of \fref{toric_diagrams_example_2}
are labelled by the associated brick matchings in \eref{p_sppxc}.

The volume of the SE$_7$ base manifold is given by
\begin{align} \label{vol-sppxc}
    {\rm Vol} ({\rm SPP} \times \IC) = 12.4976 \,,
\end{align}
according to the computation in Section \ref{sec:vol}. 

The charges under the $U(1)_R$ symmetry are summarized in \tref{tab_20}.
\begin{table}[ht!]
\centering
\begin{tabular}{|c|c|l|}
\hline
\; &  $U(1)_R$ & fugacity \\
\hline
$p_1$ & $ r_2 $ &  $t_1 = \overline{t}_2 $ \\
$p_2$ & $ r_2 $ &  $t_2 = \overline{t}_2 $ \\
$p_3$ & $ r_1 $ &  $t_3 = \overline{t}_1 $ \\
$p_4$ & $ r_1 $ &  $t_4 = \overline{t}_1 $ \\
$p_5$ & $ 2 r_2 $ &  $t_5 = \overline{t}_1^2 $ \\
\hline
\end{tabular}
\caption{Charges under the $U(1)_R$ symmetry of the SPP$\times \IC $ model of the extremal GLSM fields $p_a$.
Here, $U(1)_R$ charges $r_1$ and $r_2$ are chosen such that
the $J$- and $E$-terms coupled to Fermi fields have an overall $U(1)_R$ charge of $2$
with $2r_1 + 4 r_2 =2$.
 \label{tab_20}}
\end{table}
The Hilbert series 
refined only under the $U(1)_R$ symmetry with fugacities summarized in \tref{tab_20} takes the following form,
\beal{HS20}
g(\overline{t}_1, \overline{t}_2 ; \textrm{SPP} \times \IC ) = \frac{1+\overline{t}_1 \overline{t}_2^2 }{(1-\overline{t}_1^2) (1-\overline{t}^2_2)^2 (1- \overline{t}_1 \overline{t}^2_2) } \,.
\eea

\subsubsection*{Relevant Deformation to $D_3$}

Let us now consider the following deformation to the $J$-terms, indicated in blue in \eqref{SPP_deform}. In contrast to the previous example, this deformation simultaneously involves both mass and cubic terms.
\begin{align}
\begin{array}{rcrclcrcl}
& &  \ \ \ \ \ \ \ \ \ \ \ \ \ \ \ \ \ & J & + \, \textcolor{blue}{\Delta J} &&& E&  \ \ \ \ \ \ \ \ \ \ \ \ \ \  \\
 \Lambda_{11} : & \ \ \  & X_{13} \cdot X_{31} & - & X_{12} \cdot X_{21} +\textcolor{blue}{\mu \,\Phi_{11}}\ & \ \ \ \  & \Phi_{11} \cdot X_{11}& -& X_{11}\cdot \Phi_{11}\\
  \Lambda_{32} : & \ \ \  & X_{21} \cdot X_{12}\cdot X_{23} & - & X_{23}\cdot X_{31} \cdot X_{13} -\textcolor{blue}{\mu \,\Phi_{22}\cdot X_{23}}& \ \ \ \  & \Phi_{33} \cdot X_{32}& -& X_{32}\cdot \Phi_{22}\\
  \Lambda_{23} : & \ \ \  & X_{32} \cdot X_{21}\cdot X_{12} & - & X_{31} \cdot X_{13}\cdot X_{32}-\textcolor{blue}{\mu \,\Phi_{33}\cdot X_{32}} & \ \ \ \  & \Phi_{22} \cdot X_{23}& -& X_{23}\cdot \Phi_{33}\\
\end{array}
~.~ 
\label{SPP_deform}
\end{align}

Using the extended $\bar{P}$-matrix in \eqref{p_sppxc}, we find that all the deformation plaquettes have the same extremal brick matching content; $p_3\cdot p_4 \cdot p_5^2 \,.$ Its scaling dimension obtained by \eqref{scaledefm} is $\Delta [\Lambda \cdot \Delta J] \simeq 1.63397<2$, which implies that the deformation is relevant. This is of course to be expected, since one of the deformation terms is a mass term. However, it is more interesting to note that all the other deformation terms that share the same extremal perfect matching content must be turned on simultaneously to deform a theory to another brane brick model. 

\begin{figure}[ht!]
\begin{center}
\resizebox{0.4\hsize}{!}{
\includegraphics[height=6cm]{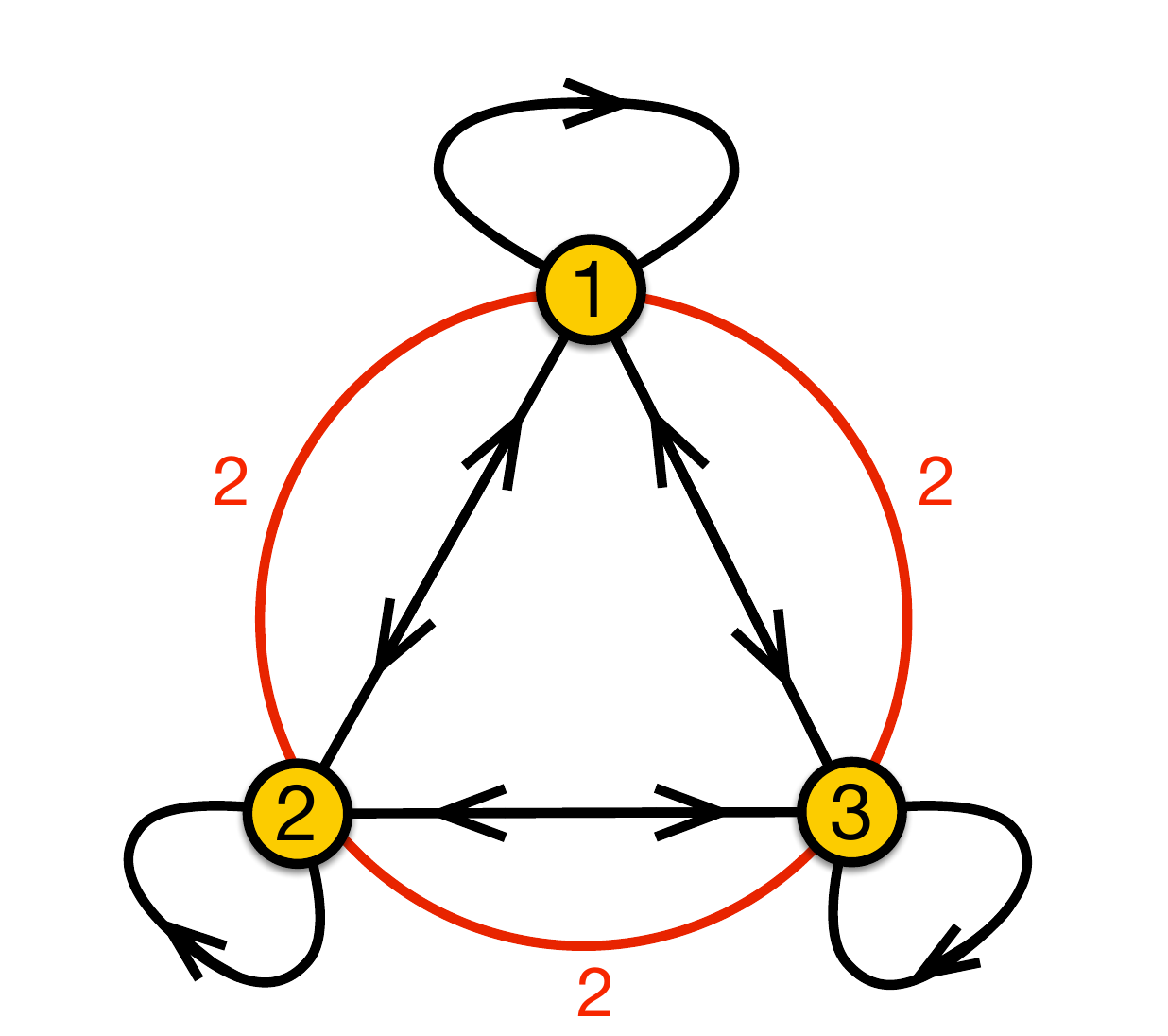}
}
\caption{
Quiver diagram for $D_3$.
\label{fig_q_d3}}
 \end{center}
 \end{figure}

Upon integrating out massive fields $\Lambda_{11}$ and $\Phi_{11}$, we proceed with the following replacement,
\begin{align}
\Phi_{11} = \frac{1}{\mu}(X_{12}\cdot X_{21}-X_{13}\cdot X_{31}) \,. 
\end{align}
The resulting quiver is shown in \fref{fig_q_d3}.
Furthermore, we perform the following change of variables,
\begin{align} \label{d3redefs}
\begin{split}
  \Phi_{22} \rightarrow & -\frac{1}{\mu}\Phi_{22} + \frac{1}{\mu}X_{21}\cdot X_{12} \,, \\
  \Phi_{33} \rightarrow & \frac{1}{\mu}\Phi_{33} -\frac{1}{\mu} X_{31}\cdot X_{13} \,. \\
\end{split}
\end{align}
Next, we relabel some of the Fermi fields as follows: $\Lambda^1_{23}=\bar{\Lambda}_{32}$ and $\Lambda^2_{23}=\Lambda_{23}$. 
The relabeling must be accompanied by the swap of $J$ and $E$-terms associated to $\Lambda_{32}$. This leads to the following $J$- and $E$-terms,
{\footnotesize
\begin{align}
\begin{array}{rcrclcrcl}
& &  \ \ \ \ \ \ \ \ \ \ \ \ \ \ \ \ \ & J & & & & E&  \ \ \ \ \ \ \ \ \ \ \ \ \ \  \\
 \Lambda_{21} : & \ \ \  & X_{12} \cdot X_{23}\cdot X_{32} & - & X_{11} \cdot X_{12} & \ \ \ \  &\frac{1}{\mu}( X_{21}\cdot X_{13}\cdot X_{31}& -& \Phi_{22} \cdot X_{21})\\
 \Lambda_{12} : & \ \ \  & X_{23} \cdot X_{32}\cdot X_{21} & - & X_{21} \cdot X_{11} & \ \ \ \  & \frac{1}{\mu}(X_{12}\cdot \Phi_{22}& -& X_{13}\cdot X_{31}\cdot X_{12})\\
 \Lambda_{31} : & \ \ \  & X_{11} \cdot X_{13} & - & X_{13} \cdot X_{32}\cdot X_{23}  & \ \ \ \  & \frac{1}{\mu}(\Phi_{33} \cdot X_{31}& -& X_{31}\cdot X_{12}\cdot X_{21})\\
 \Lambda_{13} : & \ \ \  & X_{31} \cdot X_{11} & - & X_{32} \cdot X_{23}\cdot X_{31}   & \ \ \ \  & \frac{1}{\mu}(X_{12}\cdot X_{21}\cdot X_{13}& -& X_{13}\cdot \Phi_{33}) \\
 \Lambda^1_{23} : & \ \ \  & \makecell{ \frac{1}{\mu}(\textcolor{BrickRed}{\Phi_{33}\cdot X_{32}} \\ +X_{32}\cdot\Phi_{22} } & \makecell{ - \\ - } & \makecell{ \textcolor{BrickRed}{ X_{32} \cdot X_{21} \cdot X_{12}} \\ X_{31} \cdot X_{13} \cdot X_{32} ) }  & \ \ \ \  & \Phi_{22} \cdot X_{23}& -& X_{23}\cdot X_{31}\cdot X_{13}\\
 \Lambda^2_{23} : & \ \ \  & X_{32}\cdot X_{21}\cdot X_{12} & - & \Phi_{33}\cdot X_{32} & \ \ \ \  &
 \makecell{ \frac{1}{\mu}(X_{21}\cdot X_{12}\cdot X_{23} \\ +\textcolor{BrickRed}{X_{23}\cdot X_{31}\cdot X_{13}} } & \makecell{ - \\ - } & \makecell{ X_{23}\cdot\Phi_{33} \\ \textcolor{BrickRed}{ \Phi_{22}\cdot X_{23} } ) }  \\
\end{array}
~.~ \label{SPP2}
\end{align}
}

The $J$- and $E$-terms in \eqref{SPP2} still violate the toric condition by the contributions shown in red. We can remedy this issue by an extra rotation of the Fermi fields. Specifically, we introduce fields $\Lambda^{1'}_{23}$, $\Lambda^{2'}_{23}$ which are connected to $\Lambda_{23}^1 \,, \Lambda_{23}^2$ via
\begin{align}\label{Fermirotation}
  \begin{pmatrix}
    \Lambda^1_{23} \\
    \Lambda^2_{23}
  \end{pmatrix}
  =\begin{pmatrix}
     \mu & 1 \\
     1 & -\frac{1}{\mu}
   \end{pmatrix}
   \begin{pmatrix}
          \Lambda^{1'}_{23} \\
          \Lambda^{2'}_{23}
        \end{pmatrix} \,.
\end{align}
Let us now determine the $J$- and $E$-terms for these new Fermi fields. Recall that the original Fermi fields couple to terms $J^1_{32}$ and $J^2_{32}$ via $\mathcal{L}_J=-\int d^2yd\theta^+(\Lambda_{23}^1\cdot J^1_{32}+\Lambda^2_{23} \cdot J^2_{32})$. Implementing the substitution in \eqref{Fermirotation}, we arrive at
\begin{align}\label{J-termsrotation}
\begin{split}
    J^{1'}_{32} & =\mu J^1_{32} + J^2_{32}  \,, \\
    J^{2'}_{32} & = J_{32}^1-\frac{1}{\mu} J^2_{32} \,.
\end{split}
\end{align}
Since the $E$-term is defined by the deformation on the chirality condition, we have
\begin{align}\label{E-termsrotation}
\begin{split}
    E^{1'}_{23} & =\frac{1}{2}(\frac{1}{\mu} E^1_{23} + E^2_{23}) \,, \\
    E^{2'}_{23} & =\frac{1}{2}(E^1_{23}-\mu E^2_{23}) \,.
\end{split}
\end{align}
Finally, we rescale the chiral and Fermi fields in the following way,
\begin{align} \label{FermisD3}
\begin{split}
X_{11} \rightarrow \mu X_{11} \,, \quad X_{23} \rightarrow \mu X_{23} \,, & \\
\Lambda_{21} \rightarrow \frac{1}{\mu}\Lambda_{21} \,, \quad
\Lambda_{12} \rightarrow \frac{1}{\mu}\Lambda_{12} \,,  \quad &
\Lambda_{31} \rightarrow \frac{1}{\mu} \Lambda_{31}  \,, \quad
\Lambda^{2'}_{23} \rightarrow  -\mu  \Lambda^{2}_{23} \,, \quad
\Lambda_{13} \rightarrow \frac{1}{\mu}\Lambda_{13} \,.
\end{split}
\end{align}
By renaming $\Lambda^{1'}_{23}$ as $\Lambda^1_{23}$, we obtain the following $J$- and $E$-terms, 
{\footnotesize
\begin{align}
\begin{array}{rcrclcrcl}
& &  \ \ \ \ \ \ \ \ \ \ \ \ \ \ \ \ \ & J& &&& E&  \ \ \ \ \ \ \ \ \ \ \ \ \ \  \\
 \Lambda_{21} : & \ \ \  & X_{12} \cdot X_{23}\cdot X_{32} & - & X_{11} \cdot X_{12} & \ \ \ \  & X_{21}\cdot X_{13}\cdot X_{31}& -& \Phi_{22} \cdot X_{21}\\
 \Lambda_{12} : & \ \ \  & X_{23} \cdot X_{32}\cdot X_{21} & - & X_{21} \cdot X_{11} & \ \ \ \  & X_{12}\cdot \Phi_{22}& -& X_{13}\cdot X_{31}\cdot X_{12}\\
 \Lambda_{31} : & \ \ \  & X_{11} \cdot X_{13}  & - & X_{13} \cdot X_{32}\cdot X_{23} & \ \ \ \  & \Phi_{33} \cdot X_{31}& -& X_{31}\cdot X_{12}\cdot X_{21}\\
 \Lambda_{13} : & \ \ \  & X_{31} \cdot X_{11}  & - & X_{32} \cdot X_{23}\cdot X_{31} & \ \ \ \  & X_{12}\cdot X_{21}\cdot X_{13}& -& X_{13}\cdot \Phi_{33}\\
 \Lambda^{1}_{23} : & \ \ \  & X_{32}\cdot\Phi_{22} & - & X_{31}\cdot X_{13}\cdot X_{32} & \ \ \ \  & \frac{1}{2}(X_{21}\cdot X_{12}\cdot X_{23}& -& X_{23}\cdot \Phi_{33})\\
 \Lambda^{2}_{23} : & \ \ \  & 2(X_{32}\cdot X_{21}\cdot X_{12} & - & \Phi_{33}\cdot X_{32})& \ \ \ \  & X_{23}\cdot X_{31}\cdot X_{13}& -&\Phi_{22}\cdot X_{23}\\
                     & \ \ \ &  +\textcolor{BrickRed}{X_{32}\cdot\Phi_{22}} & - & \textcolor{BrickRed}{X_{31}\cdot X_{13}\cdot X_{32}}& \ \ \ \  & +\frac{1}{2}(\textcolor{BrickRed}{X_{21}\cdot X_{12}\cdot X_{23}}& -& \textcolor{BrickRed}{X_{23}\cdot \Phi_{33}})
\end{array}
\label{SPP3} \,.
\end{align}
}
The terms in red correspond to the $J$- and $E$-terms above them, and therefore vanish on-shell. Thus, on the moduli space we have the following independent set of $J$- and $E$-terms. By introducing additional constant scaling of factor $2$, we have
{\footnotesize
\begin{align}
\begin{array}{rcrclcrcl}
& &  \ \ \ \ \ \ \ \ \ \ \ \ \ \ \ \ \ & J& &&& E&  \ \ \ \ \ \ \ \ \ \ \ \ \ \  \\
 \Lambda_{21} : & \ \ \  & X_{12} \cdot X_{23}\cdot X_{32} & - & X_{11} \cdot X_{12} & \ \ \ \  & X_{21}\cdot X_{13}\cdot X_{31}& -& \Phi_{22} \cdot X_{21}\\
 \Lambda_{12} : & \ \ \  & X_{23} \cdot X_{32}\cdot X_{21} & - & X_{21} \cdot X_{11} & \ \ \ \  & X_{12}\cdot \Phi_{22}& -& X_{13}\cdot X_{31}\cdot X_{12}\\
 \Lambda_{31} : & \ \ \  & X_{11} \cdot X_{13}  & - & X_{13} \cdot X_{32}\cdot X_{23} & \ \ \ \  & \Phi_{33} \cdot X_{31}& -& X_{31}\cdot X_{12}\cdot X_{21}\\
 \Lambda_{13} : & \ \ \  & X_{31} \cdot X_{11}  & - & X_{32} \cdot X_{23}\cdot X_{31} & \ \ \ \  & X_{12}\cdot X_{21}\cdot X_{13}& -& X_{13}\cdot \Phi_{33}\\
 \Lambda^{1}_{23} : & \ \ \  & X_{32}\cdot\Phi_{22} & - & X_{31}\cdot X_{13}\cdot X_{32} & \ \ \ \  & X_{21}\cdot X_{12}\cdot X_{23}& -& X_{23}\cdot \Phi_{33}\\
 \Lambda^{2}_{23} : & \ \ \  & X_{32}\cdot X_{21}\cdot X_{12} & - & \Phi_{33}\cdot X_{32}& \ \ \ \  & X_{23}\cdot X_{31}\cdot X_{13}& -&\Phi_{22}\cdot X_{23}\\
\end{array} \,,
\label{SPP4}
\end{align}
}
which are the known $J$- and $E$-terms for the $D_3$ model \cite{Franco:2015tna}. 
The extended $\bar{P}$-matrix for the $D_3$ theory is
{\small
\begin{align}
\left(\begin{array}{c|cccccc|c}
 	\; & p_2  &  p_3 &  p_4 &  p_5 &  p_6 &  p_7 &   \Delta  \\ 
	\hline
	X_{11} &  1 &  0 &  0 &  0 &  0 &  1 &  0.666667 \\ \
	\Phi_{22} &  0 &  0 &  0 &  1 &  1 &  0 &  0.666667 \\ 
	\Phi_{33} &  0 &  1 &  1 &  0 &  0 &  0 &  0.666667 \\ 
	X_{13} &  0 &  0 &  0 &  0 &  1 &  0 &  0.333333 \\ 
	X_{31} &  0 &  0 &  0 &  1 &  0 &  0 &  0.333333 \\ 
	X_{23} &  1 &  0 &  0 &  0 &  0 &  0 &  0.333333 \\ 
	X_{32} &  0 &  0 &  0 &  0 &  0 &  1 &  0.333333 \\ 
	X_{12} &  0 &  1 &  0 &  0 &  0 &  0 &  0.333333 \\ 
	X_{21} &  0 &  0 &  1 &  0 &  0 &  0 &  0.333333 \\ 
	\hline
	\Lambda_{21} &  0 &  0 &  1 &  1 &  1 &  0 &  1 \\ 
	\Lambda_{12} &  0 &  1 &  0 &  1 &  1 &  0 &  1 \\ 
	\Lambda_{31} &  0 &  1 &  1 &  1 &  0 &  0 &  1 \\ 
	\Lambda_{13} &  0 &  1 &  1 &  0 &  1 &  0 &  1 \\ 
	\Lambda^1_{23} &  1 &  1 &  1 &  0 &  0 &  0 &  1 \\ 
	\Lambda^2 _{23} &  1 &  0 &  0 &  1 &  1 &  0 &  1 \\
      \end{array}
      \right)~.~
\label{p_D3}
\end{align}
}
The vertices in the toric diagram for the $D_3$ model in \fref{toric_diagrams_example_2}
have been labelled by the associated brick matchings in \eref{p_D3}.

The volume of the base SE$_7$ manifold reads
\begin{align} \label{vol-D3}
    {\rm Vol} (D_3) = 13.6982 \,.
\end{align}
Comparing to \eqref{vol-sppxc}, we see that
\begin{align}
\frac{ {\rm Vol } (D_3) }{ {\rm Vol} ( {\rm SPP} \times \IC ) } = \frac{13.6982 }{12.4976 } \simeq 1.10 > 1  \,.
\end{align}
As expected, the volume of the corresponding SE$_7$ manifold increases as the gauge theory is deformed.

\begin{table}[ht!]
\centering
\begin{tabular}{|c|c|l|}
\hline
\; &  $U(1)_R$ & fugacity \\
\hline
$p_2$ & $ r_2 $ &  $t_2 = \overline{t}_2 $ \\
$p_3$ & $ r_1 $ &  $t_3 = \overline{t}_1 $ \\
$p_4$ & $ r_1 $ &  $t_4 = \overline{t}_1 $ \\
$p_5$ & $ r_2 $ &  $t_5 = \overline{t}_2 $ \\
$p_6$ & $ r_2 $ &  $t_6 = \overline{t}_2 $ \\
$p_7$ & $ r_2 $ &  $t_7 = \overline{t}_2 $ \\
\hline
\end{tabular}
\caption{Charges under the $U(1)_R$ symmetry of the $D_3$ model of the extremal GLSM fields $p_a$.
Here, $U(1)_R$ charges $r_1$ and $r_2$ are chosen such that
the $J$- and $E$-terms coupled to Fermi fields have an overall $U(1)_R$ charge of $2$
with $2r_1 + 4 r_2 =2$.
 \label{tab_21}}
\end{table}

The Hilbert series refined only under the $U(1)_R$ symmetry with fugacities summarized in \tref{tab_21} takes the following form,
\beal{HS21}
g(\bar{t}_1, \bar{t}_2 ; D_3 ) = \frac{1+\overline{t}_1 \overline{t}_2^2 }{(1-\overline{t}_1^2) (1-\overline{t}^2_2)^2 (1- \overline{t}_1 \overline{t}^2_2) } \,.
\eea
Note that the Hilbert series in \eqref{HS21} is identical with \eqref{HS20}.
This extends the observation made in \cite{Ghim:2024asj, Ghim:2025zhs} for mass deformed brane brick models
corresponding to birational transformations of the associated toric Calabi-Yau 4-folds, which leave the Hilbert series of the mesonic moduli space refined only under the $U(1)_R$ symmetry
invariant. 
\\

\section{Connecting Non-Mass and Mass Deformations via Triality} \label{section_non_mass_to_mass_via_triality}

In this section, we examine the interplay between triality and relevant deformations. In particular, we show how non-mass deformations in one toric phase can be mapped to mass deformations in another phase related by triality.\footnote{An interesting question in both brane brick models and brane tilings is whether every relevant deformation connecting two toric geometries can be realized as a mass deformation in at least one toric phase.} Below, we illustrate this phenomenon with an explicit example.

\subsection{From $P_{+-} (\textrm{PdP}_{3b})$ to $P^2_{+-}(\textrm{dP}_3)$ \label{sec:pdp3}}

\begin{figure}[h]
	\centering
	\includegraphics[height=5cm]{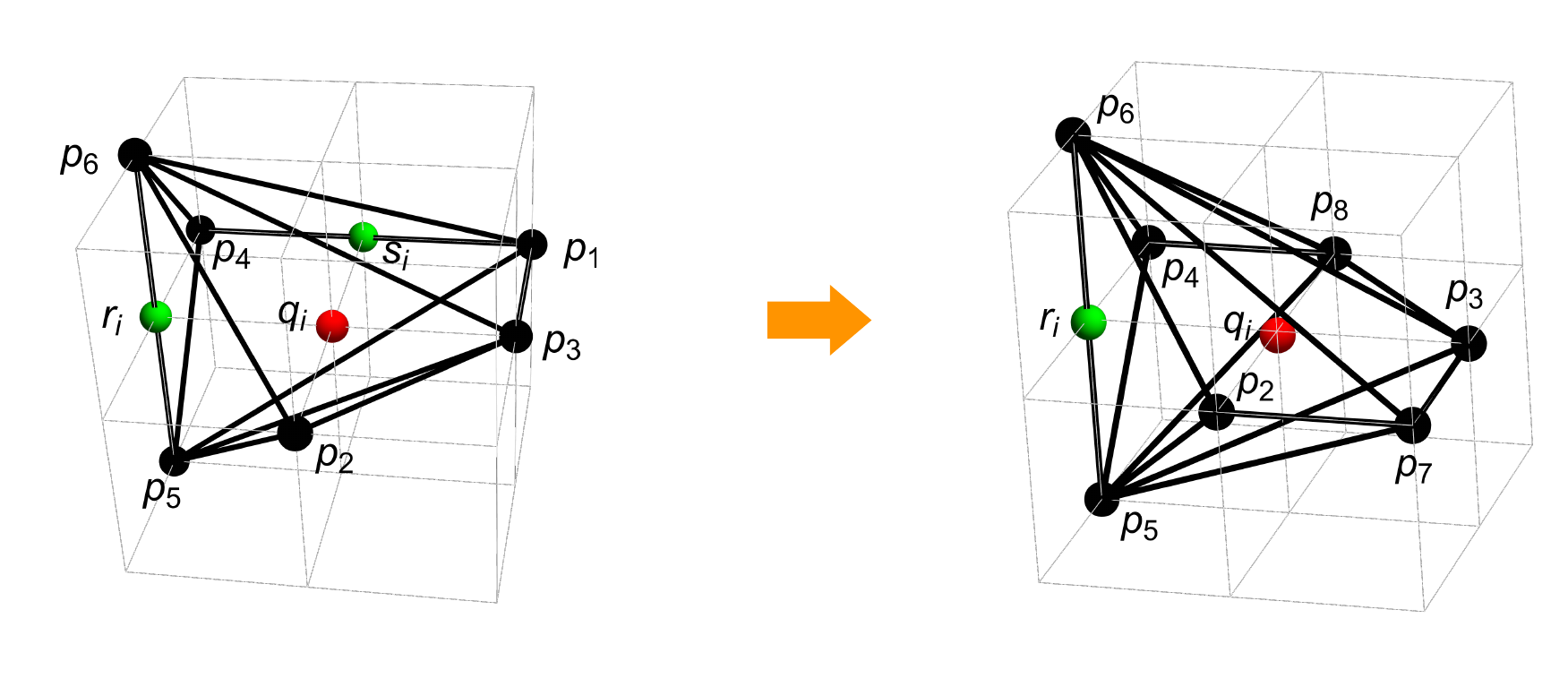}
\caption{Toric diagrams for $P_{+-} (\textrm{PdP}_{3b})$ and $P^2_{+-}(\textrm{dP}_3)$. We have labeled points anticipating the corresponding brick matchings.
}
	\label{toric_diagrams_example_3}
\end{figure}

We will study relevant deformations from $P_{+-} (\textrm{PdP}_{3b})$ to $P^2_{+-}(\textrm{dP}_3)$, whose toric diagrams are shown in \fref{toric_diagrams_example_3}. 
Like the examples we studied in Section \ref{sec:p2sppz2}, orbifold reduction techniques \cite{Franco:2016fxm} and the more general
$3d$ printing algorithm \cite{Franco:2018qsc} can be used to construct $2d$ $(0,2)$ theories corresponding to these two toric CY 4-folds from the $4d$ $\cN=1$ gauge theories 
associated to the toric CY 3-folds  $\textrm{dP}_3$ and $\textrm{PdP}_{3b}$, respectively.

The cone over the del Pezzo surface $\textrm{dP}_3$ has played a prominent role in the study of gauge theories on D3-branes probing toric CY 3-folds. For its phase structure and the corresponding brane tilings, we refer the readers to, e.g. \cite{Beasley:2001zp,Feng:2001bn,Feng:2002zw,Franco:2005rj}. The notion of pseudo del Pezzos was introduced in \cite{Feng:2002fv}. In short, $\textrm{PdP}_{n}$ corresponds to a blow-up of $\textrm{dP}_{0}$ at $n$ non-generic points. As a result, $\textrm{PdP}_{n}$’s contain non-isolated singularities, namely more than two collinear points on the boundary of their toric diagrams, which is a central feature of theories that admit deformations of the type studied in this paper. In general, there are multiple inequivalent ways of blowing-up three non-generic points in $\textrm{dP}_{0}$. In this section, we will focus on the one studied in \cite{Feng:2002fv}, which was dubbed $\textrm{PdP}_{3b}$ in \cite{Hanany:2012hi}.

\begin{figure}[ht!]
	\centering
	\includegraphics[width=0.8 \textwidth]{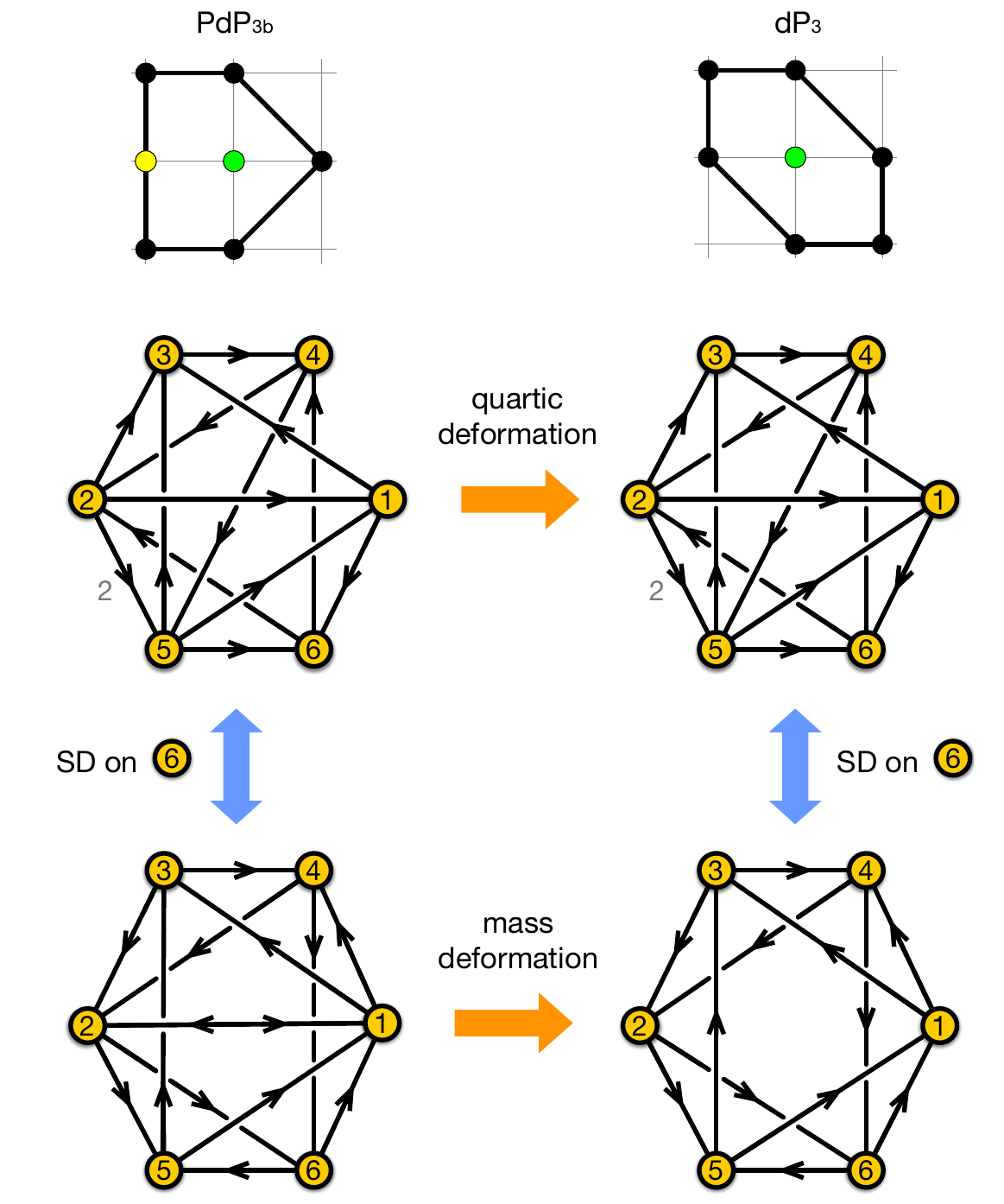}
\caption{Web of connections between some toric phases of $\textrm{PdP}_{3b}$ and $\textrm{dP}_3$ via Seiberg duality and relevant deformations. For the superpotentials for these theories, see e.g. \cite{Franco:2005rj}.}
	\label{PdP3_and_dP3}
\end{figure}

Before considering these CY 4-folds and their associated $2d$ gauge theories, it is instructive to first review the underlying $\textrm{PdP}_{3b}$ and $\textrm{dP}_3$ geometries and the corresponding $4d$ theories. Interestingly, these geometries can be connected by either non-mass or mass deformations, with the two types of deformations related by Seiberg duality. While each geometry admits multiple toric phases, we focus on those summarized in \fref{PdP3_and_dP3}. For phase labeling, we adopt the notation introduced in \cite{Hanany:2012hi}, to which we refer the reader for additional details.\footnote{This notation is not standard in the literature, but we follow it for consistency.}

In the second row of \fref{PdP3_and_dP3}, we observe that Phase (a) of $\textrm{PdP}_{3b}$ is connected to Phase (a) of $\textrm{dP}_3$ by a non-mass relevant deformation. The interesting fact that both theories share the same quiver but have different superpotentials satisfying the toric condition was first noted in \cite{Franco:2005rj}. A detailed analysis of the deformations was later presented in \cite{Cremonesi:2023psg}.

Applying Seiberg duality to Phase (a) of $\textrm{PdP}_{3b}$ leads to Phase (b), depicted in the third row of Figure \fref{PdP3_and_dP3}, where the original non-mass deformation is mapped to a mass term. Integrating out the massive fields yields Phase (b) of $\textrm{dP}_3$, which is itself related to Phase (a) of the same geometry by another Seiberg duality.

Below, we will show that the CY 4-folds constructed from these geometries exhibit a similarly rich structure of RG flows, with non-mass deformations turning into masses under triality.

\subsubsection*{A Non-Mass Deformation}

\begin{figure}[H]
\begin{center}
\resizebox{0.55\hsize}{!}{
\includegraphics[height=6cm]{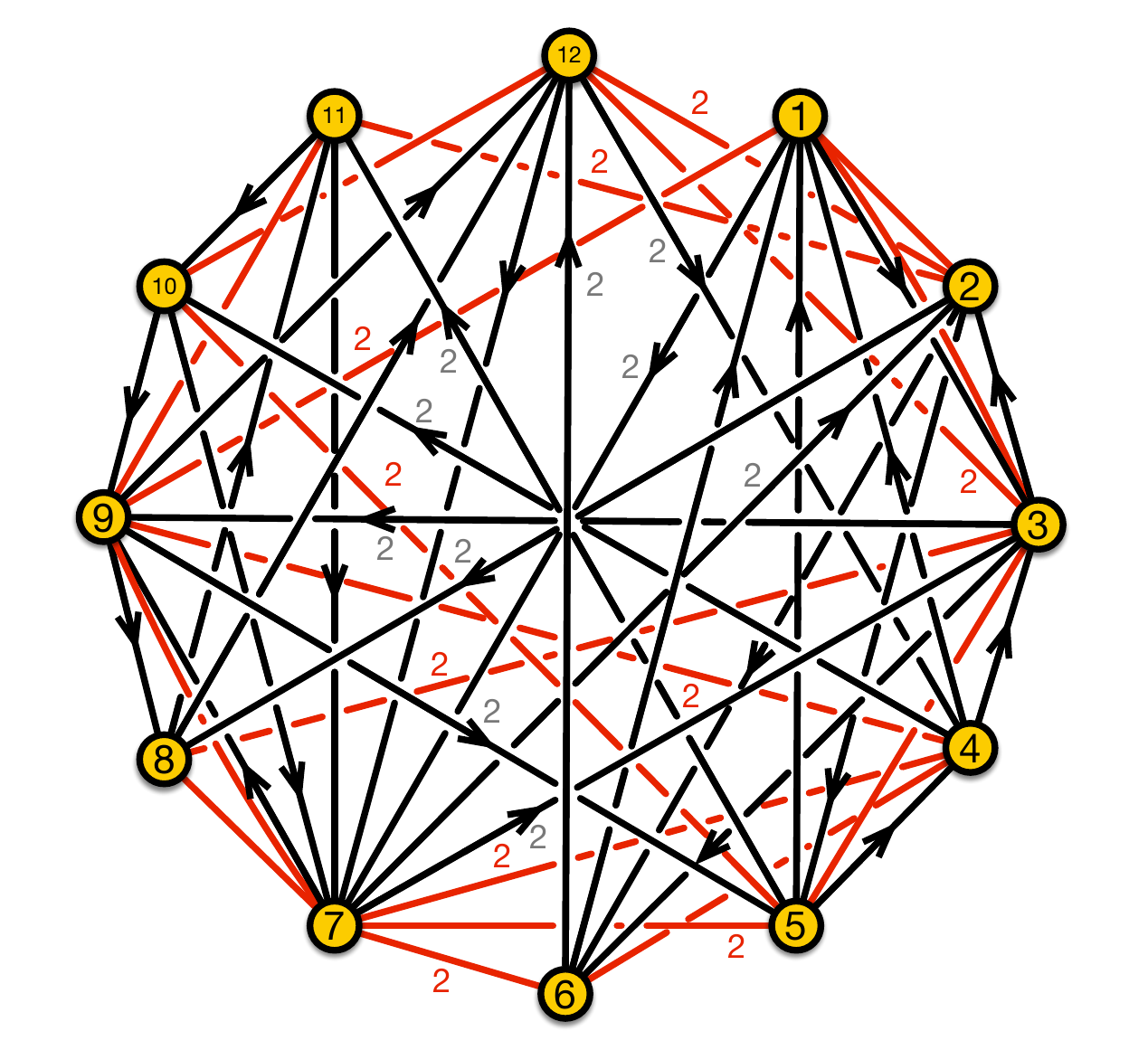}
}
\caption{Quiver diagram for Phase (a) of $P_{+-} (\textrm{PdP}_{3b})$. We will later consider a relevant deformation of this theory to a gauge theory for $P^2_{+-}(\textrm{dP}_3)$. Since this will be a non-mass deformation, the new theory will share the same quiver.
\label{fig_q_ppdp3a}}
 \end{center}
 \end{figure}

Let us consider $P_{+-} (\textrm{PdP}_{3b})$, whose toric diagram is given in \fref{toric_diagrams_example_3}. Our starting point is a $2d$ theory for this geometry constructed using orbifold reduction from Phase (b) of $\textrm{PdP}_{3b}$ in \cite{Hanany:2012hi}. \fref{fig_q_ppdp3a} shows the quiver diagram for this theory. We will call this theory Phase (a).

The $J$- and $E$-terms of this theory are given by
{\footnotesize
\begin{align}
\begin{array}{rcrclcrcl}
& &  \ \ \ \ \ \ \ \ \ \ \ \ \ \ \ \ \ & J& &&& E&  \ \ \ \ \ \ \ \ \ \ \ \ \ \  \\
 \Lambda_{12} : & \ \ \  & X_{25}\cdot X_{54}\cdot X_{41} & - & X_{26}\cdot X_{61} & \ \ \ \  & P_{17}\cdot U_{72}& -& Q_{17}\cdot W_{72}\\
 \Lambda_{64} : & \ \ \  & X_{43}\cdot X_{32}\cdot X_{26} & - & X_{41}\cdot X_{13}\cdot X_{36} & \ \ \ \  & P_{6,12}\cdot U_{12,4}& -& Q_{6,12}\cdot W_{12,4}\\
 \Lambda_{13} : & \ \ \  & X_{36}\cdot X_{61} & - & X_{32}\cdot X_{25}\cdot X_{51}& \ \ \ \  & P_{17}\cdot U_{73}& -& Q_{17}\cdot W_{73}\\
\Lambda_{35} : & \ \ \  & X_{51}\cdot X_{13} & - & X_{54}\cdot X_{43} & \ \ \ \  & P_{39}\cdot U_{95}& -& Q_{39}\cdot W_{95}\\
\Lambda_{78} : & \ \ \  & X_{8,11}\cdot X_{11,10}\cdot X_{10,7} & - & X_{8,12}\cdot X_{12,7} & \ \ \ \  &W_{72}\cdot Q_{28}& -&  U_{72}\cdot P_{28} \\
\Lambda_{12,10} : & \ \ \  & X_{10,9}\cdot X_{98}\cdot X_{8,12} & - & X_{10,7}\cdot X_{79}\cdot X_{9,12} & \ \ \ \  &W_{12,4}\cdot Q_{4,10} & -& U_{12,4}\cdot P_{4,10}\\
\Lambda_{79} : & \ \ \  & X_{9,12}\cdot X_{12,7} & - & X_{98}\cdot X_{8,11}\cdot X_{11,7} & \ \ \ \  &W_{73}\cdot Q_{39} & -&U_{73}\cdot P_{39} \\
\Lambda_{9,11} : & \ \ \  & X_{11,7}\cdot X_{79} & - & X_{11,10}\cdot X_{10,9} & \ \ \ \  &W_{95}\cdot Q_{5,11} & -&U_{95}\cdot P_{5,11} \\
\Lambda^1_{2,12} : & \ \ \  & U_{12,4}\cdot X_{43}\cdot X_{32} & - & X_{12,7}\cdot U_{72} & \ \ \ \  & P_{28}\cdot X_{8,12}& -& X_{26}\cdot P_{6,12}\\
\Lambda^2_{2,12} : & \ \ \  & W_{12,4}\cdot X_{43}\cdot X_{32} & - & X_{12,7}\cdot W_{72} & \ \ \ \  & X_{26}\cdot Q_{6,12}& -& Q_{28}\cdot X_{8,12}\\
\Lambda^1_{67} : & \ \ \  & U_{73}\cdot X_{36}& - & U_{72}\cdot X_{26} & \ \ \ \  & P_{6,12}\cdot X_{12,7}& -& X_{61}\cdot P_{17}\\
\Lambda^2_{67} : & \ \ \  & W_{73}\cdot X_{36}& - & W_{72}\cdot X_{26} & \ \ \ \  & X_{61}\cdot Q_{17}& -& Q_{6,12}\cdot X_{12,7}\\
\Lambda^1_{49} : & \ \ \  & X_{98}\cdot X_{8,12}\cdot U_{12,4} & - & U_{95}\cdot X_{54} & \ \ \ \  & P_{4,10}\cdot X_{10,9} & -& X_{43}\cdot P_{39}\\
\Lambda^2_{49} : & \ \ \  & X_{98}\cdot X_{8,12}\cdot W_{12,4} & - & W_{95}\cdot X_{54} & \ \ \ \  & X_{43}\cdot Q_{39} & -& Q_{4,10}\cdot X_{10,9}\\
\Lambda^1_{3,12} : & \ \ \  & X_{12,7}\cdot U_{73} & - & U_{12,4}\cdot X_{41}\cdot X_{13} & \ \ \ \  & P_{39}\cdot X_{9,12} & -& X_{36}\cdot P_{6,12}\\
\Lambda^2_{3,12} : & \ \ \  & X_{12,7}\cdot W_{73} & - & W_{12,4}\cdot X_{41}\cdot X_{13} & \ \ \ \  & X_{36}\cdot Q_{6,12} & -& Q_{39}\cdot X_{9,12}\\
\Lambda^1_{2,11} : & \ \ \  & X_{11,10}\cdot X_{10,7}\cdot U_{72} & - & X_{11,7}\cdot U_{73}\cdot X_{32} & \ \ \ \  & P_{28}\cdot X_{8,11} & -& X_{25}\cdot P_{5,11}\\
\Lambda^2_{2,11} : & \ \ \  & X_{11,10}\cdot X_{10,7}\cdot W_{72} & - & X_{11,7}\cdot W_{73}\cdot X_{32} & \ \ \ \  & X_{25}\cdot Q_{5,11} & -& Q_{28}\cdot X_{8,11}\\
\end{array}
\nn
\end{align}
}
{\footnotesize
\begin{align}
\begin{array}{rcrclcrcl}
\Lambda^1_{47} : & \ \ \  & U_{72}\cdot X_{25}\cdot X_{54} & - & X_{79}\cdot X_{9,12}\cdot U_{12,4}& \ \ \ \  & P_{4,10}\cdot X_{10,7} & -& X_{41}\cdot P_{17}\\
\Lambda^2_{47} : & \ \ \  & W_{72}\cdot X_{25}\cdot X_{54} & - & X_{79}\cdot X_{9,12}\cdot W_{12,4}& \ \ \ \  & X_{41}\cdot Q_{17} & -&Q_{4,10}\cdot X_{10,7} \\
\Lambda^1_{38} : & \ \ \  & X_{8,12}\cdot U_{12,4}\cdot X_{43} & - & X_{8,11}\cdot X_{11,7}\cdot U_{73}& \ \ \ \  & P_{39}\cdot X_{98} & -& X_{32}\cdot P_{28}\\
\Lambda^2_{38} : & \ \ \  & X_{8,12}\cdot W_{12,4}\cdot X_{43} & - & X_{8,11}\cdot X_{11,7}\cdot W_{73}& \ \ \ \  & X_{32}\cdot Q_{28} & -& Q_{39}\cdot X_{98}\\
\Lambda^1_{5,10} : & \ \ \  & X_{10,7}\cdot U_{72}\cdot X_{25} & - & X_{10,9}\cdot U_{95}& \ \ \ \  & P_{5,11}\cdot X_{11,10} & -& X_{54}\cdot P_{4,10}\\
\Lambda^2_{5,10} : & \ \ \  & X_{10,7}\cdot W_{72}\cdot X_{25} & - & X_{10,9}\cdot W_{95}& \ \ \ \  & X_{54}\cdot Q_{4,10} & -& Q_{5,11}\cdot X_{11,10}\\
\Lambda^1_{57} : & \ \ \  & X_{79}\cdot U_{95} & - & U_{73}\cdot X_{32}\cdot X_{25} & \ \ \ \  & P_{5,11}\cdot X_{11,7} & -& X_{51}\cdot P_{17}\\
\Lambda^2_{57} : & \ \ \  & X_{79}\cdot W_{95} & - & W_{73}\cdot X_{32}\cdot X_{25} & \ \ \ \  & X_{51}\cdot Q_{17} & -& Q_{5,11}\cdot X_{11,7}\\
\Lambda^1_{19} : & \ \ \  & U_{95}\cdot X_{51} & - & X_{9,12}\cdot U_{12,4}\cdot X_{41} & \ \ \ \  & P_{17}\cdot X_{79} & -& X_{13}\cdot P_{39}\\
\Lambda^2_{19} : & \ \ \  & W_{95}\cdot X_{51} & - & X_{9,12}\cdot W_{12,4}\cdot X_{41} & \ \ \ \  & X_{13}\cdot Q_{39} & -& Q_{17}\cdot X_{79}\\
\end{array}
~.~ 
\label{PdP3(+-)}
\end{align}
}

Let us now consider the following deformation on $J$-terms, where the new contributions are indicated in blue. Note that all the deformation terms correspond to cubic plaquettes, hence, they are not mass terms.
\begin{align}
\resizebox{0.85\textwidth}{!}{$
\begin{array}{rcrclcrcl}
& &  \ \ \ \ \ \ \ \ \ \ \ \ \ \ \ \ \ & J& + \, \textcolor{blue}{\Delta J} &&& E&  \ \ \ \ \ \ \ \ \ \ \ \ \ \  \\
 \Lambda_{12} : & \ \ \  & X_{25}\cdot X_{54}\cdot X_{41} & - & X_{26}\cdot X_{61}+\textcolor{blue}{\mu \, X_{25}\cdot X_{51}} & \ \ \ \  & P_{17}\cdot U_{72}& -& Q_{17}\cdot W_{72}\\
 \Lambda_{64} : & \ \ \  & X_{43}\cdot X_{32}\cdot X_{26} & - & X_{41}\cdot X_{13}\cdot X_{36}-\textcolor{blue}{\mu \, X_{43}\cdot X_{36}} & \ \ \ \  & P_{6,12}\cdot U_{12,4}& -& Q_{6,12}\cdot W_{12,4}\\
 \Lambda_{78} : & \ \ \  & X_{8,11}\cdot X_{11,10}\cdot X_{10,7} & - & X_{8,12}\cdot X_{12,7}+\textcolor{blue}{\mu \, X_{8,11}\cdot X_{11,7}} & \ \ \ \  &W_{72}\cdot Q_{28}& -&  U_{72}\cdot P_{28}\\
\Lambda_{12,10} : & \ \ \  & X_{10,9}\cdot X_{98}\cdot X_{8,12} & - & X_{10,7}\cdot X_{79}\cdot X_{9,12}-\textcolor{blue}{\mu \, X_{10,9}\cdot X_{9,12}} & \ \ \ \  &W_{12,4}\cdot Q_{4,10}& -&  U_{12,4}\cdot P_{4,10}\\
\Lambda^1_{49} : & \ \ \  & X_{98}\cdot X_{8,12}\cdot U_{12,4} & - & U_{95}\cdot X_{54}-\textcolor{blue}{\mu \, X_{9,12}\cdot U_{12,4}} & \ \ \ \  & P_{4,10}\cdot X_{10,9} & -& X_{43}\cdot P_{39}\\
\Lambda^2_{49} : & \ \ \  & X_{98}\cdot X_{8,12}\cdot W_{12,4} & - & W_{95}\cdot X_{54}-\textcolor{blue}{\mu \, X_{9,12}\cdot W_{12,4}} & \ \ \ \  & X_{43}\cdot Q_{39} & -& Q_{4,10}\cdot X_{10,9}\\
\Lambda^1_{3,12} : & \ \ \  & X_{12,7}\cdot U_{73} & - & U_{12,4}\cdot X_{41}\cdot X_{13}-\textcolor{blue}{\mu \, U_{12,4}\cdot X_{43}} & \ \ \ \  & P_{39}\cdot X_{9,12} & -& X_{36}\cdot P_{6,12}\\
\Lambda^2_{3,12} : & \ \ \  & X_{12,7}\cdot W_{73} & - & W_{12,4}\cdot X_{41}\cdot X_{13}-\textcolor{blue}{\mu \, W_{12,4}\cdot X_{43}} & \ \ \ \  & X_{36}\cdot Q_{6,12} & -& Q_{39}\cdot X_{9,12}\\
\Lambda^1_{2,11} : & \ \ \  & X_{11,10}\cdot X_{10,7}\cdot U_{72} & - & X_{11,7}\cdot U_{73}\cdot X_{32}+\textcolor{blue}{\mu \, X_{11,7}\cdot U_{72}} & \ \ \ \  & P_{28}\cdot X_{8,11} & -& X_{25}\cdot P_{5,11}\\
\Lambda^2_{2,11} : & \ \ \  & X_{11,10}\cdot X_{10,7}\cdot W_{72} & - & X_{11,7}\cdot W_{73}\cdot X_{32}+\textcolor{blue}{\mu \, X_{11,7}\cdot W_{72}} & \ \ \ \  & X_{25}\cdot Q_{5,11} & -& Q_{28}\cdot X_{8,11}\\
\Lambda^1_{57} : & \ \ \  & X_{79}\cdot U_{95} & - & U_{73}\cdot X_{32}\cdot X_{25}+\textcolor{blue}{\mu \, U_{72}\cdot X_{25}}& \ \ \ \  & P_{5,11}\cdot X_{11,7} & -& X_{51}\cdot P_{17}\\
\Lambda^2_{57} : & \ \ \  & X_{79}\cdot W_{95} & - & W_{73}\cdot X_{32}\cdot X_{25}+\textcolor{blue}{\mu \, W_{72}\cdot X_{25}}& \ \ \ \  & X_{51}\cdot Q_{17} & -& Q_{5,11}\cdot X_{11,7}\\
\end{array}
~.~$}
\label{PdP3(+-)_deform}
\end{align}

Additional higher-order couplings should be introduced by the change of variables in the form of \eqref{def-cov}. First, we change chiral fields as follows,
{\footnotesize
\begin{align} \label{dP32redefs}
\begin{split}
   & U_{72} \rightarrow -\frac{1}{\mu}U_{72} + \frac{1}{\mu}U_{73}\cdot X_{32} \,, \\
   & W_{72} \rightarrow -\frac{1}{\mu}W_{72} + \frac{1}{\mu}W_{73}\cdot X_{32} \,, \\
   & X_{43} \rightarrow \frac{1}{\mu}X_{43} - \frac{1}{\mu}X_{41}\cdot X_{13} \,, \\
   & X_{51} \rightarrow \frac{1}{\mu}X_{51} - \frac{1}{\mu}X_{54}\cdot X_{41} \,, \\
   & X_{36} \rightarrow -\frac{1}{\mu}X_{36} + \frac{1}{\mu}X_{32}\cdot X_{26} \,, \\
   & X_{10,9} \rightarrow \frac{1}{\mu}X_{10,9} -\frac{1}{\mu} X_{10,7}\cdot X_{79} \,, \\
   & X_{11,7} \rightarrow\frac{1}{\mu}X_{11,7} - \frac{1}{\mu}X_{11,10}\cdot X_{10,7} \,, \\
   & X_{9,12} \rightarrow -\frac{1}{\mu}X_{9,12} + \frac{1}{\mu}X_{98}\cdot X_{8,12} \,.
\end{split}
\end{align}
}
Next, we modify the $J$- and $E$-terms by changing the Fermi field variables, following the rule in \eqref{cov-summary},  
{\footnotesize
\begin{align} \label{dP32redefsF}
\begin{split}
  & \Lambda_{12} \rightarrow -\frac{1}{\mu}\Lambda_{12} + \frac{1}{\mu}\Lambda_{13}\cdot X_{32} \,, \\
  & \Lambda_{78} \rightarrow -\frac{1}{\mu}\Lambda_{78} + \frac{1}{\mu}\Lambda_{79}\cdot X_{98} + \frac{1}{\mu}U_{73}\cdot\Lambda^1_{38} + \frac{1}{\mu}W_{73}\cdot\Lambda^2_{38} \,, \\
  & \Lambda^1_{49} \rightarrow \frac{1}{\mu}\Lambda^{1}_{49} - \frac{1}{\mu}\Lambda^1_{47}\cdot X_{79} - \frac{1}{\mu}X_{41}\cdot\Lambda^1_{19} \,, \\
  & \Lambda^2_{49} \rightarrow \frac{1}{\mu}\Lambda^{2}_{49} - \frac{1}{\mu}\Lambda^2_{47}\cdot X_{79} - \frac{1}{\mu}X_{41}\cdot\Lambda^2_{19} \,, \\
  & \Lambda^1_{3,12} \rightarrow -\frac{1}{\mu}\Lambda^{1}_{3,12} + \frac{1}{\mu}\Lambda^1_{38}\cdot X_{8,12} + \frac{1}{\mu}X_{32}\cdot\Lambda^1_{2,12} \,, \\
  & \Lambda^2_{3,12} \rightarrow -\frac{1}{\mu}\Lambda^{2}_{3,12} + \frac{1}{\mu}\Lambda^2_{38}\cdot X_{8,12} + \frac{1}{\mu}X_{32}\cdot\Lambda^2_{2,12} \,, \\
  & \Lambda^1_{57} \rightarrow \frac{1}{\mu}\Lambda^{1}_{57} - \frac{1}{\mu}\Lambda^1_{5,10}\cdot X_{10,7} + \frac{1}{\mu}X_{54}\cdot\Lambda^1_{47} \,, \\
  & \Lambda^2_{57} \rightarrow \frac{1}{\mu}\Lambda^{2}_{57} - \frac{1}{\mu}\Lambda^2_{5,10}\cdot X_{10,7} + \frac{1}{\mu}X_{54}\cdot\Lambda^2_{47} \,.\\
\end{split}
\end{align}
}

After change of variables followed by the rescaling of $J$-terms to remove the overall $\frac{1}{\mu}$ factor, we obtain a theory with the same quiver diagram in \fref{fig_q_ppdp3a} but with the following $J$- and $E$-terms,
{\footnotesize
\begin{align}
\begin{array}{rcrclcrcl}
& &  \ \ \ \ \ \ \ \ \ \ \ \ \ \ \ \ \ & J& &&& E&  \ \ \ \ \ \ \ \ \ \ \ \ \ \  \\
 \Lambda_{12} : & \ \ \  & X_{26}\cdot X_{61} & - & X_{25}\cdot X_{51}& \ \ \ \  & P_{17}\cdot U_{72}& -& Q_{17}\cdot W_{72}\\
 \Lambda_{64} : & \ \ \  & X_{43}\cdot X_{36} & - & X_{41}\cdot X_{13}\cdot X_{32}\cdot X_{26} & \ \ \ \  & P_{6,12}\cdot U_{12,4}& -& Q_{6,12}\cdot W_{12,4}\\
 \Lambda_{13} : & \ \ \  & X_{32}\cdot X_{25}\cdot X_{54}\cdot X_{41} & - & X_{36}\cdot X_{61}& \ \ \ \  & P_{17}\cdot U_{73}& -& Q_{17}\cdot W_{73}\\
\Lambda_{35} : & \ \ \  & X_{51}\cdot X_{13} & - & X_{54}\cdot X_{43} & \ \ \ \  & P_{39}\cdot U_{95}& -& Q_{39}\cdot W_{95}\\
\Lambda_{78} : & \ \ \  & X_{8,12}\cdot X_{12,7} & - & X_{8,11}\cdot X_{11,7} & \ \ \ \  &W_{72}\cdot Q_{28} & -& U_{72}\cdot P_{28}\\
\Lambda_{12,10} : & \ \ \  & X_{10,9}\cdot X_{9,12}& - & X_{10,7}\cdot X_{79}\cdot X_{98}\cdot X_{8,12} & \ \ \ \  &W_{12,4}\cdot Q_{4,10}  & -&U_{12,4}\cdot P_{4,10}\\
\Lambda_{79} : & \ \ \  & X_{98}\cdot X_{8,11}\cdot X_{11,10}\cdot X_{10,7} & - & X_{9,12}\cdot X_{12,7} & \ \ \ \  &W_{73}\cdot Q_{39} & -&U_{73}\cdot P_{39} \\
\Lambda_{9,11} : & \ \ \  & X_{11,7}\cdot X_{79} & - & X_{11,10}\cdot X_{10,9} & \ \ \ \  & W_{95}\cdot Q_{5,11}& -& U_{95}\cdot P_{5,11} \\
\Lambda^1_{2,12} : & \ \ \  & X_{12,7}\cdot U_{72} & - & U_{12,4}\cdot X_{41}\cdot X_{13}\cdot X_{32} & \ \ \ \  & P_{28}\cdot X_{8,12}& -& X_{26}\cdot P_{6,12}\\
\Lambda^2_{2,12} : & \ \ \  & X_{12,7}\cdot W_{72} & - & W_{12,4}\cdot X_{41}\cdot X_{13}\cdot X_{32} & \ \ \ \  & X_{26}\cdot Q_{6,12}& -& Q_{28}\cdot X_{8,12}\\
\Lambda^1_{67} : & \ \ \  & U_{72}\cdot X_{26}& - & U_{73}\cdot X_{36} & \ \ \ \  & P_{6,12}\cdot X_{12,7}& -& X_{61}\cdot P_{17}\\
\Lambda^2_{67} : & \ \ \  & W_{72}\cdot X_{26}& - & W_{73}\cdot X_{36} & \ \ \ \  & X_{61}\cdot Q_{17}& -& Q_{6,12}\cdot X_{12,7}\\
\Lambda^{1}_{49} : & \ \ \  & X_{9,12}\cdot U_{12,4} & - & U_{95}\cdot X_{54} & \ \ \ \  & P_{4,10}\cdot X_{10,9} & -& X_{43}\cdot P_{39}\\
\Lambda^{2}_{49} : & \ \ \  & X_{9,12}\cdot W_{12,4} & - & W_{95}\cdot X_{54} & \ \ \ \  & X_{43}\cdot Q_{39} & -& Q_{4,10}\cdot X_{10,9}\\
\Lambda^{1}_{3,12} : & \ \ \  & U_{12,4}\cdot X_{43} & - & X_{12,7}\cdot U_{73} & \ \ \ \  & P_{39}\cdot X_{9,12} & -& X_{36}\cdot P_{6,12}\\
\Lambda^{2}_{3,12} : & \ \ \  & W_{12,4}\cdot X_{43} & - & X_{12,7}\cdot W_{73} & \ \ \ \  & X_{36}\cdot Q_{6,12} & -& Q_{39}\cdot X_{9,12}\\
\Lambda^1_{2,11} : & \ \ \  & X_{11,10}\cdot X_{10,7}\cdot U_{73}\cdot X_{32} & - & X_{11,7}\cdot U_{72}& \ \ \ \  & P_{28}\cdot X_{8,11} & -& X_{25}\cdot P_{5,11}\\
\Lambda^2_{2,11} : & \ \ \  & X_{11,10}\cdot X_{10,7}\cdot W_{73}\cdot X_{32} & - & X_{11,7}\cdot W_{72}& \ \ \ \  & X_{25}\cdot Q_{5,11} & -& Q_{28}\cdot X_{8,11}\\
\Lambda^1_{47} : & \ \ \  & U_{73}\cdot X_{32}\cdot X_{25}\cdot X_{54} & - & X_{79}\cdot X_{98}\cdot X_{8,12}\cdot U_{12,4}& \ \ \ \  & P_{4,10}\cdot X_{10,7} & -& X_{41}\cdot P_{17}\\
\Lambda^2_{47} : & \ \ \  & W_{73}\cdot X_{32}\cdot X_{25}\cdot X_{54} & - & X_{79}\cdot X_{98}\cdot X_{8,12}\cdot W_{12,4}& \ \ \ \  & X_{41}\cdot Q_{17} & -& Q_{4,10}\cdot X_{10,7}\\
\Lambda^1_{38} : & \ \ \  & X_{8,11}\cdot X_{11,10}\cdot X_{10,7}\cdot U_{73} & - & X_{8,12}\cdot U_{12,4}\cdot X_{41}\cdot X_{13}& \ \ \ \  & P_{39}\cdot X_{98} & -& X_{32}\cdot P_{28}\\
\Lambda^2_{38} : & \ \ \  & X_{8,11}\cdot X_{11,10}\cdot X_{10,7}\cdot W_{73} & - & X_{8,12}\cdot W_{12,4}\cdot X_{41}\cdot X_{13}& \ \ \ \  & X_{32}\cdot Q_{28} & -& Q_{39}\cdot X_{98}\\
\end{array}
\nn
\end{align}
}
{\footnotesize
\begin{align}
\begin{array}{rcrclcrcl}
\Lambda^1_{5,10} : & \ \ \  & X_{10,7}\cdot U_{73}\cdot X_{32}\cdot X_{25} & - & X_{10,9}\cdot U_{95}& \ \ \ \  & P_{5,11}\cdot X_{11,10} & -& X_{54}\cdot P_{4,10}\\
\Lambda^2_{5,10} : & \ \ \  & X_{10,7}\cdot W_{73}\cdot X_{32}\cdot X_{25} & - & X_{10,9}\cdot W_{95}& \ \ \ \  & X_{54}\cdot Q_{4,10} & -& Q_{5,11}\cdot X_{11,10}\\
\Lambda^{1}_{57} : & \ \ \  & X_{79}\cdot U_{95} & - & U_{72}\cdot X_{25}& \ \ \ \  & P_{5,11}\cdot X_{11,7} & -& X_{51}\cdot P_{17}\\
\Lambda^{2}_{57} : & \ \ \  & X_{79}\cdot W_{95} & - & W_{72}\cdot X_{25}& \ \ \ \  & X_{51}\cdot Q_{17} & -& Q_{5,11}\cdot X_{11,7}\\
\Lambda^1_{19} : & \ \ \  & U_{95}\cdot X_{51} & - & X_{98}\cdot X_{8,12}\cdot U_{12,4}\cdot X_{41} & \ \ \ \  & P_{17}\cdot X_{79} & -& X_{13}\cdot P_{39}\\
\Lambda^2_{19} : & \ \ \  & W_{95}\cdot X_{51} & - & X_{98}\cdot X_{8,12}\cdot W_{12,4}\cdot X_{41} & \ \ \ \  & X_{13}\cdot Q_{39} & -& Q_{17}\cdot X_{79}\\
\end{array}
~.~ 
\label{dP3(+-)}
\end{align}
}

The forward algorithm confirms that this theory corresponds to $P^2_{+-}(\textrm{dP}_3)$, whose toric diagram is presented in the right-hand side of \fref{toric_diagrams_example_3}. The extended $\bar{P}$-matrices for these theories are collected in Appendix \ref{sec:pmdata}. 
We call this theory Phase (a) of $P^2_{+-} (\textrm{dP}_3) $ to distinguish its other phases to be mentioned later. 

As the attentive reader may have realized—much like in other examples discussed in this paper—the choice of the theory to deform and the specific deformation implemented are not arbitrary, but are instead guided by a clear understanding of the desired final result. In fact, the final theory we obtained can be directly constructed via $3d$ printing, starting from Phase (b) of $\textrm{dP}_3$ in \cite{Hanany:2012hi}. In other words, the non-mass deformation we presented descends from the theories in the first row of \fref{PdP3_and_dP3} through $3d$ printing.

\begin{figure}[H]
\begin{center}
\resizebox{0.55\hsize}{!}{
\includegraphics[height=6cm]{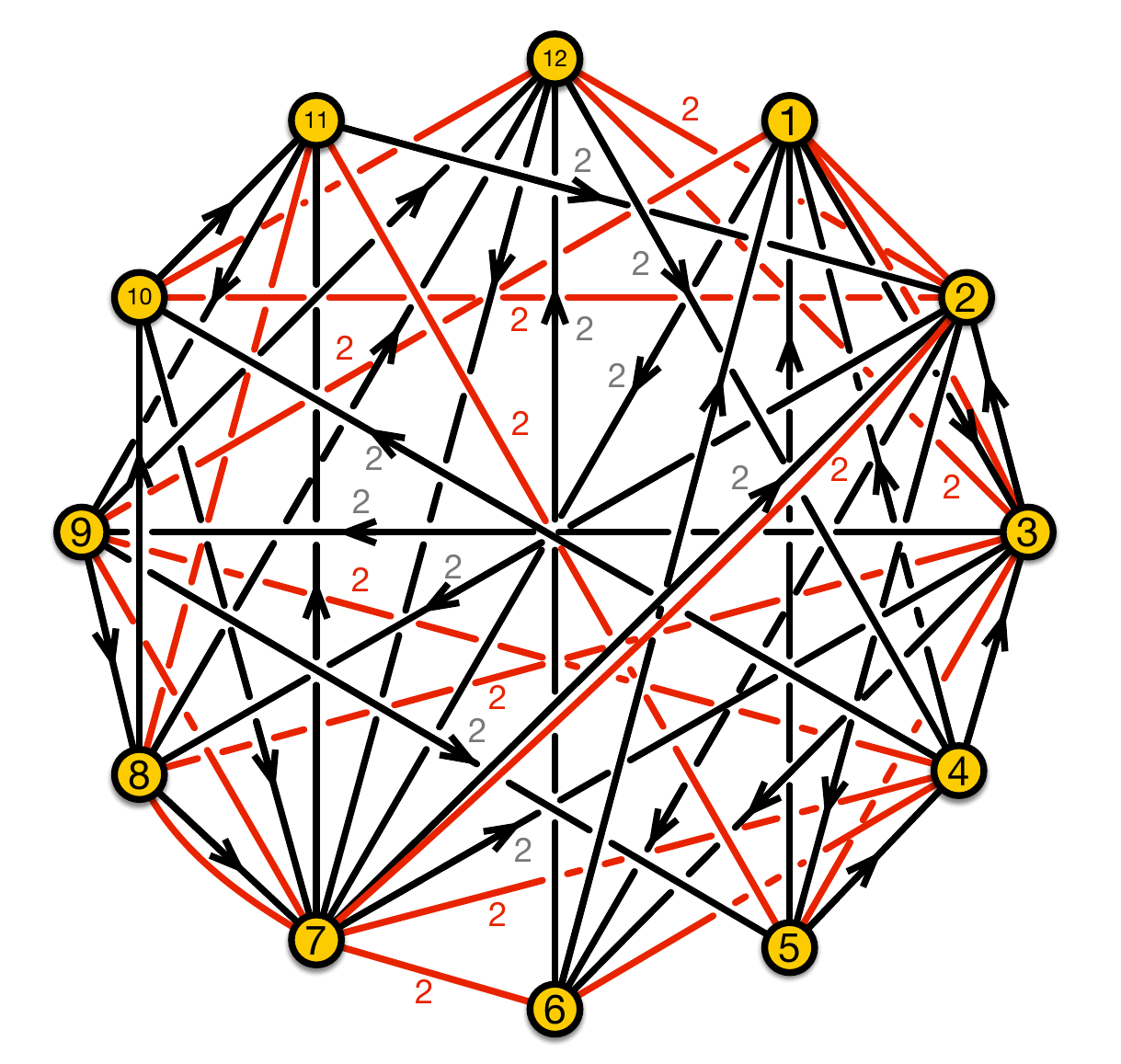}
}
\caption{Quiver diagram for Phase (b) of $P_{+-} (\textrm{PdP}_{3b})$, obtained from the one in \fref{fig_q_ppdp3a} by acting the inverse triality on node 11.
\label{fig_q_ppdp3b}}
 \end{center}
 \end{figure}

\subsubsection*{Mass Deformation in a Triality Dual Phase}

Let us return to the starting point of the deformation considered above, Phase (a) for $P_{+-} (\textrm{PdP}_{3b})$, whose quiver is shown in \fref{fig_q_ppdp3a}. Node 11 has two outgoing arrows, so the inverse triality operation on it results in another toric phase, which we call Phase (b). \fref{fig_q_ppdp3b} shows the quiver for this phase.

The $J$- and $E$-terms for this phase are
\begin{align}
\resizebox{0.95\textwidth}{!}{$
\begin{array}{rcrclcrcl}
& &  \ \ \ \ \ \ \ \ \ \ \ \ \ \ \ \ \ & J& &&& E&  \ \ \ \ \ \ \ \ \ \ \ \ \ \  \\
 \Lambda_{12} : & \ \ \  & X_{25}\cdot X_{54}\cdot X_{41} & - & X_{26}\cdot X_{61} & \  \ \  & P_{17}\cdot U_{72}& - &Q_{17}\cdot W_{72}\\
 \Lambda_{64} : & \ \ \  & X_{43}\cdot X_{32}\cdot X_{26} & - & X_{41}\cdot X_{13}\cdot X_{36} & \  \ \  & P_{6,12}\cdot U_{12,4} &-& Q_{6,12}\cdot W_{12,4}\\
 \Lambda_{13} : & \ \ \  & X_{36}\cdot X_{61} & - & X_{32}\cdot X_{25}\cdot X_{51}& \ \  \  & P_{17}\cdot U_{73}& -& Q_{17}\cdot W_{73}\\
\Lambda_{35} : & \ \ \  & X_{51}\cdot X_{13} & - & X_{54}\cdot X_{43} & \ \ \ \  & P_{39}\cdot U_{95} & - & Q_{39}\cdot W_{95}\\
\Lambda_{78} : & \ \ \  & R_{8,10}\cdot X_{10,7} & - & X_{8,12}\cdot X_{12,7}  & \ \ \ \  & W_{72}\cdot Q_{28} &- & U_{72}\cdot P_{28}\\
\Lambda_{12,10} : & \ \ \  & X_{10,11}\cdot X_{11,9}\cdot X_{98}\cdot X_{8,12} & - & X_{10,7}\cdot X_{7,11}\cdot X_{11,9}\cdot X_{9,12} & \ \ \ \  & W_{12,4}\cdot Q_{4,10} &- &U_{12,4}\cdot P_{4,10}\\
\Lambda_{79} : & \ \ \  & X_{9,12}\cdot X_{12,7} & - & X_{98}\cdot V_{87} & \ \ \ \  &  W_{73}\cdot Q_{39} &-&U_{73}\cdot P_{39}\\
\Lambda^1_{2,12} : & \ \ \  & U_{12,4}\cdot X_{43}\cdot X_{32} & - & X_{12,7}\cdot U_{72} & \ \ \ \  & P_{28}\cdot X_{8,12} & - &X_{26}\cdot P_{6,12}\\
\Lambda^2_{2,12} : & \ \ \  & W_{12,4}\cdot X_{43}\cdot X_{32} & - & X_{12,7}\cdot W_{72} & \ \ \ \  & X_{26}\cdot Q_{6,12} & - & Q_{28}\cdot X_{8,12}\\
\Lambda^1_{67} : & \ \ \  & U_{73}\cdot X_{36} & - & U_{72}\cdot X_{26} & \ \ \ \  & P_{6,12}\cdot X_{12,7}& - & X_{61}\cdot P_{17}\\
\Lambda^2_{67} : & \ \ \  & W_{73}\cdot X_{36} & - & W_{72}\cdot X_{26} & \ \ \ \  & X_{61}\cdot Q_{17} &- &Q_{6,12}\cdot X_{12,7}\\
\Lambda^1_{49} : & \ \ \  & X_{98}\cdot X_{8,12}\cdot U_{12,4} & - & U_{95}\cdot X_{54} & \ \ \ \  & P_{4,10}\cdot X_{10,11}\cdot X_{11,9} & - & X_{43}\cdot P_{39}\\
\Lambda^2_{49} : & \ \ \  & X_{98}\cdot X_{8,12}\cdot W_{12,4} & -& W_{95}\cdot X_{54} & \ \ \ \  & X_{43}\cdot Q_{39} & - & Q_{4,10}\cdot X_{10,11}\cdot X_{11,9}\\
\Lambda^1_{3,12} : & \ \ \  & X_{12,7}\cdot U_{73} & - & U_{12,4}\cdot X_{41}\cdot X_{13} & \ \ \ \  & P_{39}\cdot X_{9,12} & - & X_{36}\cdot P_{6,12}\\
\Lambda^2_{3,12} : & \ \ \  & X_{12,7}\cdot W_{73} & - & W_{12,4}\cdot X_{41}\cdot X_{13} & \ \ \ \  & X_{36}\cdot Q_{6,12} & - &Q_{39}\cdot X_{9,12}\\
\Lambda^1_{47} : & \ \ \  & U_{72}\cdot X_{25}\cdot X_{54} & - & X_{7,11}\cdot X_{11,9}\cdot X_{9,12}\cdot U_{12,4}& \ \ \ \  & P_{4,10}\cdot X_{10,7} & - & X_{41}\cdot P_{17}\\
\Lambda^2_{47} : & \ \ \  & W_{72}\cdot X_{25}\cdot X_{54} & - &X_{7,11}\cdot X_{11,9}\cdot X_{9,12}\cdot W_{12,4}& \ \ \ \  & X_{41}\cdot Q_{17} & - & Q_{4,10}\cdot X_{10,7}\\
\Lambda^1_{38} : & \ \ \  & X_{8,12}\cdot U_{12,4}\cdot X_{43} & - & V_{87}\cdot U_{73}& \ \ \ \  & P_{39}\cdot X_{98} & - & X_{32}\cdot P_{28}\\
\Lambda^2_{38} : & \ \ \  & X_{8,12}\cdot W_{12,4}\cdot X_{43} & - & V_{87}\cdot W_{73}& \ \ \ \  & X_{32}\cdot Q_{28} & - & Q_{39}\cdot X_{98}\\
\Lambda^1_{19} : & \ \ \  & U_{95}\cdot X_{51} & - & X_{9,12}\cdot U_{12,4}\cdot X_{41} & \ \ \ \  & P_{17}\cdot X_{7,11}\cdot X_{11,9} & -  &X_{13}\cdot P_{39}\\
\Lambda^2_{19} : & \ \ \  & W_{95}\cdot X_{51} & - & X_{9,12}\cdot W_{12,4}\cdot X_{41} & \ \ \ \  & X_{13}\cdot Q_{39} & - & Q_{17}\cdot X_{7,11}\cdot X_{11,9}\\
\Lambda_{11,8} : & \ \ \  & R_{8,10}\cdot X_{10,11} &  - & V_{87}\cdot X_{7,11} & \ \ \ \  & X_{11,2}\cdot P_{28} & - & Y_{11,2}\cdot Q_{28}\\
\Lambda^1_{5,11} : & \ \ \  & X_{11,2}\cdot X_{25}  & - & X_{11,9}\cdot U_{95} & \ \ \ \  & X_{51}\cdot P_{17}\cdot X_{7,11}  & - & X_{54}\cdot P_{4,10}\cdot X_{10,11} \\
\Lambda^2_{5,11} : & \ \ \  & Y_{11,2}\cdot X_{25}  & - & X_{11,9}\cdot W_{95} & \ \ \ \  & X_{54}\cdot Q_{4,10}\cdot X_{10,11} & - & X_{51}\cdot Q_{17}\cdot X_{7,11}\\
\Lambda^1_{2,10} : & \ \ \  & X_{10,7}\cdot U_{72}  & - & X_{10,11}\cdot X_{11,2} & \ \ \ \  & P_{28}\cdot R_{8,10} & - & X_{25}\cdot X_{54}\cdot P_{4,10}\\
\Lambda^2_{2,10} : & \ \ \  & X_{10,7}\cdot W_{72} & - & X_{10,11}\cdot Y_{11,2} & \ \ \ \  & X_{25}\cdot X_{54}\cdot Q_{4,10} & - & Q_{28}\cdot R_{8,10}\\
\Lambda^1_{27} : & \ \ \  & X_{7,11}\cdot X_{11,2} & - & U_{73}\cdot X_{32} & \ \ \ \  & P_{28}\cdot V_{87}& - & X_{25}\cdot X_{51}\cdot P_{17}\\
\Lambda^2_{27} : & \ \ \  & X_{7,11}\cdot Y_{11,2} & - & W_{73}\cdot X_{32} & \ \ \ \  & X_{25}\cdot X_{51}\cdot Q_{17} & - & Q_{28}\cdot V_{87} \\
\end{array}
~.~ $}
\label{PdP3(+-)2}
\end{align}

Under triality, the cubic deformation applied to Phase (a) of $P_{+-}(\textrm{PdP}_{3b})$, as given in \eqref{PdP3(+-)_deform}, is mapped to a combination of mass and cubic deformations in Phase (b), as follows.
\beal{PdP3(+-)2-deform}
\resizebox{0.95\textwidth}{!}{$
\begin{array}{rcrclcrcl}
& &  \ \ \ \ \ \ \ \ \ \ \ \ \ \ \ \ \ J & +  & \textcolor{blue}{\Delta J} &&& E&  \ \ \ \ \ \ \ \ \ \ \ \ \ \  \\
 \Lambda_{12} : &  & X_{25}\cdot X_{54}\cdot X_{41} - X_{26}\cdot X_{61}& + &\textcolor{blue}{\mu \, X_{25}\cdot X_{51}} & \  \ \  & P_{17}\cdot U_{72}& - &Q_{17}\cdot W_{72}\\
 \Lambda_{64} : &  & X_{43}\cdot X_{32}\cdot X_{26} - X_{41}\cdot X_{13}\cdot X_{36} & - & \textcolor{blue}{\mu \, X_{43}\cdot X_{36}} & \  \ \  & P_{6,12}\cdot U_{12,4} &-& Q_{6,12}\cdot W_{12,4}\\
\Lambda_{78} : &  & R_{8,10}\cdot X_{10,7} - X_{8,12}\cdot X_{12,7} & + & \textcolor{blue}{\mu \, V_{87}} & \ \ \ \  &W_{72}\cdot Q_{28} &- &U_{72}\cdot P_{28}\\
\Lambda_{12,10} : & & X_{10,11}\cdot X_{11,9}\cdot X_{98}\cdot X_{8,12} & - & X_{10,7}\cdot X_{7,11}\cdot X_{11,9}\cdot X_{9,12} & & W_{12,4}\cdot Q_{4,10} &- &U_{12,4}\cdot P_{4,10}\\
& &  & - & \textcolor{blue}{\mu \, X_{10,11}\cdot X_{11,9}\cdot X_{9,12}}    \\
\end{array}
\nn
$}
\eea
\beal{PdP3(+-)2-deform}
\resizebox{0.95\textwidth}{!}{$
\begin{array}{rcrclcrcl}
 \Lambda^1_{49} : &  & X_{98}\cdot X_{8,12}\cdot U_{12,4} - U_{95}\cdot X_{54}  & - & \textcolor{blue}{\mu \, X_{9,12}\cdot U_{12,4}} & \ \ \ \  & P_{4,10}\cdot X_{10,11}\cdot X_{11,9} & - & X_{43}\cdot P_{39}\\
\Lambda^2_{49} : & & X_{98}\cdot X_{8,12}\cdot W_{12,4} - W_{95}\cdot X_{54}
 & -& \textcolor{blue}{\mu \, X_{9,12}\cdot W_{12,4}} & \ \ \ \  & X_{43}\cdot Q_{39} & - & Q_{4,10}\cdot X_{10,11}\cdot X_{11,9}\\
\Lambda^1_{3,12} : & & X_{12,7}\cdot U_{73} - U_{12,4}\cdot X_{41}\cdot X_{13} & - & \textcolor{blue}{\mu \, U_{12,4}\cdot X_{43}} & \ \ \ \  & P_{39}\cdot X_{9,12} & - & X_{36}\cdot P_{6,12}\\
\Lambda^2_{3,12} : & & X_{12,7}\cdot W_{73} - W_{12,4}\cdot X_{41}\cdot X_{13} & - & \textcolor{blue}{\mu \, W_{12,4}\cdot X_{43}} & \ \ \ \  & X_{36}\cdot Q_{6,12} & - &Q_{39}\cdot X_{9,12}\\
\Lambda^1_{27} : & & X_{7,11}\cdot X_{11,2} - U_{73}\cdot X_{32} & + & \textcolor{blue}{\mu \, U_{72}} & \ \ \ \  & P_{28}\cdot V_{87}& - & X_{25}\cdot X_{51}\cdot P_{17}\\
\Lambda^2_{27} : & & X_{7,11}\cdot Y_{11,2} - W_{73}\cdot X_{32} & + & \textcolor{blue}{\mu \, W_{72}} & \ \ \ \  & X_{25}\cdot X_{51}\cdot Q_{17} & - & Q_{28}\cdot V_{87}
\end{array}
$}
\eea
There are three chiral–Fermi massive pairs—($\Lambda_{78}$, $V_{87}$), ($\Lambda_{27}^1$, $U_{72}$), and ($\Lambda_{78}^2$, $W_{87}$)—which are shown in \fref{fig_q_ppdp3b}. Integrating them out, we replace the massive chiral fields as follows,
\begin{align}\label{dP3mass}
\begin{split}
  V_{87}= & \frac{1}{\mu}(X_{8,12}\cdot X_{12,7} - R_{8,10}\cdot X_{10,7}) \,, \\
  U_{72}= & \frac{1}{\mu}(U_{73}\cdot X_{32}-X_{7,11}\cdot X_{11,2}) \,, \\
  W_{72}= & \frac{1}{\mu}(W_{73}\cdot X_{32}-X_{7,11}\cdot Y_{11,2}) \,.
\end{split}
\end{align}

Introducing the following changes of variables,
{\footnotesize
\begin{equation}\label{dP33redefs}
\begin{aligned}
  & X_{43} \rightarrow \frac{1}{\mu}X_{43} - \frac{1}{\mu}X_{41}\cdot X_{13} \,, \\
  & X_{51} \rightarrow \frac{1}{\mu}X_{51} - \frac{1}{\mu}X_{54}\cdot X_{41} \,, \\
  & X_{36} \rightarrow -\frac{1}{\mu}X_{36} + \frac{1}{\mu}X_{32}\cdot X_{26} \,, \\
  & X_{10,11} \rightarrow \frac{1}{\mu}X_{10,11} - \frac{1}{\mu}X_{10,7}\cdot X_{7,11}  \,,\\
  & X_{9,12} \rightarrow -\frac{1}{\mu}X_{9,12} + \frac{1}{\mu}X_{98}\cdot X_{8,12} \,, \\
  & \Lambda_{12} \rightarrow -\frac{1}{\mu}\Lambda_{12} + \frac{1}{\mu}\Lambda_{13}\cdot X_{32} \,, \\
  & \Lambda^1_{49} \rightarrow \frac{1}{\mu}\Lambda^{1}_{49} - \frac{1}{\mu}\Lambda^1_{47}\cdot X_{7,11}\cdot X_{11,9} - \frac{1}{\mu}X_{41}\cdot\Lambda^1_{19} \,, \\
  & \Lambda^2_{49} \rightarrow  \frac{1}{\mu}\Lambda^{2}_{49} - \frac{1}{\mu}\Lambda^2_{47}\cdot X_{7,11}\cdot X_{11,9} - \frac{1}{\mu}X_{41}\cdot\Lambda^2_{19} \,, \\
  & \Lambda^1_{3,12} \rightarrow -\frac{1}{\mu}\Lambda^{1}_{3,12} + \frac{1}{\mu}\Lambda^1_{38}\cdot X_{8,12} + \frac{1}{\mu}X_{32}\cdot\Lambda^1_{2,12} \,, \\
  & \Lambda^2_{3,12} \rightarrow -\frac{1}{\mu}\Lambda^{2}_{3,12} + \frac{1}{\mu}\Lambda^2_{38}\cdot X_{8,12} + \frac{1}{\mu}X_{32}\cdot\Lambda^2_{2,12} \,, \\
  & \Lambda^1_{5,11} \rightarrow \frac{1}{\mu}\Lambda^{1}_{5,11} - \frac{1}{\mu}X_{54}\cdot\Lambda^1_{47}\cdot X_{7,11} \,, \\
  & \Lambda^2_{5,11} \rightarrow \frac{1}{\mu}\Lambda^{2}_{5,11} - \frac{1}{\mu}X_{54}\cdot\Lambda^2_{47}\cdot X_{7,11} \,,
\end{aligned}
\end{equation}
}
we obtain the following $J$- and $E$-terms,
\begin{align}
\resizebox{0.95\textwidth}{!}{$
\begin{array}{rcrclcrcl}
& &  \ \ \ \ \ \ \ \ \ \ \ \ \ \ \ \ \ & J& &&& E&  \ \ \ \ \ \ \ \ \ \ \ \ \ \  \\
 \Lambda_{12} : & \ \ \  & X_{26}\cdot X_{61} & - & X_{25}\cdot X_{51}& \ \ \ \  & P_{17}\cdot X_{7,11} \cdot X_{11,2} & -& Q_{17}\cdot X_{7,11} \cdot Y_{11,2} \\
 \Lambda_{64} : & \ \ \  & X_{43}\cdot X_{36} & - & X_{41}\cdot X_{13}\cdot X_{32}\cdot X_{26} & \ \ \ \  & P_{6,12}\cdot U_{12,4}& -& Q_{6,12}\cdot W_{12,4}\\
 \Lambda_{13} : & \ \ \  & X_{32}\cdot X_{25}\cdot X_{54}\cdot X_{41} & - & X_{36}\cdot X_{61}& \ \ \ \  & P_{17}\cdot U_{73}& -& Q_{17}\cdot W_{73}\\
\Lambda_{35} : & \ \ \  & X_{51}\cdot X_{13} & - & X_{54}\cdot X_{43} & \ \ \ \  & P_{39}\cdot U_{95}& -& Q_{39}\cdot W_{95}\\
\Lambda_{12,10} : & \ \ \  & X_{10,11}\cdot X_{11,9}\cdot X_{9,12}& - & X_{10,7}\cdot X_{7,11}\cdot X_{11,9}\cdot X_{98}\cdot X_{8,12} & \ \ \ \  & W_{12,4}\cdot Q_{4,10} & -&U_{12,4}\cdot P_{4,10} \\
\Lambda_{79} : & \ \ \  & X_{98}\cdot R_{8,10}\cdot X_{10,7} & - & X_{9,12}\cdot X_{12,7} & \ \ \ \  &W_{73}\cdot Q_{39} & -& U_{73}\cdot P_{39}\\
\Lambda^1_{2,12} : & \ \ \  & X_{12,7}\cdot X_{7,11}\cdot X_{11,2} & - & U_{12,4}\cdot X_{41}\cdot X_{13}\cdot X_{32} & \ \ \ \  & P_{28}\cdot X_{8,12}& -& X_{26}\cdot P_{6,12}\\
\Lambda^2_{2,12} : & \ \ \  & X_{12,7}\cdot X_{7,11}\cdot Y_{11,2} & - & W_{12,4}\cdot X_{41}\cdot X_{13}\cdot X_{32} & \ \ \ \  & X_{26}\cdot Q_{6,12}& -& Q_{28}\cdot X_{8,12}\\
\Lambda^1_{67} : & \ \ \  & X_{7,11}\cdot X_{11,2}\cdot X_{26}& - & U_{73}\cdot X_{36} & \ \ \ \  & P_{6,12}\cdot X_{12,7}& -& X_{61}\cdot P_{17}\\
\Lambda^2_{67} : & \ \ \  & X_{7,11}\cdot Y_{11,2}\cdot X_{26}& - & W_{73}\cdot X_{36} & \ \ \ \  & X_{61}\cdot Q_{17}& -& Q_{6,12}\cdot X_{12,7}\\
\Lambda^{1}_{49} : & \ \ \  & X_{9,12}\cdot U_{12,4} & - & U_{95}\cdot X_{54} & \ \ \ \  & P_{4,10}\cdot X_{10,11}\cdot X_{11,9} & -& X_{43}\cdot P_{39}\\
\Lambda^{2}_{49} : & \ \ \  & X_{9,12}\cdot W_{12,4} & - & W_{95}\cdot X_{54} & \ \ \ \  & X_{43}\cdot Q_{39} & -& Q_{4,10}\cdot X_{10,11}\cdot X_{11,9}\\
\Lambda^{1}_{3,12} : & \ \ \  & U_{12,4}\cdot X_{43} & - & X_{12,7}\cdot U_{73} & \ \ \ \  & P_{39}\cdot X_{9,12} & -& X_{36}\cdot P_{6,12}\\
\Lambda^{2}_{3,12} : & \ \ \  & W_{12,4}\cdot X_{43} & - & X_{12,7}\cdot W_{73} & \ \ \ \  & X_{36}\cdot Q_{6,12} & -& Q_{39}\cdot X_{9,12}\\
\Lambda^1_{47} : & \ \ \  & U_{73}\cdot X_{32}\cdot X_{25}\cdot X_{54} & - & X_{7,11}\cdot X_{11,9}\cdot X_{98}\cdot X_{8,12}\cdot U_{12,4}& \ \ \ \  & P_{4,10}\cdot X_{10,7} & -& X_{41}\cdot P_{17}\\
\Lambda^2_{47} : & \ \ \  & W_{73}\cdot X_{32}\cdot X_{25}\cdot X_{54} & - & X_{7,11}\cdot X_{11,9}\cdot X_{98}\cdot X_{8,12}\cdot W_{12,4}& \ \ \ \  & X_{41}\cdot Q_{17} & -& Q_{4,10}\cdot X_{10,7}\\
\Lambda^1_{38} : & \ \ \  & R_{8,10}\cdot X_{10,7}\cdot U_{73} & - & X_{8,12}\cdot U_{12,4}\cdot X_{41}\cdot X_{13}& \ \ \ \  & P_{39}\cdot X_{98} & -& X_{32}\cdot P_{28}\\
\Lambda^2_{38} : & \ \ \  & R_{8,10}\cdot X_{10,7}\cdot W_{73} & - & X_{8,12}\cdot W_{12,4}\cdot X_{41}\cdot X_{13}& \ \ \ \  & X_{32}\cdot Q_{28} & - & Q_{39}\cdot X_{98}\\
\Lambda^1_{19} : & \ \ \  & U_{95}\cdot X_{51} & - & X_{98}\cdot X_{8,12}\cdot U_{12,4}\cdot X_{41} & \ \ \ \  & P_{17}\cdot X_{7,11}\cdot X_{11,9} & -& X_{13}\cdot P_{39}\\
\Lambda^2_{19} : & \ \ \  & W_{95}\cdot X_{51} & - & X_{98}\cdot X_{8,12}\cdot W_{12,4}\cdot X_{41} & \ \ \ \  & X_{13}\cdot Q_{39} & -& Q_{17}\cdot X_{7,11}\cdot X_{11,9}\\
\Lambda_{11,8} : & \ \ \  & R_{8,10}\cdot X_{10,11} &  - & X_{8,12}\cdot X_{12,7}\cdot X_{7,11} & \ \ \ \  & X_{11,2}\cdot P_{28} & - & Y_{11,2}\cdot Q_{28}\\
\Lambda^{1}_{5,11} : & \ \ \  & X_{11,2}\cdot X_{25}  & - & X_{11,9}\cdot U_{95} & \ \ \ \  & X_{51}\cdot P_{17}\cdot X_{7,11}  & - & X_{54}\cdot P_{4,10}\cdot X_{10,11} \\
\Lambda^{2}_{5,11} : & \ \ \  & Y_{11,2}\cdot X_{25}  & - & X_{11,9}\cdot W_{95} & \ \ \ \  & X_{54}\cdot Q_{4,10}\cdot X_{10,11} & - & X_{51}\cdot Q_{17}\cdot X_{7,11}\\
\Lambda^1_{2,10} : & \ \ \  & X_{10,7}\cdot U_{73}\cdot X_{32}  & - &  X_{10,11}\cdot X_{11,2} & \ \ \ \  & P_{28}\cdot R_{8,10} & - & X_{25}\cdot X_{54}\cdot P_{4,10}\\
\Lambda^2_{2,10} : & \ \ \  & X_{10,7}\cdot W_{73}\cdot X_{32}  & - &  X_{10,11}\cdot Y_{11,2} & \ \ \ \  & X_{25}\cdot X_{54}\cdot Q_{4,10} & - & Q_{28}\cdot R_{8,10}\\
\end{array}
~.~$}
\label{dP3triality(+-)}
\end{align}
The complete forward algorithm verifies that this theory corresponds to another phase of $P^2_{+-}(\textrm{dP}_3)$. We therefore call it as Phase (b) of $P^2_{+-}(\textrm{dP}_3)$. The extended $\bar{P}$-matrix for this theory, along with the scaling dimension information, is presented in Appendix \ref{sec:pmdata}.
Moreover, it is straightforward to show that Phase (b) of $P^2_{+-}(\textrm{dP}_3)$ is connected to Phase (a) of $P^2_{+-}(\textrm{dP}_3)$ by triality on node 11.

\subsubsection*{Deformation and its Relevance}

Appendix \ref{sec:pmdata} contains the extended $\bar{P}$-matrices for all the models in this section. Based on them, one can verify that in both Phase (a) and (b) of $P_{+-}(\textrm{PdP}_{3b})$, the deformation plaquettes have the common extremal perfect matching structure; $p_2\cdot p^2_3\cdot p_4\cdot p_6$ which yields a scaling dimension $\Delta [  \Lambda \cdot \Delta J  ] \simeq 1.67662<2$, confirming that the deformation is relevant.

Moreover, we can calculate the volume of the SE$_7$ bases of the initial and final geometry, obtaining
\begin{align}
\frac{ {\rm Vol } (P^2_{+-}(\textrm{dP}_3)) }{ {\rm Vol} (P_{+-}(\textrm{PdP}_{3b})) } = \frac{3.94587}{3.70274} \simeq 1.07  > 1  \, .
\end{align}
This is consistent with the growth in the volume of base ${\rm SE}_7$ under the deformation, providing additional evidence that the deformation under consideration is a relevant deformation.
\\

\section{Triality and Masses From Seemingly Non-Holomorphic Couplings} \label{section_additional_masses}

In Section \ref{section_non_mass_to_mass_via_triality}, we observed that triality can transform a non-mass relevant deformation into a mass deformation. In this section, we explore another intriguing consequence of the interplay between relevant deformations and triality: a set of mass terms in one phase can generate additional mass terms in a triality dual. Interestingly, these new mass terms may appear, at first glance, to arise from non-holomorphic couplings in the original theory.

\begin{figure}[h]
	\centering
	\includegraphics[width=13.5cm]{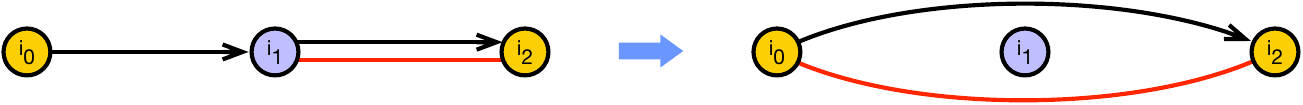}
\caption{Triality generates a new mass term from a mass term in the original theory via rule (3.d).}
	\label{triality_rule_3d_masses}
\end{figure}

Before turning to an explicit example, let us first discuss the general mechanism responsible for the appearance of these mass terms. Their origin can be traced to step \textbf{(d)} of the process of how triality affects the $J$- and $E$-terms as explained in Section \ref{sec:triality}. \fref{triality_rule_3d_masses} shows the portion of the quiver responsible for generating this term. Let us consider triality on node $i_1$ when the original theory contains a mass term of the form $X_{i_1 i_2} \Lambda_{i_2 i_1}$, along with an incoming chiral field $X_{i_0 i_1}$. Triality generates mesons $X_{i_0 i_2}$ and $\Lambda_{i_2 i_0}$ and a mass term $X_{i_0 i_2} \Lambda_{i_2 i_0}$. From the point of view of the original theory, this term would naively seem to follow from a non-holomorphic quadratic coupling of the form $(X_{i_0 i_1} X_{i_1 i_2}) (\Lambda_{i_2 i_1} \overline{X}_{i_0 i_1})$, where the fields in parentheses combine to give rise to the mesons.

There is an analogue of step \textbf{(d)} for inverse triality, in which the additional chiral field, incoming into the dualized node in \fref{triality_rule_3d}, 
is replaced by an outgoing chiral field. As a result, there is a discussion similar to the one above in the case of inverse triality, in which $X_{i_0i_1}$ is replaced by $X_{i_1i_0}$ in \fref{triality_rule_3d_masses}. Once again, inverse triality in a theory with mass terms generates masses that would seem to arise from non-holomorphic couplings in the original theory. This inverse version of the rule will play an important role in Section \ref{sec:c3z2z2}, where we analyze how deformations are mapped between two triality-related theories.

\subsection{From $P_{+-} (\IC^3 / \IZ_2 \times \IZ_2)$ to $P_{+-} (\cC / \IZ_2)$ \label{sec:c3z2z2}}

\begin{figure}[h]
	\centering
	\includegraphics[height=5cm]{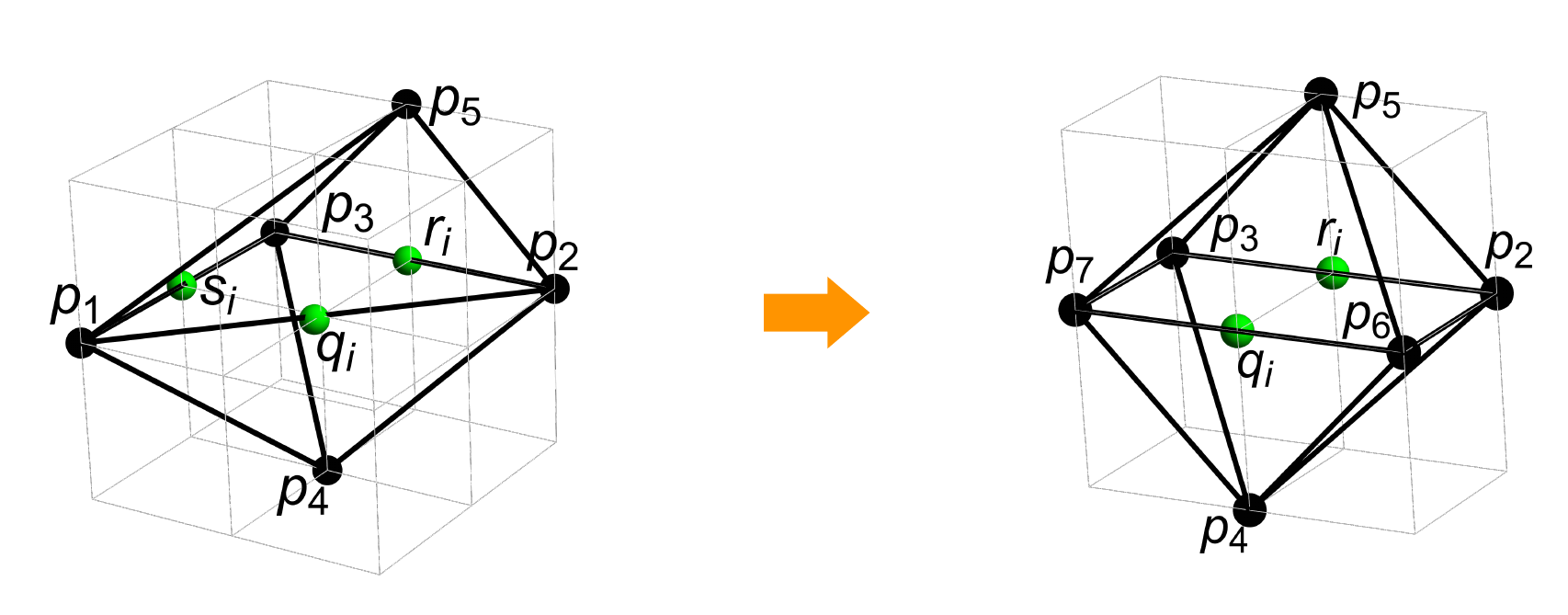}
\caption{Toric diagrams for $P_{+-} (\IC^3 / \IZ_2 \times \IZ_2)$ and $P_{+-} (\cC / \IZ_2)$. We have labeled points anticipating the corresponding brick matchings.
}
	\label{toric_diagrams_example_4}
\end{figure}

To illustrate the phenomenon discussed above, we will study a deformation from $P_{+-} (\IC^3 / \IZ_2 \times \IZ_2)$ to $P_{+-} (\cC / \IZ_2)$, whose toric diagrams are shown in \fref{toric_diagrams_example_4}.

\begin{figure}[ht!]
\begin{center}
\resizebox{0.5\hsize}{!}{
\includegraphics[height=5cm]{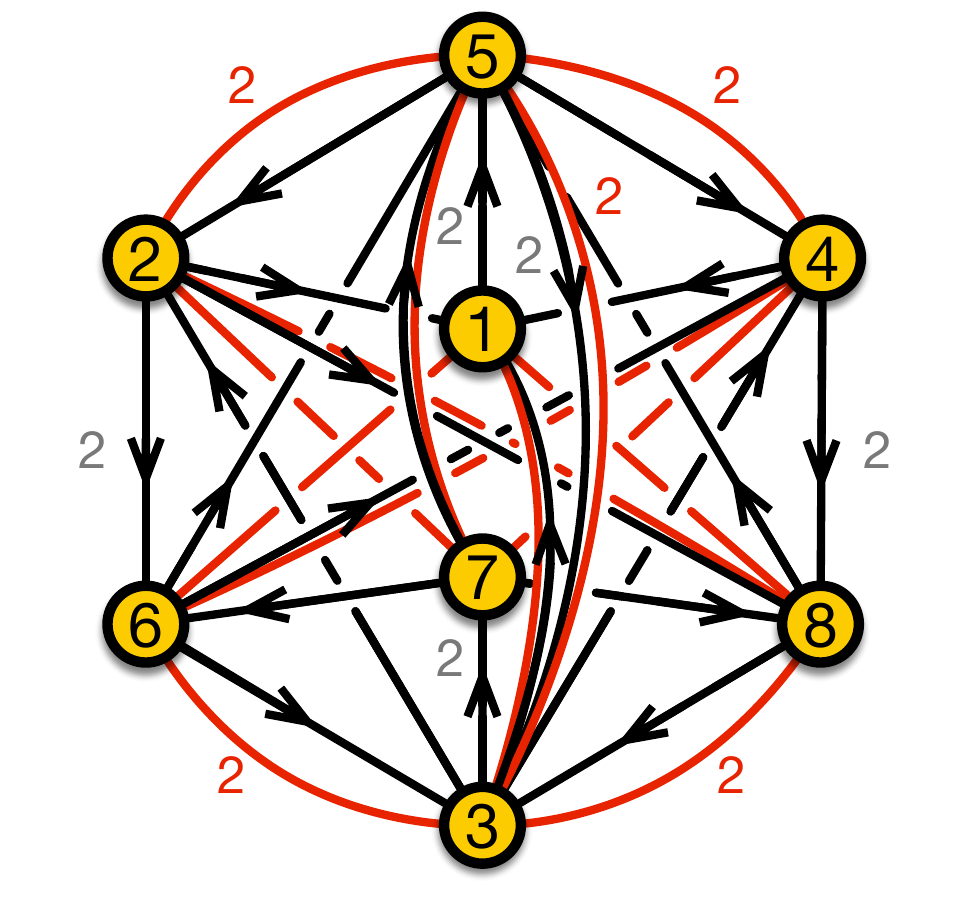}
}
\caption{Quiver diagram for Phase (a) of $P_{+-} (\IC^3 / \IZ_2 \times \IZ_2)$.
\label{fig_quiver_PC3Z2Z2_phaseA}}
 \end{center}
 \end{figure}

\subsubsection*{Mass deformation of Phase A of $P_{+-} (\IC^3 / \IZ_2 \times \IZ_2)$} 

The starting point is $P_{+-} (\IC^3 / \IZ_2 \times \IZ_2)$, for which a gauge theory can be obtained from the $4d$ $\cN=1$ gauge theory corresponding to $\IC^3 / \IZ_2 \times \IZ_2$ using $3d$ printing. We will refer to this theory as Phase (a) of  $P_{+-} (\IC^3 / \IZ_2 \times \IZ_2)$. Its quiver diagram is shown in \fref{fig_quiver_PC3Z2Z2_phaseA}.

Its $J$- and $E$-terms are
{\footnotesize
\begin{align}
\begin{array}{rcrclcrcl}
& &  \ \ \ \ \ \ \ \ \ \ \ \ \ \ \ \ \ & J& &&& E&  \ \ \ \ \ \ \ \ \ \ \ \ \ \  \\
 \Lambda_{13} : & \ \ \  & X_{34} \cdot X_{41}& -&  X_{32} \cdot X_{21}  & \ \ \ \  & P_{15} \cdot Y_{53}& -& Q_{15} \cdot Z_{53}\\
 \Lambda_{57} : & \ \ \  & X_{78} \cdot X_{85}& -&  X_{76} \cdot X_{65}  & \ \ \ \  & Y_{53} \cdot P_{37}& -& Z_{53} \cdot Q_{37}\\
 \Lambda_{16} : & \ \ \  & Y_{63} \cdot X_{31}& -&  Y_{64} \cdot X_{41}  & \ \ \ \  & Q_{15} \cdot Z_{52}\cdot P_{26}& -& P_{15} \cdot Z_{52}\cdot Q_{26}\\
 \Lambda_{46} : & \ \ \  & Y_{63} \cdot X_{34}& -&  X_{65} \cdot Y_{54}  & \ \ \ \  & P_{48} \cdot Z_{82}\cdot Q_{26}& -& Q_{48} \cdot Z_{82}\cdot P_{26}\\
 \Lambda_{27} : & \ \ \  & X_{78} \cdot Z_{82}& -&  X_{75} \cdot Z_{52}  & \ \ \ \  & Q_{26} \cdot Y_{63}\cdot P_{37}& -& P_{26} \cdot Y_{63}\cdot Q_{37}\\
 \Lambda_{47} : & \ \ \  & X_{76} \cdot Y_{64}& -&  X_{75} \cdot Y_{54}  & \ \ \ \  & P_{48} \cdot Z_{83}\cdot Q_{37}& -& Q_{48} \cdot Z_{83}\cdot P_{37}\\
\Lambda_{18} : & \ \ \  & Z_{82} \cdot X_{21}& -&  Z_{83} \cdot X_{31}  & \ \ \ \  & Q_{15} \cdot Y_{54}\cdot P_{48}& -& P_{15} \cdot Y_{54}\cdot Q_{48}\\
\Lambda_{28} : & \ \ \  & X_{85} \cdot Z_{52}& -&  Z_{83} \cdot X_{32}  & \ \ \ \  & Q_{26} \cdot Y_{64}\cdot P_{48}& -& P_{26} \cdot Y_{64}\cdot Q_{48}\\
\Lambda^1_{25} : & \ \ \  & Y_{54} \cdot Q_{48}\cdot Z_{82}& -&  Y_{53} \cdot X_{32}  & \ \ \ \  & P_{26} \cdot X_{65}& -& X_{21} \cdot P_{15}\\
\Lambda^2_{25} : & \ \ \  & Y_{54} \cdot P_{48}\cdot Z_{82}& -&  Z_{53} \cdot X_{32}  & \ \ \ \  & Q_{26} \cdot X_{65}& -& X_{21} \cdot Q_{15}\\
\Lambda^1_{38} : & \ \ \  & Z_{82} \cdot Q_{26}\cdot Y_{63}& -&  X_{85} \cdot Y_{53}  & \ \ \ \  & P_{37} \cdot X_{78}& -& X_{34} \cdot P_{48}\\
\Lambda^2_{38} : & \ \ \  & Z_{82} \cdot P_{26}\cdot Y_{63}& -&  X_{85} \cdot Z_{53}  & \ \ \ \  & Q_{37} \cdot X_{78}& -& X_{34} \cdot Q_{48}\\
\Lambda^1_{45} : & \ \ \  & Z_{52} \cdot Q_{26}\cdot Y_{64}& -&  Y_{53} \cdot X_{34}  & \ \ \ \  & P_{48} \cdot X_{85}& -& X_{41} \cdot P_{15}\\
\Lambda^2_{45} : & \ \ \  & Z_{52} \cdot P_{26}\cdot Y_{64}& -&  Z_{53} \cdot X_{34}  & \ \ \ \  & Q_{48} \cdot X_{85}& -& X_{41} \cdot Q_{15}\\
\Lambda^1_{36} : & \ \ \  & Y_{64} \cdot Q_{48}\cdot Z_{83}& -&  X_{65} \cdot Y_{53}  & \ \ \ \  & P_{37} \cdot X_{76}& -& X_{32} \cdot P_{26}\\
\Lambda^2_{36} : & \ \ \  & Y_{64} \cdot P_{48}\cdot Z_{83}& -&  X_{65} \cdot Z_{53}  & \ \ \ \  & Q_{37} \cdot X_{76}& -& X_{32} \cdot Q_{26}\\
\Lambda^1_{35} : & \ \ \  & Y_{54} \cdot Q_{48}\cdot Z_{83}& -&  Z_{52} \cdot Q_{26}\cdot Y_{63}  & \ \ \ \  & P_{37} \cdot X_{75}& -& X_{31} \cdot P_{15}\\
\Lambda^2_{35} : & \ \ \  & Y_{54} \cdot P_{48}\cdot Z_{83}& -&  Z_{52} \cdot P_{26}\cdot Y_{63}  & \ \ \ \  & Q_{37} \cdot X_{75}& -& X_{31} \cdot Q_{15}\\
 \end{array}
 \,.
\label{hpmC3Z2Z2_JE}
\end{align}
}

\begin{figure}[H]
\begin{center}
\resizebox{0.5\hsize}{!}{
\includegraphics[height=5cm]{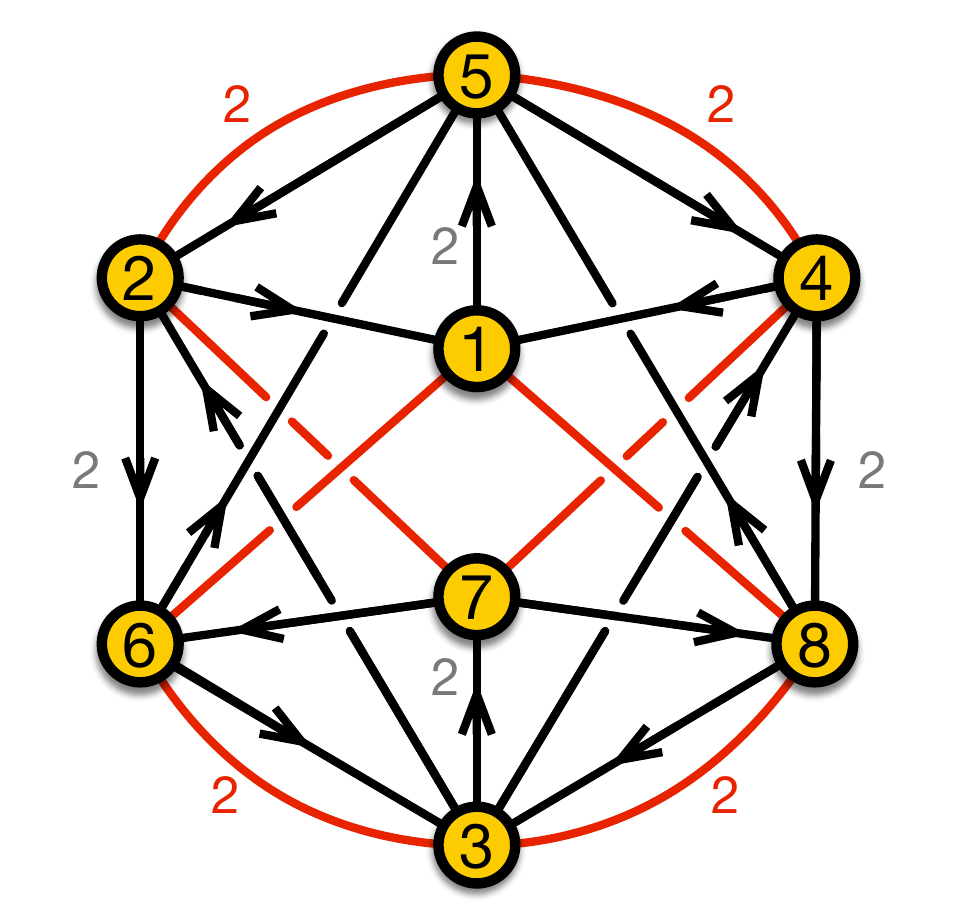}
}
\caption{Quiver diagram for Phase (a) of $P_{+-}(\cC / \IZ_2)$, obtained from Phase (a) of $P_{+-} (\IC^3 / \IZ_2 \times \IZ_2)$ via the mass deformation in \eqref{hpmC3Z2Z2_deform}.
\label{fig_q_pcoz2a}}
 \end{center}
 \end{figure}

We deform this theory by the following mass deformation in which the new terms are marked in blue,
\begin{align}
\begin{array}{rcrclcrcl}
& &  \ \ \ \ \ \ \ \ \ \ \ \ \ \ \ \ \ & J & + \, \textcolor{blue}{\Delta J}  &&& E &  \ \ \ \ \ \ \ \ \ \ \ \ \ \  \\
 \Lambda_{13} : & \ \ \  & X_{34} \cdot X_{41}& -&  X_{32} \cdot X_{21} + \textcolor{blue}{ \mu X_{31}} & \ \ \ \  & P_{15} \cdot Y_{53}& -& Q_{15} \cdot Z_{53}\\
  \Lambda_{57} : & \ \ \  & X_{78} \cdot X_{85}& -&  X_{76} \cdot X_{65} + \textcolor{blue}{ \mu X_{75}} & \ \ \ \  & Y_{53} \cdot P_{37}& -& Z_{53} \cdot Q_{37}\\
  \Lambda_{46} : & \ \ \  & Y_{63} \cdot X_{34}& -&  X_{65} \cdot Y_{54}  + \textcolor{blue}{ \mu Y_{64}}& \ \ \ \  & P_{48} \cdot Z_{82}\cdot Q_{26}& -& Q_{48} \cdot Z_{82}\cdot P_{26}\\
  \Lambda_{28} : & \ \ \  & X_{85} \cdot Z_{52}& -&  Z_{83} \cdot X_{32}  + \textcolor{blue}{ \mu Z_{82}}& \ \ \ \  & Q_{26} \cdot Y_{64}\cdot P_{48}& -& P_{26} \cdot Y_{64}\cdot Q_{48}\\
  \Lambda^1_{35} : & \ \ \  & Y_{54} \cdot Q_{48}\cdot Z_{83}& -&  Z_{52} \cdot Q_{26}\cdot Y_{63} -\textcolor{blue}{ \mu Y_{53}} & \ \ \ \  & P_{37} \cdot X_{75}& -& X_{31} \cdot P_{15}\\
\Lambda^2_{35} : & \ \ \  & Y_{54} \cdot P_{48}\cdot Z_{83}& -&  Z_{52} \cdot P_{26}\cdot Y_{63}  -\textcolor{blue}{ \mu Z_{53}}& \ \ \ \  & Q_{37} \cdot X_{75}& -& X_{31} \cdot Q_{15}\\
 \end{array} \,,
\label{hpmC3Z2Z2_deform}
\end{align}
Integrating out the massive chiral-Fermi pairs, we obtain the quiver diagram in \fref{fig_q_pcoz2a}.

Its $J$- and $E$-terms take the following form after rescaling in terms of $\mu$,
{\footnotesize
\begin{align}
\begin{array}{rcrclcrcl}
& &  \ \ \ \ \ \ \ \ \ \ \ \ \ \ \ \ \ & J& &&& E&  \ \ \ \ \ \ \ \ \ \ \ \ \ \  \\
\Lambda_{16} : & \ \ \  & Y_{63} \cdot X_{32}\cdot X_{21}& -&  X_{65}\cdot Y_{54} \cdot X_{41}  & \ \ \ \  & Q_{15} \cdot Z_{52}\cdot P_{26}& -& P_{15} \cdot Z_{52}\cdot Q_{26} \\
\Lambda_{27} : & \ \ \  & X_{78} \cdot Z_{83}\cdot X_{32} & -&  X_{76}\cdot X_{65} \cdot Z_{52}  & \ \ \ \  & Q_{26} \cdot Y_{63}\cdot P_{37}& -& P_{26} \cdot Y_{63}\cdot Q_{37} \\
\Lambda_{47} : & \ \ \  & X_{76} \cdot Y_{63}\cdot X_{34}& -&  X_{78}\cdot X_{85} \cdot Y_{54}  & \ \ \ \  & P_{48} \cdot Z_{83}\cdot Q_{37}& -& Q_{48} \cdot Z_{83}\cdot P_{37} \\
\Lambda_{18} : & \ \ \  & Z_{83}\cdot X_{34} \cdot X_{41}& -&  X_{85} \cdot Z_{52}\cdot X_{21}  & \ \ \ \  & Q_{15} \cdot Y_{54}\cdot P_{48}& -& P_{15} \cdot Y_{54}\cdot Q_{48} \\
\Lambda^1_{25} : & \ \ \  & Y_{54} \cdot Q_{48}\cdot X_{85}\cdot Z_{52}& -&  Z_{52}\cdot Q_{26}\cdot Y_{63} \cdot X_{32}  & \ \ \ \  & P_{26} \cdot X_{65}& -& X_{21} \cdot P_{15} \\
\Lambda^2_{25} : & \ \ \  & Y_{54} \cdot P_{48}\cdot X_{85}\cdot Z_{52}& -&  Z_{52}\cdot P_{26}\cdot Y_{63} \cdot X_{32}  & \ \ \ \  & Q_{26} \cdot X_{65}& -& X_{21} \cdot Q_{15} \\
\Lambda^1_{38} : & \ \ \  & Z_{83}\cdot X_{32} \cdot Q_{26}\cdot Y_{63}& -&  X_{85} \cdot Y_{54}\cdot Q_{48}\cdot Z_{83}  & \ \ \ \  & P_{37} \cdot X_{78}& -& X_{34} \cdot P_{48} \\
\Lambda^2_{38} : & \ \ \  & Z_{83}\cdot X_{32} \cdot P_{26}\cdot Y_{63}& -&  X_{85} \cdot Y_{54}\cdot P_{48}\cdot Z_{83}  & \ \ \ \  & Q_{37} \cdot X_{78}& -& X_{34} \cdot Q_{48} \\
\Lambda^1_{45} : & \ \ \  & Z_{52} \cdot Q_{26}\cdot X_{65}\cdot Y_{54}& -&  Y_{54}\cdot Q_{48}\cdot Z_{83} \cdot X_{34}  & \ \ \ \  & P_{48} \cdot X_{85}& -& X_{41} \cdot P_{15} \\
\Lambda^2_{45} : & \ \ \  & Z_{52} \cdot P_{26}\cdot X_{65}\cdot Y_{54}& -&  Y_{54}\cdot P_{48}\cdot Z_{83} \cdot X_{34}  & \ \ \ \  & Q_{48} \cdot X_{85}& -& X_{41} \cdot Q_{15} \\
\Lambda^1_{36} : & \ \ \  & Y_{63}\cdot X_{34} \cdot Q_{48}\cdot Z_{83}& -&  X_{65} \cdot Z_{52}\cdot Q_{26}\cdot Y_{63}  & \ \ \ \  & P_{37} \cdot X_{76}& -& X_{32} \cdot P_{26} \\
\Lambda^2_{36} : & \ \ \  & Y_{63}\cdot X_{34} \cdot P_{48}\cdot Z_{83}& -&  X_{65} \cdot Z_{52}\cdot P_{26}\cdot Y_{63}  & \ \ \ \  & Q_{37} \cdot X_{76}& -& X_{32} \cdot Q_{26} \\
 \end{array}
 \,.
\label{hpmcoz2_JE}
\end{align}
}

Using the complete forward algorithm, we verify that this theory has $P_{+-}(\cC / \IZ_2)$ as its mesonic moduli space, whose toric diagram is shown in the right-hand side of \fref{toric_diagrams_example_4}. We will refer to this model as Phase (a). Its extended $\bar{P}$-matrix is given in Appendix \ref{sec:pmdata}.

\subsubsection*{The Same Deformation in a Triality Dual Phase}

\begin{figure}[H]
\begin{center}
\resizebox{0.5\hsize}{!}{
\includegraphics[height=6cm]{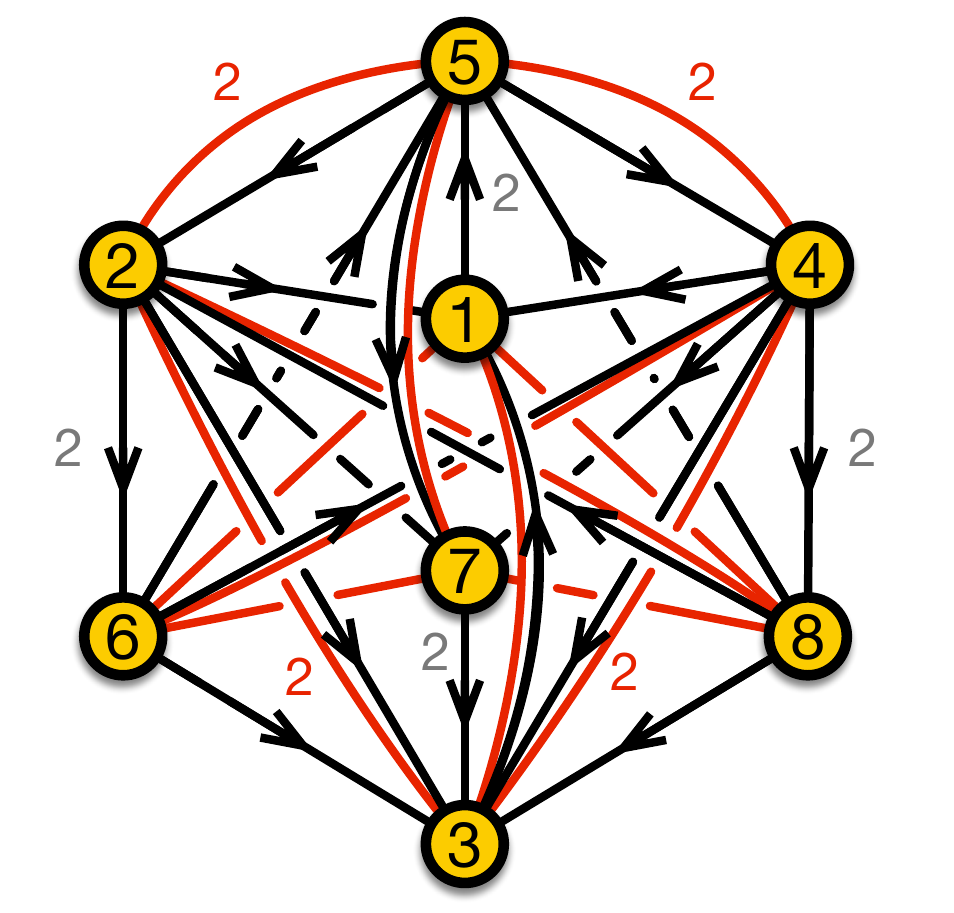}
}
\caption{Quiver diagram for Phase (b) of $P_{+-} (\IC^3 / \IZ_2 \times \IZ_2)$.
\label{fig_q_pc3z3z3b}}
 \end{center}
 \end{figure}

As shown in \fref{fig_quiver_PC3Z2Z2_phaseA}, node 7 of the quiver for Phase A of $P_{+-} (\IC^3 / \IZ_2 \times \IZ_2)$ contains two incoming chiral fields. Therefore, triality action on this node results in another toric phase, which we call Phase (b). Its quiver diagram is shown in \fref{fig_q_pc3z3z3b}.

The $J$- and $E$-terms for Phase (b) are
{\footnotesize
\begin{align}
\begin{array}{rcrclcrcl}
& &  \ \ \ \ \ \ \ \ \ \ \ \ \ \ \ \ \ & J& &&& E&  \ \ \ \ \ \ \ \ \ \ \ \ \ \  \\
 \Lambda_{13} : & \ \ \  & X_{34} \cdot X_{41}& -&  X_{32} \cdot X_{21}  & \ \ \ \  & P_{15} \cdot X_{57}\cdot P_{73}& -& Q_{15} \cdot X_{57}\cdot Q_{73}\\
 \Lambda_{16} : & \ \ \  & Y_{63} \cdot X_{31}& -&  Y_{64} \cdot X_{41}  & \ \ \ \  & Q_{15} \cdot Z_{52}\cdot P_{26}& -& P_{15} \cdot Z_{52}\cdot Q_{26}\\
 \Lambda_{46} : & \ \ \  & Y_{63} \cdot X_{34}& -&  X_{65} \cdot Y_{54}  & \ \ \ \  & P_{48} \cdot Z_{82}\cdot Q_{26}& -& Q_{48} \cdot Z_{82}\cdot P_{26}\\
\Lambda_{18} : & \ \ \  & Z_{82} \cdot X_{21}& -&  Z_{83} \cdot X_{31}  & \ \ \ \  & Q_{15} \cdot Y_{54}\cdot P_{48}& -& P_{15} \cdot Y_{54}\cdot Q_{48}\\
\Lambda_{28} : & \ \ \  & X_{85} \cdot Z_{52}& -&  Z_{83} \cdot X_{32}  & \ \ \ \  & Q_{26} \cdot Y_{64}\cdot P_{48}& -& P_{26} \cdot Y_{64}\cdot Q_{48}\\
\Lambda^1_{25} : & \ \ \  & Y_{54} \cdot Q_{48}\cdot Z_{82}& -&  X_{57}\cdot P_{73} \cdot X_{32}  & \ \ \ \  & P_{26} \cdot X_{65}& -& X_{21} \cdot P_{15}\\
\Lambda^2_{25} : & \ \ \  & Y_{54} \cdot P_{48}\cdot Z_{82}& -&  X_{57}\cdot Q_{73} \cdot X_{32}  & \ \ \ \  & Q_{26} \cdot X_{65}& -& X_{21} \cdot Q_{15}\\
\Lambda^1_{45} : & \ \ \  & Z_{52} \cdot Q_{26}\cdot Y_{64}& -&  X_{57}\cdot P_{73} \cdot X_{34}  & \ \ \ \  & P_{48} \cdot X_{85}& -& X_{41} \cdot P_{15}\\
\Lambda^2_{45} : & \ \ \  & Z_{52} \cdot P_{26}\cdot Y_{64}& -&  X_{57}\cdot Q_{73} \cdot X_{34}  & \ \ \ \  & Q_{48} \cdot X_{85}& -& X_{41} \cdot Q_{15}\\
\Lambda^1_{34} : & \ \ \  & Q_{48} \cdot Z_{83}& -&  X_{47} \cdot P_{73}  & \ \ \ \  & X_{32} \cdot P_{26}\cdot Y_{64}& -& X_{31} \cdot P_{15}\cdot Y_{54}\\
\Lambda^2_{34} : & \ \ \  & P_{48} \cdot Z_{83}& -&  X_{47} \cdot Q_{73}  & \ \ \ \  & X_{32} \cdot Q_{26}\cdot Y_{64}& -& X_{31} \cdot Q_{15}\cdot Y_{54}\\
\Lambda^1_{32} : & \ \ \  & Q_{26} \cdot Y_{63}& -&  X_{27} \cdot P_{73}  & \ \ \ \  & X_{34} \cdot P_{48}\cdot Z_{82}& -& X_{31} \cdot P_{15}\cdot Z_{52}\\
\Lambda^2_{32} : & \ \ \  & P_{26} \cdot Y_{63}& -&  X_{27} \cdot Q_{73}  & \ \ \ \  & X_{34} \cdot Q_{48}\cdot Z_{82}& -& X_{31} \cdot Q_{15}\cdot Z_{52}\\
\Lambda_{76} : & \ \ \  & Y_{64} \cdot X_{47}& -&  X_{65} \cdot X_{57}  & \ \ \ \  & P_{73} \cdot X_{32}\cdot P_{26}& -& Q_{73} \cdot X_{32}\cdot Q_{26}\\
\Lambda_{78} : & \ \ \  & Z_{82} \cdot X_{27}& -&  X_{85} \cdot X_{57}  & \ \ \ \  & P_{73} \cdot X_{34}\cdot P_{48}& -& Q_{73} \cdot X_{34}\cdot Q_{48}\\
\Lambda_{75} : & \ \ \  & Z_{52} \cdot X_{27}& -&  Y_{54} \cdot X_{47}  & \ \ \ \  & P_{73} \cdot X_{31}\cdot P_{15}& -& Q_{73} \cdot X_{31}\cdot Q_{15}\\
 \end{array} \,.
\label{hpmC3Z2Z2_tri_JE}
\end{align}
}
Using the forward algorithm, we confirm that this theory corresponds to $P_{+-} (\IC^3 / \IZ_2 \times \IZ_2)$, whose toric diagram is shown in \fref{toric_diagrams_example_4}.

\subsubsection*{Mass Deformation of Phase (b) of $P_{+-} (\IC^3 / \IZ_2 \times \IZ_2)$}

\begin{figure}[H]
\begin{center}
\resizebox{0.5\hsize}{!}{
\includegraphics[height=5cm]{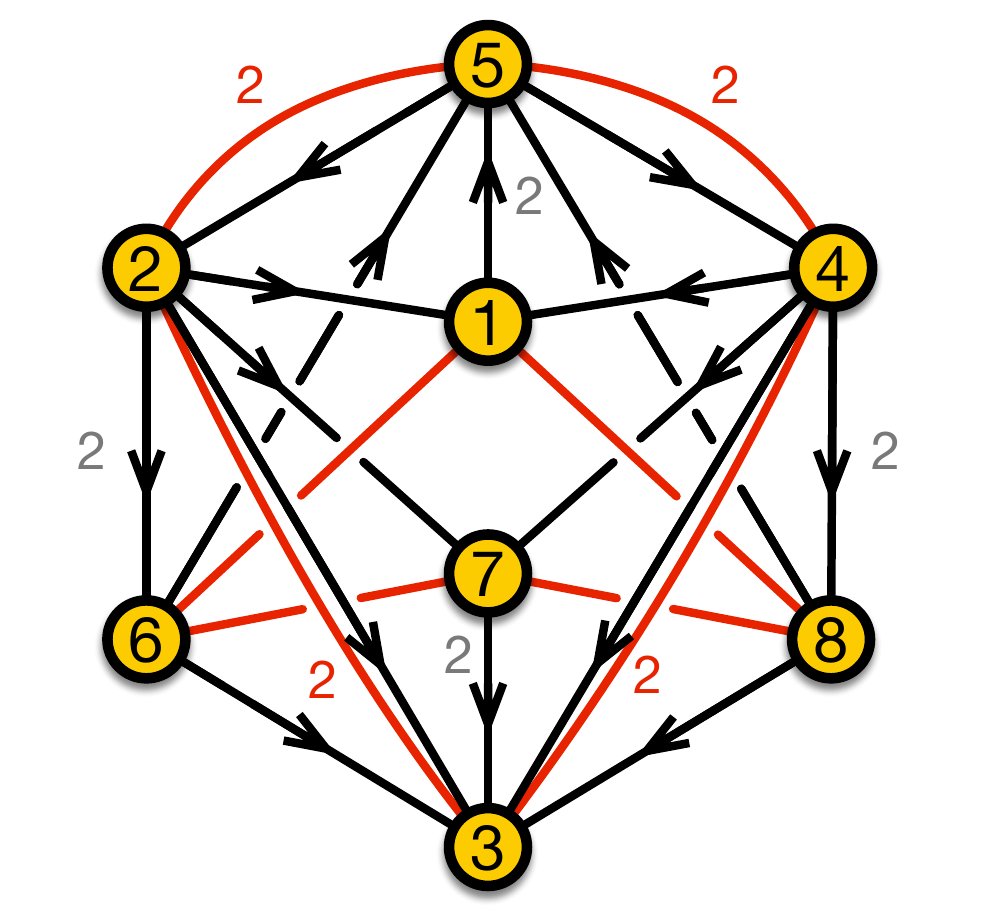}
}
\caption{
Quiver diagram for Phase (b) of $P_{+-}(\cC / \IZ_2)$, obtained from Phase (b) of $P_{+-} (\IC^3 / \IZ_2 \times \IZ_2)$ via the mass deformation in \eqref{hpmC3Z2Z2_tri_deform}.
\label{fig_q_pcoz2b}}
 \end{center}
 \end{figure}

Let us now consider the following mass deformation of Phase (b) of $P_{+-} (\IC^3 / \IZ_2 \times \IZ_2)$,
\begin{align}
\begin{array}{rcccc}
& &  J + \, \textcolor{blue}{\Delta J} && E   \ \ \ \ \ \ \ \ \ \ \ \ \ \  \\
 \Lambda_{13} : & \ \ \  & X_{34} \cdot X_{41} -  X_{32} \cdot X_{21} - \textcolor{blue}{\mu X_{31}} & \ \ \ \  & P_{15} \cdot Y_{53}- Q_{15} \cdot Z_{53} \\
 \Lambda_{46} : & \ \ \  & Y_{63} \cdot X_{34} -  X_{65} \cdot Y_{54} - \textcolor{blue}{\mu Y_{64}}  & \ \ \ \  & P_{48} \cdot Z_{82}\cdot Q_{26} - Q_{48} \cdot Z_{82}\cdot P_{26}\\
 \Lambda_{28} : & \ \ \  & X_{85} \cdot Z_{52} -  Z_{83} \cdot X_{32} - \textcolor{blue}{\mu Z_{82}} & \ \ \ \  & Q_{26} \cdot Y_{64}\cdot P_{48}- P_{26} \cdot Y_{64}\cdot Q_{48}\\
 \Lambda_{75} : & \ \ \  & Z_{52} \cdot X_{27} -  Y_{54} \cdot X_{47} - \textcolor{blue}{\mu X_{57}} & \ \ \ \  & P_{73} \cdot X_{31}\cdot P_{15}- Q_{73} \cdot X_{31}\cdot Q_{15}\\
 \end{array}
 \,.
\label{hpmC3Z2Z2_tri_deform}
\end{align}
Integrating out the massive fields, we obtain the quiver shown in \fref{fig_q_pcoz2b}.

The $J$- and $E$-terms for this theory after rescaling in terms of $\mu$ are
\begin{align}
\resizebox{0.95\textwidth}{!}{$
\begin{array}{rcrclcrcl}
& &  \ \ \ \ \ \ \ \ \ \ \ \ \ \ \ \ \ & J& &&& E&  \ \ \ \ \ \ \ \ \ \ \ \ \ \  \\
 \Lambda_{16} : & \ \ \  & X_{65} \cdot Y_{54} \cdot X_{41}& -&  Y_{63} \cdot X_{32} \cdot X_{21} & \ \ \ \  & Q_{15} \cdot Z_{52}\cdot P_{26}& -& P_{15} \cdot Z_{52}\cdot Q_{26}\\
\Lambda_{18} : & \ \ \  & X_{85} \cdot Z_{52} \cdot X_{21}& -&  Z_{83} \cdot X_{34} \cdot X_{41}  & \ \ \ \  & Q_{15} \cdot Y_{54}\cdot P_{48}& -& P_{15} \cdot Y_{54}\cdot Q_{48}\\
\Lambda^1_{25} : & \ \ \  & Y_{54} \cdot Q_{48}\cdot X_{85} \cdot Z_{52}& -&  Z_{52} \cdot X_{27}\cdot P_{73} \cdot X_{32}  & \ \ \ \  & P_{26} \cdot X_{65}& -& X_{21} \cdot P_{15}\\
\Lambda^2_{25} : & \ \ \  & Y_{54} \cdot P_{48}\cdot X_{85} \cdot Z_{52} & -&  Z_{52} \cdot X_{27} \cdot Q_{73} \cdot X_{32}  & \ \ \ \  & Q_{26} \cdot X_{65}& -& X_{21} \cdot Q_{15}\\
\Lambda^1_{45} : & \ \ \  & Y_{54}\cdot X_{47} \cdot P_{73} \cdot X_{34} & -& Z_{52} \cdot Q_{26} \cdot X_{65} \cdot Y_{54}   & \ \ \ \  & P_{48} \cdot X_{85}& -& X_{41} \cdot P_{15}\\
\Lambda^2_{45} : & \ \ \  & Y_{54} \cdot X_{47} \cdot Q_{73} \cdot X_{34} & - & Z_{52} \cdot P_{26} \cdot X_{65} \cdot Y_{54} & \ \ \ \  & Q_{48} \cdot X_{85}& -& X_{41} \cdot Q_{15}\\
 \end{array}\nn
 $}
\end{align}
\begin{align}
\resizebox{0.95\textwidth}{!}{$
\begin{array}{rcrclcrcl}
\Lambda^1_{34} : & \ \ \  & Q_{48} \cdot Z_{83}& -&  X_{47} \cdot P_{73}  & \ \ \ \  & X_{32} \cdot P_{26}\cdot Y_{63} \cdot X_{34} & -& X_{34} \cdot X_{41} \cdot P_{15}\cdot Y_{54}\\
\Lambda^2_{34} : & \ \ \  & P_{48} \cdot Z_{83}& -&  X_{47} \cdot Q_{73}  & \ \ \ \  & X_{32} \cdot Q_{26}\cdot Y_{63} \cdot X_{34} & -& X_{34} \cdot X_{41} \cdot Q_{15}\cdot Y_{54}\\
\Lambda^1_{32} : & \ \ \  & Q_{26} \cdot Y_{63}& -&  X_{27} \cdot P_{73}  & \ \ \ \  & X_{32} \cdot X_{21} \cdot P_{15} \cdot Z_{52} & -& X_{34} \cdot P_{48}\cdot Z_{83} \cdot X_{32} \\
\Lambda^2_{32} : & \ \ \  & P_{26} \cdot Y_{63}& -&  X_{27} \cdot Q_{73}  & \ \ \ \ & X_{32} \cdot X_{21} \cdot Q_{15} \cdot Z_{52} & -& X_{34} \cdot Q_{48}\cdot Z_{83} \cdot X_{32} \\
\Lambda_{76} : & \ \ \  & Y_{63} \cdot X_{34} \cdot X_{47}& -&  X_{65} \cdot Z_{52} \cdot X_{27}  & \ \ \ \  & P_{73} \cdot X_{32}\cdot P_{26}& -& Q_{73} \cdot X_{32}\cdot Q_{26}\\
\Lambda_{78} : & \ \ \  & X_{85} \cdot Y_{54} \cdot X_{47}& -&  Z_{83} \cdot X_{32} \cdot X_{27}  & \ \ \ \  & P_{73} \cdot X_{34}\cdot P_{48}& -& Q_{73} \cdot X_{34}\cdot Q_{48}\\
 \end{array}
 $}
\label{hpmcoz2_tri_JE}
\end{align}
The forward algorithm shows that the corresponding geometry of this theory is $P_{+-}(\cC / \IZ_2)$, whose toric diagram is shown in the right-hand side of \fref{toric_diagrams_example_4}. We therefore refer to it as Phase (b) of $P_{+-}(\cC / \IZ_2)$. As for the two toric phases we considered for $P_{+-} (\IC^3 / \IZ_2 \times \IZ_2)$, Phase (b) of $P_{+-}(\cC / \IZ_2)$ is obtained from Phase (a) of $P_{+-}(\cC / \IZ_2)$ by triality on node 7. This confirms an expectation: the relevant deformations and triality mutually commute.

\subsubsection*{Connecting the Deformations of Phases (a) and (b) of $P_{+-} (\IC^3 / \IZ_2 \times \IZ_2)$}

Above, we have determined mass deformations that take the $P_{+-} (\IC^3 / \IZ_2 \times \IZ_2)$ model to $P_{+-}(\cC / \IZ_2)$ starting from two triality-related phases. Let us now discuss in further detail how they are related to each other. As mentioned earlier, we can go from Phase (a) to Phase (b) of $P_{+-} (\IC^3 / \IZ_2 \times \IZ_2)$ by acting with triality on node 7. Conversely, we go from Phase (b) to Phase (a) by inverse triality on node 7.

The deformations in Phases (b) and (a) involve four and six mass terms, respectively. Let us examine how they are related. The mass terms for each phase are as follows,
\begin{align}
\begin{array}{ccc}
\mbox{{\bf Phase (b)}} & & \mbox{{\bf Phase (a)}} \\[1 mm]
\Lambda_{13} \cdot X_{31} & \ \ \ \to \ \ \ & \Lambda_{13} \cdot X_{31} \\
\Lambda_{28} \cdot Z_{82} & \ \ \ \to \ \ \ & \Lambda_{28} \cdot Z_{82} \\
\Lambda_{46} \cdot Y_{64} & \ \ \ \to \ \ \ & \Lambda_{46} \cdot Y_{64} \\
\Lambda_{75} \cdot X_{57} & \ \ \ \to \ \ \ & \Lambda_{57} \cdot X_{75} \\
? & \ \ \ \to \ \ \ & \Lambda^1_{35} \cdot Y_{53} \\
? & \ \ \ \to \ \ \ & \Lambda^2_{35} \cdot Z_{53} 
\end{array}
\label{mass_deformations_Phases_A_and_B}
\end{align}
Starting from Phase (b) and acting with inverse triality on Phase (a), the arrows in \eqref{mass_deformations_Phases_A_and_B} indicate which deformation terms in Phase (b) generate each term in Phase (a). For the first four terms, the map is trivial. The first three mass terms simply remain unperturbed by triality. The fourth term contains fields changed under the dualized node and the action of triality exchanges the roles of Fermis and chirals.

But, as we noticed, the deformation of Phase (a) involves two more terms, which are listed in the last two rows of \eqref{mass_deformations_Phases_A_and_B}. At first glance, it may appear that additional relevant deformations are required in Phase (b) to generate these masses. Moreover, such deformations would correspond to quartic, non-holomorphic couplings as follows,
\begin{align}
\begin{array}{ccc}
\mbox{{\bf Phase (b)}} & & \mbox{{\bf Phase (a)}} \\[1 mm]
X_{57} \cdot \textcolor{red}{P_{73} \cdot \overline{P}_{73}} \cdot \Lambda_{75} & \ \ \ \to \ \ \ & \Lambda^1_{35} \cdot Y_{53} \\
X_{57} \cdot \textcolor{red}{Q_{73} \cdot \overline{Q}_{73}} \cdot \Lambda_{75} & \ \ \ \to \ \ \ & \Lambda^2_{35} \cdot Z_{53}
\end{array}
\end{align}
As shown in red, the quartic couplings involve products of chiral fields coming out of node 7 and their conjugates. Under more careful consideration, it is easy to realize that we do not have to turn on such quartic terms {\it independently} of the four mass terms in Phase (b). In fact, the two additional mass terms arise from $\Lambda_{57} \cdot X_{75}$ under triality, via rule \textbf{(d)} as outlined in Section \sref{sec:triality}.

\subsubsection*{Deformation, Brick Matchings and Toric Geometry}

As in previous examples, brick matchings provide an elegant connection between deformations and geometry. From the extended $\bar{P}$-matrix in \eqref{C3 3d print Phase B}, we determine the extremal brick matching content of the fields that become massive in Phase (b), which is given by
\beq
\begin{array}{cclcccl}
\Lambda_{13} & = & p_1 \, p_2 \, p_4 \,, & \ \ \ \ & X_{31} & = & p_4 \\
\Lambda_{28} & = & p_1 \, p_2 \, p_4 \,, & \ \ \ \ & Z_{82} & = & p_4 \\
\Lambda_{46} & = &  p_1 \, p_2 \, p_4 \,,  & \ \ \ \ & Y_{64} & = & p_4 \\
\Lambda_{75} & = & p_1 \, p_2 \, p_4 \,, & \ \ \ \ & X_{57} & = & p_4
\label{fields_ in_mass_deformations_Phases_B_and_pms}
\end{array}
\eeq
This implies that the extremal brick matching content of the mass terms for Phase (b) are therefore given by 
\beq
\begin{array}{ccc}
\mbox{{\bf Phase B}} \\[1 mm]
\Lambda_{13} \cdot X_{31} & \ \ \ \to \ \ \ & p_1 \, p_2 \, p_4^2 \\ 
\Lambda_{28} \cdot Z_{82} & \ \ \ \to \ \ \ & p_1 \, p_2 \, p_4^2 \\
\Lambda_{46} \cdot Y_{64} & \ \ \ \to \ \ \ & p_1 \, p_2 \, p_4^2 \\
\Lambda_{75} \cdot X_{57} & \ \ \ \to \ \ \ & p_1 \, p_2 \, p_4^2
\end{array}
\label{mass_deformations_Phases_B_and_pms}
\eeq
As noted in \cite{Franco:2023tyf}, all terms involved in the deformation share the same expression in terms of extremal brick matchings. It is straightforward to verify that all deformation-induced mass terms in Phase (a) also exhibit this property. This geometric perspective provides additional guidance on selecting the terms necessary to realize a given deformation. Interestingly, \eqref{fields_ in_mass_deformations_Phases_B_and_pms} shows that not only do the complete expressions for the mass terms coincide, but each individual Fermi and chiral field that become massive has the same expression as well. This property is not shared by all the examples considered in this paper, but it would be worth exploring whether it holds under special circumstances. 

The deformations considered in this section are mass terms, so there is no need to independently verify that they are relevant. For completeness, however, we can compute the ratio of the volumes of the corresponding SE$_7$ manifolds, which is
\begin{equation}
  \frac{{\rm Vol} ( P_{+-} (\cC / \IZ_2 ) )}{{\rm Vol } (P_{+-} (\IC^3 / \IZ_2 \times \IZ_2)) }=\frac{6.08807}{5.07527}\simeq 1.20 > 1 \, ,
\end{equation}
which is consistent with the expected growth as we flow to the IR.
\\

\section{Deformations to a Theory with Extra Irrelevant Terms}

\label{section_irrelevant}

In this section, we consider a relevant deformation that results in terms which appear to violate the toric condition for the $J$- and $E$-terms, even after change of variables. However, we will use the underlying geometry to determine the scaling dimensions of the fields and demonstrate that such extra terms are, in fact, irrelevant. The resulting IR theory is therefore a toric phase described by a brane brick model. While this behavior has been previously observed for deformations of toric CY 3-folds \cite{Cremonesi:2023psg}, this is the first time it is examined in the context of toric CY 4-folds.

\begin{figure}[h]
	\centering
	\includegraphics[height=4.5cm]{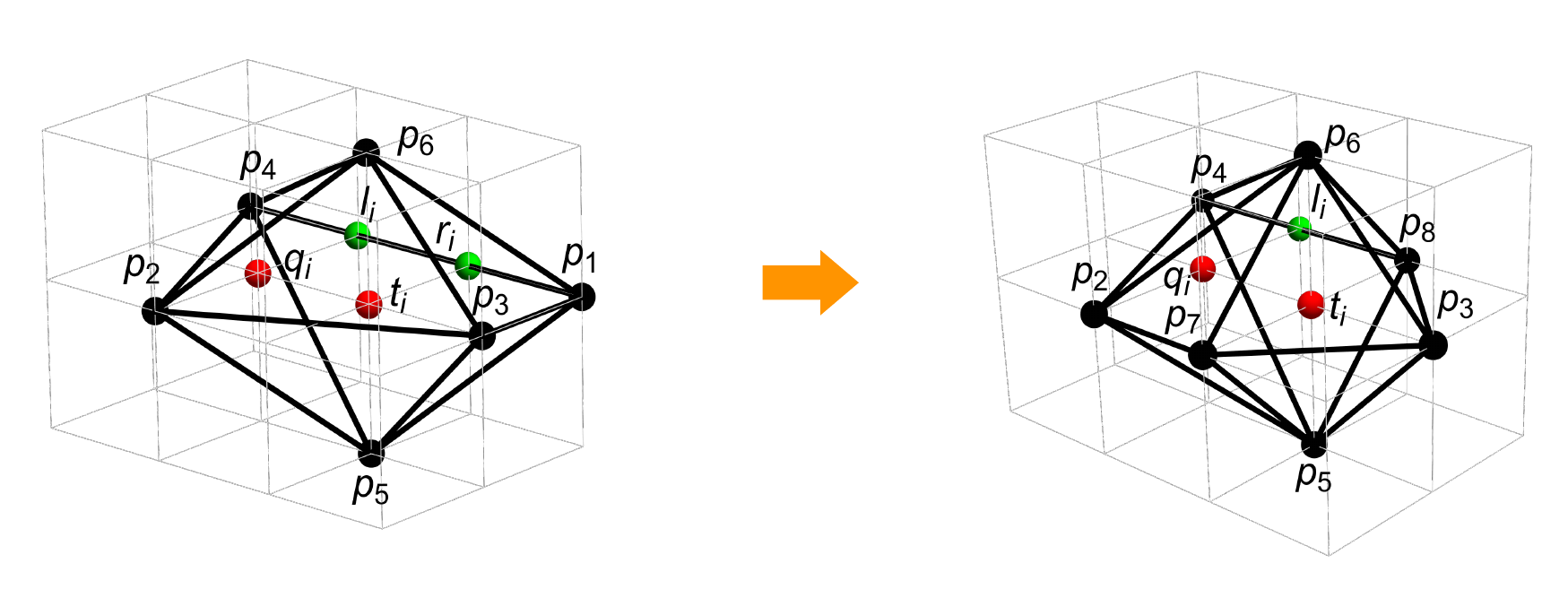}
\caption{Toric diagrams for two geometries connected by a relevant deformation. 
} 
	\label{toric_diagrams_example_5}
\end{figure}

We will consider a deformation that connects the geometries whose toric diagrams are shown in \fref{toric_diagrams_example_5}. The corresponding gauge theories can be obtained via orbifold reduction. Since these geometries are rather intricate, we will not assign them explicit names, but will instead identify them through their toric diagrams.

\subsubsection*{The Initial Theory}

\begin{figure}[ht!]
\begin{center}
\includegraphics[width=9cm]{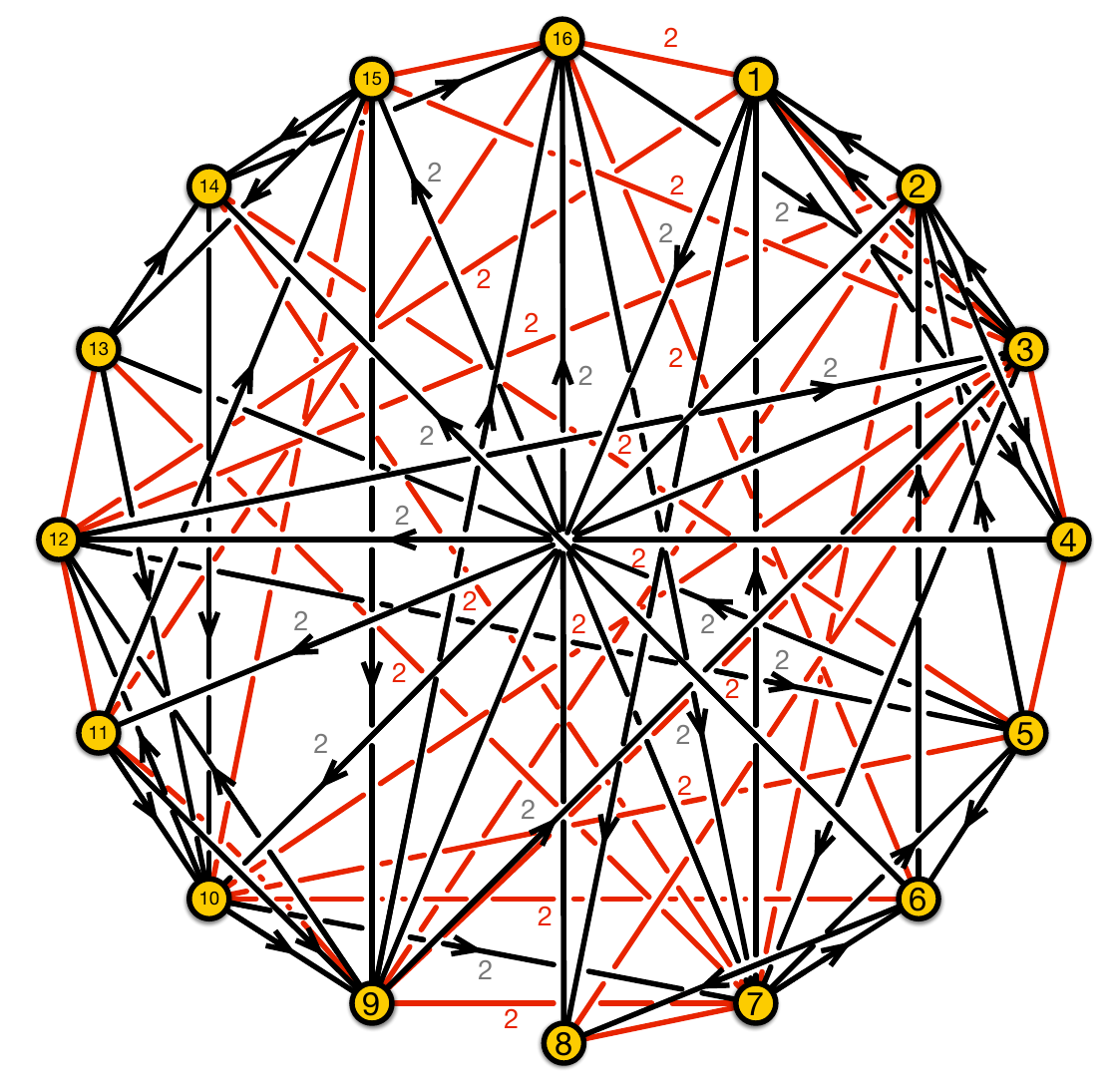}
\caption{
Quiver diagram for a toric phase for the starting geometry in \fref{toric_diagrams_example_5}.}.
\label{fig_quiver_M1_01}
 \end{center}
 \end{figure}

We can construct a gauge theory for the initial geometry using orbifold reduction. Its quiver diagram is shown in \fref{fig_quiver_M1_01}.

Its $J$- and $E$-terms are
\begin{align}
\resizebox{0.8 \textwidth}{!}{$
\begin{array}{rcrclcrcl}
& &  \ \ \ \ \ \ \ \ \ \ \ \ \ \ \ \ \ & J& &&& E&  \ \ \ \ \ \ \ \ \ \ \ \ \ \  \\
 \Lambda_{13} : & \ \ \  & X_{32}\cdot X_{21}& - & X_{37}\cdot X_{71}& \ \ \ \  & P_{19}\cdot U_{93}& -& Q_{19}\cdot W_{93}\\
 \Lambda_{9,11} :& \ \ \  & X_{11,10}\cdot X_{10,9}& - & X_{11,15}\cdot X_{15,9}& \ \ \ \  &W_{93}\cdot Q_{3,11} & -& U_{93}\cdot P_{3,11}\\
 \Lambda_{27} :& \ \ \  & X_{75}\cdot X_{52}& - & X_{76}\cdot X_{62}& \ \ \ \  & P_{2,10}\cdot U_{10,7}& -& Q_{2,10}\cdot W_{10,7}\\
 \Lambda_{10,15} :& \ \ \  & X_{15,13}\cdot X_{13,10}& - & X_{15,14}\cdot X_{14,10}& \ \ \ \  &W_{10,7}\cdot Q_{7,15} & -& U_{10,7}\cdot P_{7,15}\\
 \Lambda_{87} :& \ \ \  & X_{71}\cdot X_{18}& - & X_{75}\cdot X_{56}\cdot X_{68}& \ \ \ \  & P_{8,16}\cdot U_{16,7}& -& Q_{8,16}\cdot W_{16,7}\\
 \Lambda_{16,15} :& \ \ \  & X_{15,9}\cdot X_{9,16}& - & X_{15,13}\cdot X_{13,14}\cdot X_{14,16}& \ \ \ \  & W_{16,7}\cdot Q_{7,15} & -& U_{16,7}\cdot P_{7,15}\\
 \Lambda_{43} :& \ \ \  & X_{31}\cdot X_{14}& - & X_{32}\cdot X_{24}& \ \ \ \  & P_{4,12}\cdot U_{12,3}& -& Q_{4,12}\cdot W_{12,3}\\
 \Lambda_{12,11} :& \ \ \  & X_{11,9}\cdot X_{9,12}& - & X_{11,10}\cdot X_{10,12}& \ \ \ \  &W_{12,3}\cdot Q_{3,11} & -& U_{12,3}\cdot P_{3,11}\\
 \Lambda_{45} :& \ \ \  & X_{56}\cdot X_{62}\cdot X_{24}& - & X_{52}\cdot X_{21}\cdot X_{14}& \ \ \ \  & P_{4,12}\cdot U_{12,5}& -& Q_{4,12}\cdot W_{12,5}\\
 \Lambda_{12,13} :& \ \ \  & X_{13,14}\cdot X_{14,10}\cdot X_{10,12}& - & X_{13,10}\cdot X_{10,9}\cdot X_{9,12}& \ \ \ \  & W_{12,5}\cdot Q_{5,13} & -& U_{12,5}\cdot P_{5,13}\\
 \Lambda_{83} :& \ \ \  & X_{37}\cdot X_{76}\cdot X_{68}& - & X_{31}\cdot X_{18}& \ \ \ \  & P_{8,16}\cdot U_{16,3}& -& Q_{8,16}\cdot W_{16,3}\\
 \Lambda_{16,11} :& \ \ \  & X_{11,15}\cdot X_{15,14}\cdot X_{14,16}& - & X_{11,9}\cdot X_{9,16}& \ \ \ \  & W_{16,3}\cdot Q_{3,11}& -&  U_{16,3}\cdot P_{3,11}\\
 \Lambda^1_{3,10} :& \ \ \  & X_{10,9}\cdot U_{93}& - & X_{10,12}\cdot U_{12,3}& \ \ \ \  & P_{3,11}\cdot X_{11,10}& -& X_{32}\cdot P_{2,10}\\
 \Lambda^2_{3,10} :& \ \ \  & X_{10,9}\cdot W_{93}& - & X_{10,12}\cdot W_{12,3}& \ \ \ \  & X_{32}\cdot Q_{2,10} & -& Q_{3,11}\cdot X_{11,10}\\
 \Lambda^1_{29} : & \ \ \  & U_{93}\cdot X_{32}& - & X_{9,12}\cdot U_{12,5}\cdot X_{52}& \ \ \ \  & P_{2,10}\cdot X_{10,9}& -& X_{21}\cdot P_{19}\\
 \Lambda^2_{29} : & \ \ \  & W_{93}\cdot X_{32}& - & X_{9,12}\cdot W_{12,5}\cdot X_{52}& \ \ \ \  & X_{21}\cdot Q_{19} & -& Q_{2,10}\cdot X_{10,9}\\
 \Lambda^1_{5,10}:& \ \ \  & U_{10,7}\cdot X_{75}& - & X_{10,9}\cdot X_{9,12}\cdot U_{12,5}& \ \ \ \  & P_{5,13}\cdot X_{13,10} & -& X_{52}\cdot P_{2,10}\\
 \Lambda^2_{5,10}:& \ \ \  & W_{10,7}\cdot X_{75}& - & X_{10,9}\cdot X_{9,12}\cdot W_{12,5}& \ \ \ \  & X_{52}\cdot Q_{2,10} & -& Q_{5,13}\cdot X_{13,10} \\
 \Lambda^1_{7,13}:& \ \ \  & X_{13,10}\cdot U_{10,7} & - & X_{13,14}\cdot X_{14,16}\cdot U_{16,7}& \ \ \ \  & P_{7,15}\cdot X_{15,13} & -& X_{75}\cdot P_{5,13} \\
 \Lambda^2_{7,13}:& \ \ \  & X_{13,10}\cdot W_{10,7} & - & X_{13,14}\cdot X_{14,16}\cdot W_{16,7}& \ \ \ \  & X_{75}\cdot Q_{5,13} & - & Q_{7,15}\cdot X_{15,13} \\
 \Lambda^1_{1,16}: & \ \ \  & U_{16,7}\cdot X_{71} & - & U_{16,3}\cdot X_{31} & \ \ \ \  & P_{19}\cdot X_{9,16} & -& X_{18}\cdot P_{8,16} \\
 \Lambda^2_{1,16}: & \ \ \  & W_{16,7}\cdot X_{71} & - & W_{16,3}\cdot X_{31} & \ \ \ \  & X_{18}\cdot Q_{8,16} & - & Q_{19}\cdot X_{9,16} \\
    \end{array}
  \nn
 $}
\end{align}
 \begin{align}
\resizebox{0.8 \textwidth}{!}{$
\begin{array}{rcrclcrcl}
 \Lambda^1_{79}: & \ \ \  & X_{9,16}\cdot U_{16,7} & - & U_{93}\cdot X_{37} & \ \ \ \  & P_{7,15}\cdot X_{15,9} & - & X_{71}\cdot P_{19} \\
 \Lambda^2_{79}: & \ \ \  & X_{9,16}\cdot W_{16,7} & - & W_{93}\cdot X_{37} & \ \ \ \  & X_{71}\cdot Q_{19} & - & Q_{7,15}\cdot X_{15,9} \\
 \Lambda^1_{1,12}:& \ \ \  & U_{12,3}\cdot X_{31} & - & U_{12,5}\cdot X_{52}\cdot X_{21} & \ \ \ \  & P_{19}\cdot X_{9,12} & - & X_{14}\cdot P_{4,12} \\
 \Lambda^2_{1,12}:& \ \ \  & W_{12,3}\cdot X_{31} & - & W_{12,5}\cdot X_{52}\cdot X_{21} & \ \ \ \ & X_{14}\cdot Q_{4,12} & - & Q_{19}\cdot X_{9,12} \\
 \Lambda^1_{39}: & \ \ \  & X_{9,12}\cdot U_{12,3} & - & X_{9,16}\cdot U_{16,3}& \ \ \ \  & P_{3,11}\cdot X_{11,9}& - & X_{31}\cdot P_{19}\\
 \Lambda^2_{39}: & \ \ \  & X_{9,12}\cdot W_{12,3} & - & X_{9,16}\cdot W_{16,3}& \ \ \ \  & X_{31}\cdot Q_{19} & - & Q_{3,11}\cdot X_{11,9}\\
 \Lambda^1_{5,14}:& \ \ \  & X_{14,10}\cdot X_{10,12}\cdot U_{12,5} & - & X_{14,16}\cdot U_{16,7}\cdot X_{75} & \ \ \ \  & P_{5,13}\cdot X_{13,14} & - & X_{56}\cdot P_{6,14} \\
 \Lambda^2_{5,14}:& \ \ \  & X_{14,10}\cdot X_{10,12}\cdot W_{12,5} & - & X_{14,16}\cdot W_{16,7}\cdot X_{75} & \ \ \ \  & X_{56}\cdot Q_{6,14} & - & Q_{5,13}\cdot X_{13,14}\\
 \Lambda^1_{6,10}: & \ \ \  & X_{10,12}\cdot U_{12,5}\cdot X_{56} & - & U_{10,7}\cdot X_{76} & \ \ \ \  & P_{6,14}\cdot X_{14,10} & - & X_{62}\cdot P_{2,10} \\
 \Lambda^2_{6,10}: & \ \ \  & X_{10,12}\cdot W_{12,5}\cdot X_{56} & - & W_{10,7}\cdot X_{76} & \ \ \ \  & X_{62}\cdot Q_{2,10} & - & Q_{6,14}\cdot X_{14,10}\\
 \Lambda^1_{2,12}: & \ \ \  & U_{12,5}\cdot X_{56}\cdot X_{62} & - & U_{12,3}\cdot X_{32} & \ \ \ \  & P_{2,10}\cdot X_{10,12} & - & X_{24}\cdot P_{4,12} \\
 \Lambda^2_{2,12}:  & \ \ \  & W_{12,5}\cdot X_{56}\cdot X_{62} & - & W_{12,3}\cdot X_{32} & \ \ \ \  & X_{24}\cdot Q_{4,12} & - & Q_{2,10}\cdot X_{10,12}\\
 \Lambda^1_{3,15}: & \ \ \  & X_{15,14}\cdot X_{14,16}\cdot U_{16,3} & - & X_{15,9}\cdot U_{93} & \ \ \ \  & P_{3,11}\cdot X_{11,15} & - & X_{37}\cdot P_{7,15} \\
 \Lambda^2_{3,15}: & \ \ \  & X_{15,14}\cdot X_{14,16}\cdot W_{16,3} & - & X_{15,9}\cdot W_{93} & \ \ \ \  & X_{37}\cdot Q_{7,15} & - & Q_{3,11}\cdot X_{11,15}\\
 \Lambda^1_{7,14}: & \ \ \  & X_{14,16}\cdot U_{16,3}\cdot X_{37}& - & X_{14,10}\cdot U_{10,7}& \ \ \ \  & P_{7,15}\cdot X_{15,14}& -& X_{76}\cdot P_{6,14}\\
 \Lambda^2_{7,14}:& \ \ \  & X_{14,16}\cdot W_{16,3}\cdot X_{37}& - & X_{14,10}\cdot W_{10,7}& \ \ \ \  & X_{76}\cdot Q_{6,14} & -& Q_{7,15}\cdot X_{15,14}\\
 \Lambda^1_{6,16}:& \ \ \  & U_{16,3}\cdot X_{37}\cdot X_{76}& - & U_{16,7}\cdot X_{75}\cdot X_{56}& \ \ \ \  & P_{6,14}\cdot X_{14,16}& -& X_{68}\cdot P_{8,16}\\
 \Lambda^2_{6,16}:& \ \ \  & W_{16,3}\cdot X_{37}\cdot X_{76}& - & W_{16,7}\cdot X_{75}\cdot X_{56}& \ \ \ \  & X_{68}\cdot Q_{8,16} & -& Q_{6,14}\cdot X_{14,16}\\
 \end{array}
 $}
\label{Model1A(+-)}
\end{align}
Using the forward algorithm, we verified this theory corresponds to the initial geometry in \fref{toric_diagrams_example_5}. Its extended $\bar{P}$-matrix is given in Appendix \ref{sec:pmdata}.

\subsubsection*{Irrelevant Terms After Deformation}

Let us consider the following deformation of the  $J$-terms
\begin{align}
\resizebox{0.9\textwidth}{!}{$
\begin{array}{rcrclcrcl}
& &  \ \ \ \ \ \ \ \ \ \ \ \ \ \ \ \ \ & J& + \, \textcolor{blue}{\Delta J} &&& E&  \ \ \ \ \ \ \ \ \ \ \ \ \ \  \\
 \Lambda_{13} : & \ \ \  & X_{32}\cdot X_{21}& - & X_{37}\cdot X_{71} +\textcolor{blue}{\mu \,  X_{31}} & \ \ \ \  & P_{19}\cdot U_{93}& -& Q_{19}\cdot W_{93} \\
\Lambda_{9,11} :& \ \ \  & X_{11,10}\cdot X_{10,9}& - & X_{11,15}\cdot X_{15,9} + \textcolor{blue}{\mu \,  X_{11,9}}& \ \ \ \  &W_{93}\cdot Q_{3,11} & -& U_{93}\cdot P_{3,11}\\
\Lambda_{87} :& \ \ \  & X_{71}\cdot X_{18}& - & X_{75}\cdot X_{56}\cdot X_{68}-\textcolor{blue}{\mu \,  X_{76}\cdot X_{68}} & \ \ \ \  & P_{8,16}\cdot U_{16,7}& -& Q_{8,16}\cdot W_{16,7}\\
\Lambda_{16,15} :& \ \ \  & X_{15,9}\cdot X_{9,16}& - & X_{15,13}\cdot X_{13,14}\cdot X_{14,16} -\textcolor{blue}{\mu \,  X_{15,14}\cdot X_{14,16}} & \ \ \ \  & W_{16,7}\cdot Q_{7,15} & -& U_{16,7}\cdot P_{7,15}\\
\Lambda^1_{39}: & \ \ \  & X_{9,12}\cdot U_{12,3} & - & X_{9,16}\cdot U_{16,3} +\textcolor{blue}{\mu \,  U_{93}} & \ \ \ \  & P_{3,11}\cdot X_{11,9}& - & X_{31}\cdot P_{19}\\
\Lambda^2_{39}: & \ \ \  & X_{9,12}\cdot W_{12,3} & - & X_{9,16}\cdot W_{16,3} +\textcolor{blue}{\mu \,  W_{93}} & \ \ \ \  & X_{31}\cdot Q_{19} & - & Q_{3,11}\cdot X_{11,9}\\
\Lambda^1_{7,14}: & \ \ \  & X_{14,16}\cdot U_{16,3}\cdot X_{37}& - & X_{14,10}\cdot U_{10,7} -\textcolor{blue}{\mu \,  X_{14,16}\cdot U_{16,7}} & \ \ \ \  & P_{7,15}\cdot X_{15,14}& -& X_{76}\cdot P_{6,14}\\
\Lambda^2_{7,14}:& \ \ \  & X_{14,16}\cdot W_{16,3}\cdot X_{37}& - & X_{14,10}\cdot W_{10,7} -\textcolor{blue}{\mu \,  X_{14,16}\cdot W_{16,7}} & \ \ \ \  & X_{76}\cdot Q_{6,14} & -& Q_{7,15}\cdot X_{15,14}\\
\Lambda^1_{6,16}:& \ \ \  & U_{16,3}\cdot X_{37}\cdot X_{76}& - & U_{16,7}\cdot X_{75}\cdot X_{56} -\textcolor{blue}{\mu \,  U_{16,7}\cdot X_{76}} & \ \ \ \  & P_{6,14}\cdot X_{14,16}& -& X_{68}\cdot P_{8,16}\\
\Lambda^2_{6,16}:& \ \ \  & W_{16,3}\cdot X_{37}\cdot X_{76}& - & W_{16,7}\cdot X_{75}\cdot X_{56} -\textcolor{blue}{\mu \,  W_{16,7}\cdot X_{76}} & \ \ \ \  & X_{68}\cdot Q_{8,16} & -& Q_{6,14}\cdot X_{14,16}
\end{array}
$}
\label{m1a_deform}
\end{align}
This deformation involves mass terms and higher order terms.

\begin{figure}[ht!]
\begin{center}
\includegraphics[width=9cm]{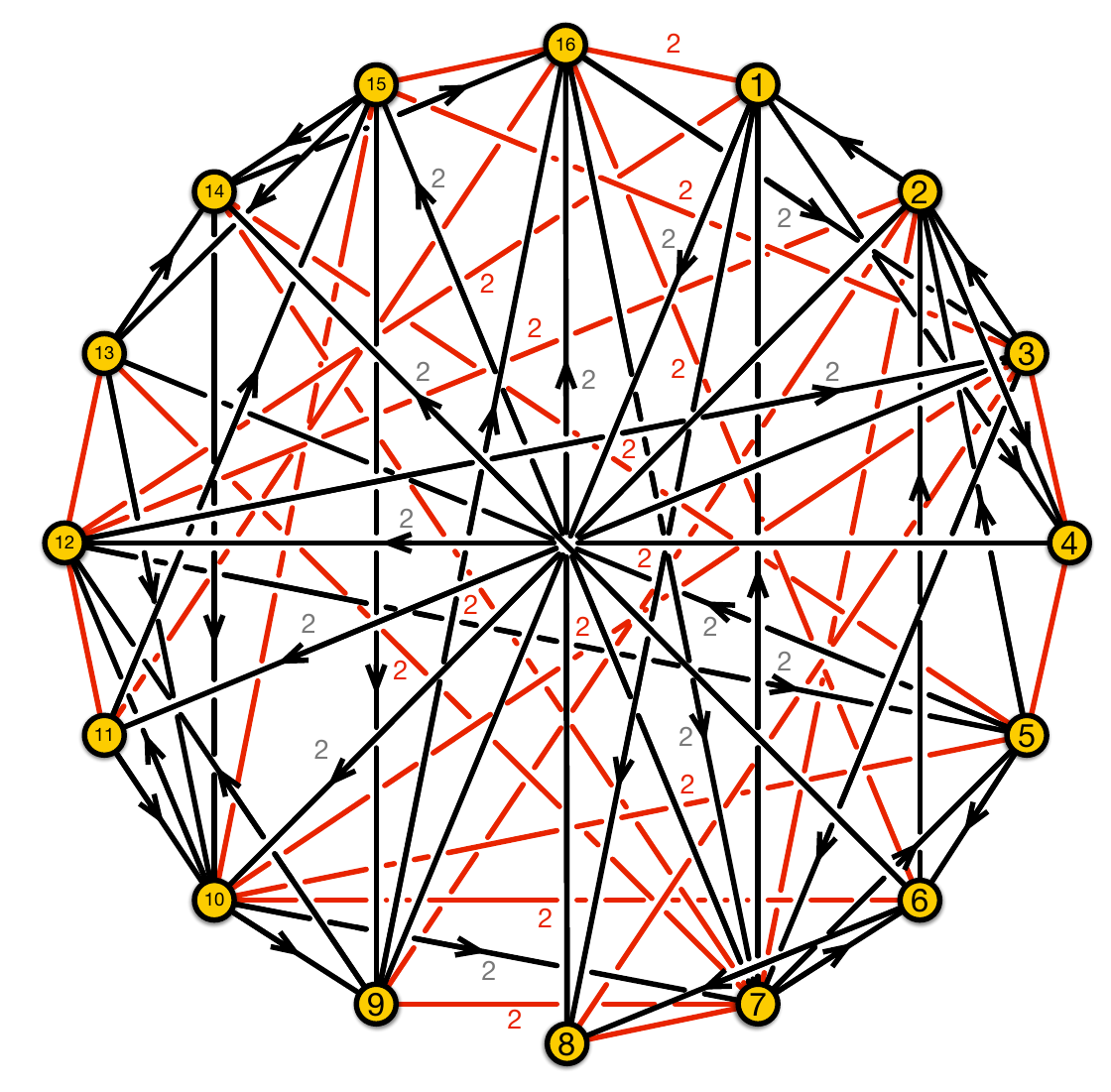}
\caption{
Quiver diagram obtained by the mass deformation of the theory in \fref{fig_quiver_M1_01}.}.
\label{fig_quiver_M1_02}
 \end{center}
 \end{figure}

Using the extended $\bar{P}$-matrix given in \eqref{pm-m1-before}, we find all the deformation plaquettes have the same extremal brick matching content: $p_3 \cdot p_4^2 \cdot p_5 \cdot p_6$. Once again, this is strong evidence that we can exploit the geometry to pick the terms in a complicated deformation like the one above, which mixes masses and higher order terms. The scaling dimension obtained by \eqref{scaledefm} is $\Delta [\Lambda \cdot \Delta J] \simeq 1.51063 <2$, which implies that the deformation is relevant. This is, of course, expected, since the deformation includes mass terms.
Integrating out the massive chiral-Fermi pairs, we obtain the quiver in \fref{fig_quiver_M1_02}.

When integrating out the massive fields, we replace the massive chiral fields as follows
\begin{align}
\begin{split}
  X_{31} = & \frac{1}{\mu}(X_{37}\cdot X_{71} - X_{32}\cdot X_{21}) \,, \\
  X_{11,9} = & \frac{1}{\mu}(X_{11,15}\cdot X_{15,9} - X_{11,10}\cdot X_{10,9}) \,, \\
  U_{93} = &\frac{1}{\mu}(X_{9,16}\cdot U_{16,3} - X_{9,12}\cdot U_{12,3}) \,, \\
  W_{93} = &\frac{1}{\mu}(X_{9,16}\cdot W_{16,3} - X_{9,12}\cdot W_{12,3}) \,. \\
\end{split}
\end{align}

Introducing the following changes of variables
\begin{align}
\begin{split}
  U_{16,7} \rightarrow &-\frac{1}{\mu}U_{16,7} + \frac{1}{\mu}U_{16,3}\cdot X_{37} \,, \\
  W_{16,7} \rightarrow & -\frac{1}{\mu}W_{16,7} + \frac{1}{\mu}W_{16,3}\cdot X_{37} \,, \\
  X_{18} \rightarrow & \, \mu X_{18} \,, \\
  X_{37} \rightarrow &  \, \mu X_{37} \,,\\
  X_{9,16} \rightarrow & \, \mu X_{9,16} \,, \\
  X_{11,15} \rightarrow & \, \mu X_{11,15} \,, \\  
  \Lambda_{87} \rightarrow & -\frac{1}{\mu}\Lambda_{87} + \frac{1}{\mu}\Lambda_{83}\cdot X_{37} \,, \\
  \Lambda_{16,15} \rightarrow & -\frac{1}{\mu}\Lambda_{16,15} - \frac{1}{\mu}U_{16,3}\cdot \Lambda^1_{3,15} 
  -\frac{1}{\mu}W_{16,3}\cdot \Lambda^2_{3,15}+\frac{1}{\mu}\Lambda_{16,11}\cdot X_{11,15} \,, \\
  \Lambda^1_{3,15} \rightarrow & \, \mu \Lambda^{1}_{3,15} \,, \\
  \Lambda^2_{3,15} \rightarrow & \, \mu \Lambda^{2}_{3,15} \,,\\
  \Lambda^1_{1,16} \rightarrow & \, \mu \Lambda^{1}_{1,16} \,, \\
  \Lambda^2_{1,16} \rightarrow & \, \mu \Lambda^{2}_{1,16} \,, \\
\end{split}    
\end{align}
renders the $J$- and $E$-terms in the following form,
\begin{align}
\resizebox{0.95\textwidth}{!}{$
\begin{array}{rcrclcrcl}
& &  \ \ \ \ \ \ \ \ \ \ \ \ \ \ \ \ \ & J& &&& E&  \ \ \ \ \ \ \ \ \ \ \ \ \ \  \\
 \Lambda_{27} :& \ \ \  & X_{75}\cdot X_{52}& - & X_{76}\cdot X_{62}& \ \ \ \  & P_{2,10}\cdot U_{10,7}& -& Q_{2,10}\cdot W_{10,7}\\
 \Lambda_{10,15} :& \ \ \  & X_{15,13}\cdot X_{13,10}& - & X_{15,14}\cdot X_{14,10}& \ \ \ \  & W_{10,7}\cdot Q_{7,15} & -& U_{10,7}\cdot P_{7,15}\\
 \Lambda_{87} :& \ \ \  & X_{76}\cdot X_{68}& - & X_{71}\cdot X_{18}+\textcolor{magenta}{\frac{1}{\mu}X_{75}\cdot X_{56}\cdot X_{68}}& \ \ \ \  & P_{8,16}\cdot U_{16,7}& -& Q_{8,16}\cdot W_{16,7}\\
 \Lambda_{16,15} :& \ \ \  & X_{15,14}\cdot X_{14,16}& - & X_{15,9}\cdot X_{9,16}+\textcolor{magenta}{\frac{1}{\mu}X_{15,13}\cdot X_{13,14}\cdot X_{14,16}}& \ \ \ \  & W_{16,7}\cdot Q_{7,15} & -& U_{16,7}\cdot P_{7,15}\\
 \Lambda_{43} :& \ \ \  & X_{37}\cdot X_{71}\cdot X_{14}& - & X_{32}\cdot X_{24}-\textcolor{magenta}{\frac{1}{\mu}X_{32}\cdot X_{21}\cdot X_{14}}& \ \ \ \  & P_{4,12}\cdot U_{12,3}& -& Q_{4,12}\cdot W_{12,3}\\
 \Lambda_{12,11} :& \ \ \  & X_{11,15}\cdot X_{15,9}\cdot X_{9,12}& - & X_{11,10}\cdot X_{10,12}+\textcolor{magenta}{\frac{1}{\mu}X_{11,10}\cdot X_{10,9}\cdot X_{9,12}}& \ \ \ \  & W_{12,3}\cdot Q_{3,11} & -& U_{12,3}\cdot P_{3,11}\\
 \Lambda_{45} :& \ \ \  & X_{56}\cdot X_{62}\cdot X_{24}& - & X_{52}\cdot X_{21}\cdot X_{14}& \ \ \ \  & P_{4,12}\cdot U_{12,5}& -& Q_{4,12}\cdot W_{12,5}\\
 \Lambda_{12,13} :& \ \ \  & X_{13,14}\cdot X_{14,10}\cdot X_{10,12}& - & X_{13,10}\cdot X_{10,9}\cdot X_{9,12}& \ \ \ \  & W_{12,5}\cdot Q_{5,13} & -& U_{12,5}\cdot P_{5,13}\\
 \Lambda_{83} :& \ \ \  & X_{32}\cdot X_{21}\cdot X_{18}& - & X_{37}\cdot X_{75}\cdot X_{56}\cdot X_{68}& \ \ \ \  & P_{8,16}\cdot U_{16,3}& -& Q_{8,16}\cdot W_{16,3}\\
 \Lambda_{16,11} :& \ \ \  & X_{11,10}\cdot X_{10,9}\cdot X_{9,16}& - & X_{11,15}\cdot X_{15,13}\cdot X_{13,14}\cdot X_{14,16}& \ \ \ \  & W_{16,3}\cdot Q_{3,11}& -& U_{16,3}\cdot P_{3,11}\\
 \Lambda^1_{3,10} :& \ \ \  & X_{10,9}\cdot X_{9,16}\cdot U_{16,3}& - & X_{10,12}\cdot U_{12,3}-\textcolor{magenta}{\frac{1}{\mu}X_{10,9}\cdot X_{9,12}\cdot U_{12,3}}& \ \ \ \  & P_{3,11}\cdot X_{11,10}& -& X_{32}\cdot P_{2,10}\\
 \Lambda^2_{3,10} :& \ \ \  & X_{10,9}\cdot X_{9,16}\cdot W_{16,3}& - & X_{10,12}\cdot W_{12,3}-\textcolor{magenta}{\frac{1}{\mu}X_{10,9}\cdot X_{9,12}\cdot W_{12,3}}& \ \ \ \  & X_{32}\cdot Q_{2,10} & -& Q_{3,11}\cdot X_{11,10}\\
 \Lambda^1_{29} : & \ \ \  & X_{9,16}\cdot U_{16,3}\cdot X_{32}& - & X_{9,12}\cdot U_{12,5}\cdot X_{52}-\textcolor{magenta}{\frac{1}{\mu}X_{9,12}\cdot U_{12,3}\cdot X_{32}}& \ \ \ \  & P_{2,10}\cdot X_{10,9}& -& X_{21}\cdot P_{19}\\
 \Lambda^2_{29} : & \ \ \  & X_{9,16}\cdot W_{16,3}\cdot X_{32}& - & X_{9,12}\cdot W_{12,5}\cdot X_{52}-\textcolor{magenta}{\frac{1}{\mu}X_{9,12}\cdot W_{12,3}\cdot X_{32}}& \ \ \ \  & X_{21}\cdot Q_{19} & -& Q_{2,10}\cdot X_{10,9}\\
 \Lambda^1_{5,10}:& \ \ \  & U_{10,7}\cdot X_{75}& - & X_{10,9}\cdot X_{9,12}\cdot U_{12,5}& \ \ \ \  & P_{5,13}\cdot X_{13,10} & -& X_{52}\cdot P_{2,10}\\
 \Lambda^2_{5,10}:& \ \ \  & W_{10,7}\cdot X_{75}& - & X_{10,9}\cdot X_{9,12}\cdot W_{12,5}& \ \ \ \  & X_{52}\cdot Q_{2,10} & -& Q_{5,13}\cdot X_{13,10} \\
 \Lambda^1_{7,13}:& \ \ \  & X_{13,10}\cdot U_{10,7} & - & X_{13,14}\cdot X_{14,16}\cdot U_{16,3}\cdot X_{37}& \ \ \ \  & P_{7,15}\cdot X_{15,13} & -& X_{75}\cdot P_{5,13} \\
 & & & &+\textcolor{magenta}{\frac{1}{\mu}X_{13,14}\cdot X_{14,16}\cdot U_{16,7}}&&& &   \\
 \Lambda^2_{7,13}:& \ \ \  & X_{13,10}\cdot W_{10,7} & - & X_{13,14}\cdot X_{14,16}\cdot W_{16,3}\cdot X_{37}& \ \ \ \  & X_{75}\cdot Q_{5,13} & - & Q_{7,15}\cdot X_{15,13} \\
  & & & &+\textcolor{magenta}{\frac{1}{\mu}X_{13,14}\cdot X_{14,16}\cdot W_{16,7}}&&& &   \\
    \end{array}
 \nn
 $}
\end{align} 
\begin{align}
\resizebox{0.95\textwidth}{!}{$
\begin{array}{rcrclcrcl}
 \Lambda^1_{1,16}: & \ \ \  & U_{16,3}\cdot X_{32}\cdot X_{21} & - & U_{16,7}\cdot X_{71} & \ \ \ \  & P_{19}\cdot X_{9,16} & -& X_{18}\cdot P_{8,16} \\
 \Lambda^2_{1,16}: & \ \ \  & W_{16,3}\cdot X_{32}\cdot X_{21} & - & W_{16,7}\cdot X_{71} & \ \ \ \  & X_{18}\cdot Q_{8,16} & - & Q_{19}\cdot X_{9,16} \\
 \Lambda^1_{79}: & \ \ \  & X_{9,12}\cdot U_{12,3}\cdot X_{37} & - & X_{9,16}\cdot U_{16,7} & \ \ \ \  & P_{7,15}\cdot X_{15,9} & - & X_{71}\cdot P_{19} \\
 \Lambda^2_{79}: & \ \ \  & X_{9,12}\cdot W_{12,3}\cdot X_{37} & - & X_{9,16}\cdot W_{16,7} & \ \ \ \  & X_{71}\cdot Q_{19} & - & Q_{7,15}\cdot X_{15,9} \\
 \Lambda^1_{1,12}:& \ \ \  & U_{12,3}\cdot X_{37}\cdot X_{71} & - & U_{12,5}\cdot X_{52}\cdot X_{21}-\textcolor{magenta}{\frac{1}{\mu}U_{12,3}\cdot X_{32}\cdot X_{21}} & \ \ \ \  & P_{19}\cdot X_{9,12} & - & X_{14}\cdot P_{4,12} \\
 \Lambda^2_{1,12}:& \ \ \  & W_{12,3}\cdot X_{37}\cdot X_{71} & - & W_{12,5}\cdot X_{52}\cdot X_{21}-\textcolor{magenta}{\frac{1}{\mu}W_{12,3}\cdot X_{32}\cdot X_{21}} & \ \ \ \ & X_{14}\cdot Q_{4,12} & - & Q_{19}\cdot X_{9,12} \\
 \Lambda^1_{5,14}:& \ \ \  & X_{14,10}\cdot X_{10,12}\cdot U_{12,5} & - & X_{14,16}\cdot U_{16,3}\cdot X_{37}\cdot X_{75} & \ \ \ \  & P_{5,13}\cdot X_{13,14} & - & X_{56}\cdot P_{6,14} \\
 & & & &+\textcolor{magenta}{\frac{1}{\mu}X_{14,16}\cdot U_{16,7}\cdot X_{75}}&&& &   \\
 \Lambda^2_{5,14}:& \ \ \  & X_{14,10}\cdot X_{10,12}\cdot W_{12,5} & - & X_{14,16}\cdot W_{16,3}\cdot X_{37}\cdot X_{75} & \ \ \ \  & X_{56}\cdot Q_{6,14} & - & Q_{5,13}\cdot X_{13,14}\\
  & & & &+\textcolor{magenta}{\frac{1}{\mu}X_{14,16}\cdot W_{16,7}\cdot X_{75}}&&& &   \\
 \Lambda^1_{6,10}: & \ \ \  & X_{10,12}\cdot U_{12,5}\cdot X_{56} & - & U_{10,7}\cdot X_{76} & \ \ \ \  & P_{6,14}\cdot X_{14,10} & - & X_{62}\cdot P_{2,10} \\
 \Lambda^2_{6,10}: & \ \ \  & X_{10,12}\cdot W_{12,5}\cdot X_{56} & - & W_{10,7}\cdot X_{76} & \ \ \ \  & X_{62}\cdot Q_{2,10} & - & Q_{6,14}\cdot X_{14,10}\\
 \Lambda^1_{2,12}: & \ \ \  & U_{12,5}\cdot X_{56}\cdot X_{62} & - & U_{12,3}\cdot X_{32} & \ \ \ \  & P_{2,10}\cdot X_{10,12} & - & X_{24}\cdot P_{4,12} \\
 \Lambda^2_{2,12}:  & \ \ \  & W_{12,5}\cdot X_{56}\cdot X_{62} & - & W_{12,3}\cdot X_{32} & \ \ \ \  & X_{24}\cdot Q_{4,12} & - & Q_{2,10}\cdot X_{10,12}\\
 \Lambda^1_{3,15}: & \ \ \  & X_{15,9}\cdot X_{9,12}\cdot U_{12,3} & - & X_{15,13}\cdot X_{13,14}\cdot X_{14,16}\cdot U_{16,3} & \ \ \ \  & P_{3,11}\cdot X_{11,15} & - & X_{37}\cdot P_{7,15} \\
 \Lambda^2_{3,15}: & \ \ \  & X_{15,9}\cdot X_{9,12}\cdot W_{12,3} & - & X_{15,13}\cdot X_{13,14}\cdot X_{14,16}\cdot W_{16,3} & \ \ \ \  & X_{37}\cdot Q_{7,15} & - & Q_{3,11}\cdot X_{11,15}\\
 \Lambda^1_{7,14}: & \ \ \  & X_{14,16}\cdot U_{16,7}& - & X_{14,10}\cdot U_{10,7}& \ \ \ \  & P_{7,15}\cdot X_{15,14}& -& X_{76}\cdot P_{6,14}\\
 \Lambda^2_{7,14}:& \ \ \  & X_{14,16}\cdot W_{16,7}& - & X_{14,10}\cdot W_{10,7}& \ \ \ \  & X_{76}\cdot Q_{6,14} & -& Q_{7,15}\cdot X_{15,14}\\
 \Lambda^1_{6,16}:& \ \ \  & U_{16,7}\cdot X_{76}& - & U_{16,3}\cdot X_{37}\cdot X_{75}\cdot X_{56}& \ \ \ \  & P_{6,14}\cdot X_{14,16}& -& X_{68}\cdot P_{8,16}\\
  & & & &+\textcolor{magenta}{\frac{1}{\mu}U_{16,7}\cdot X_{75}\cdot X_{56}}&&& &   \\
 \Lambda^2_{6,16}:& \ \ \  & W_{16,7}\cdot X_{76}& - & W_{16,3}\cdot X_{37}\cdot X_{75}\cdot X_{56}& \ \ \ \  & X_{68}\cdot Q_{8,16} & -& Q_{6,14}\cdot X_{14,16}\\
  & & & &+\textcolor{magenta}{\frac{1}{\mu}W_{16,7}\cdot X_{75}\cdot X_{56}}&&& &   \\
 \end{array}
~.~$}
\label{Model1A(+-)deformed}
\end{align}

If we momentarily ignore the magenta terms in \eqref{Model1A(+-)deformed}, the remaining $J$- and $E$-terms turn out to satisfy both the binomial property 
and the vanishing trace condition. 
Furthermore, the forward algorithm for the quiver in \fref{fig_quiver_M1_02} with $J$- and $E$-terms in \eref{Model1A(+-)deformed} with the magenta terms removed 
gives a mesonic moduli space whose toric diagram is shown on the right of \fref{toric_diagrams_example_5}. 
The corresponding extended $\bar{P}$-matrix is presented in Appendix \ref{sec:pmdata}. 

We can simply argue that such extra plaquettes can be eliminated by taking the limit $\mu \rightarrow \infty$. However, we can be much more precise than this. Using the $\bar{P}$-matrix obtained without the magenta terms, we find that the extra plaquettes share the same extremal brick matching content: $p_1^2\cdot p_3\cdot p_5^2\cdot p_6\cdot p_7$. In turn, this determines the scaling dimension of extra plaquettes as $\Delta[\Lambda \cdot \textcolor{magenta}{\delta J} ] \simeq 2.15068 \,.$ Thus, they correspond to an irrelevant deformation of the toric theory and can be neglected.

The ratio between the volumes of the final and initial SE$_7$ manifolds reads
\begin{equation}
  \frac{{\rm Vol} ( Y_7') }{{\rm Vol} (Y_7)}=\frac{3.1256}{2.8752}\simeq 1.08 > 1 \, ,
\end{equation}
which is consistent with the expected growth in the volume of SE$_7$ manifolds towards the IR.
\\

\section{Birational Transformations and Relevant Deformations \label{section_polytope_mutation}}

Birational transformations \cite{2012arXiv1212.1785A}
that relate toric Calabi-Yau 4-folds
as well as the associated brane brick models
have been studied systematically in the case of toric Fano 3-folds in \cite{Ghim:2024asj}
as well as in the more general case beyond toric Fano 3-folds in \cite{Ghim:2025zhs}.
In both cases, it was shown that a family of birational transformations that relate toric Calabi-Yau 4-folds
can be identified with mass deformations of the corresponding brane brick models.

In this paper, by studying more general families of deformations of brane brick models, 
we observe that non-mass relevant deformations also relate to birational transformations between the associated toric Calabi-Yau 4-folds.
Let us summarize our observations with the example in section \sref{sec:p2sppz2}, which introduces
a non-mass relevant deformation from the brane brick model corresponding to $P_{+-}(\textrm{SPP}/\mathbb{Z}_2)$
to the model corresponding to $P^2_{+-}(\textrm{PdP}_3)$.

Starting with the $P_{+-}(\textrm{SPP}/\mathbb{Z}_2)$ model, whose toric diagram is shown in \fref{toric_diagrams_example_1}, 
the corresponding Newton polynomial can be written as follows, 
\beal{esbt01}
P(x,y,z) = 
2 \frac{1}{x} + x + 2\frac{1}{y} + \frac{1}{x y} + \frac{x}{y} + \frac{y}{x} + x z + \frac{1}{z} + c ~,~
\eea
where we choose the internal point at the origin to have a coefficient $c \in \mathbb{C}^*$.
Using the above Newton polynomial for 
$P_{+-}(\textrm{SPP}/\mathbb{Z}_2)$
with its choice of coefficients, 
we can introduce as described in \cite{2012arXiv1212.1785A,Ghim:2024asj,Ghim:2025zhs}
a birational transformation $\varphi_A$ of the following form, 
\beal{esbt02}
\varphi_A ~:~
(x,y,z) 
\mapsto
\left(
x, y, (1+y) z
\right)
~,~
\eea
where the Laurent polynomial $A(x,y)$ here is chosen to be,
\beal{esbt03}
A(x,y) = 1+ y ~.~
\eea
Here, we note that $A(x,y)$ is only in terms of $y$, indicating that the birational transformation 
effectively acts only on the $(y,z)$-plane containing the $2$-dimensional toric diagram for $\textrm{SPP}/\mathbb{Z}_2$, which is a slice of the $3$-dimensional toric diagram for $P_{+-}(\textrm{SPP}/\mathbb{Z}_2)$.
Under the birational transformation in \eref{esbt02}, 
we obtain the Newton polynomial for $P^2_{+-}(\textrm{PdP}_3)$ as follows, 
\beal{esbt04}
P^\vee(x,y,z)
= \frac{1}{x} + 2 \frac{1}{y} + y + x y + \frac{1}{z} + \frac{1}{y z} + z + \frac{z}{y} + c~,~
\eea
where the brane brick model for $P^2_{+-}(\textrm{PdP}_3)$
is obtained by a relevant non-mass deformation from the brane brick model for $P_{+-}(\textrm{SPP}/\mathbb{Z}_2)$
in section \sref{sec:p2sppz2}.

From this example, we can also observe the following invariant quantities that 
were originally observed in \cite{Ghim:2024asj,Ghim:2025zhs}
and can now be also identified for a non-mass relevant deformations of brane brick models corresponding to birational transformations of toric Calabi-Yau 4-folds:
\begin{itemize}
\item As originally observed in \cite{Ghim:2024asj,Ghim:2025zhs} for mass deformations corresponding to birational transformations on toric Calabi-Yau 4-folds, the \textbf{number of generators} of the mesonic moduli space $\mathcal{M}^{mes}$ is preserved under more general deformations of brane brick models. In our example, both the $P_{+-}(\textrm{SPP}/\mathbb{Z}_2)$ model and 
the $P^2_{+-}(\textrm{PdP}_3)$ model have 19 generators for their respective mesonic moduli spaces $\mathcal{M}^{mes}$.
In terms of the GLSM fields corresponding to extremal vertices of the toric diagram
of $P_{+-}(\textrm{SPP}/\mathbb{Z}_2)$ model, the 19 generators take the form, 
\beal{esbt07}
&
p_1^2 p_2^2 p_5 ~,~ p_2^2 p_4 p_5^2 ~,~ p_1^2 p_2^2 p_6 ~,~ p_2^2 p_4 p_5 p_6 ~,~  p_2^2 p_4 p_6^2~,~
&
\nn\\
&
p_1^3 p_2 p_3 p_5 ~,~ p_1 p_2 p_3 p_4 p_5^2 ~,~ p_1^3 p_2 p_3 p_6 ~,~ p_1 p_2 p_3 p_4 p_5 p_6 ~,~ p_1 p_2 p_3 p_4 p_6^2 ~,~
&
\nn\\
&
p_1^4 p_3^2 p_5 ~,~ p_1^2 p_3^2 p_4 p_5^2 ~,~ p_3^2 p_4^2 p_5^3 ~,~ p_1^4 p_3^2 p_6 ~,~  p_1^2 p_3^2 p_4 p_5 p_6 ~,~ p_3^2 p_4^2 p_5^2 p_6 ~,~ 
&
\nn\\
&
p_1^2 p_3^2 p_4 p_6^2 ~,~  p_3^2 p_4^2 p_5 p_6^2 ~,~ p_3^2 p_4^2 p_6^3 ~.~
&
\eea
In comparison, in terms of the extremal GLSM fields of the $P^2_{+-}(\textrm{PdP}_3)$ model, 
the 19 generators take the form, 
\beal{esbt08}
&
p_1 p_2 p_3^2 p_4^3 ~,~ p_1 p_2 p_3^2 p_4^2 p_5 ~,~ p_1 p_2 p_3^2 p_4 p_5^2 ~,~  p_1 p_2 p_3^2 p_5^3 ~,~ p_1^2 p_3 p_4^2 p_6^2 ~,~ 
&
\nn\\
&
p_1^2 p_3 p_4 p_5 p_6^2 ~,~  p_1^2 p_3 p_5^2 p_6^2 ~,~ p_1 p_2 p_3 p_4^2 p_6 p_7 ~,~ p_1 p_2 p_3 p_4 p_5 p_6 p_7 ~,~  p_1 p_2 p_3 p_5^2 p_6 p_7 ~,~ 
&
\nn\\
&
p_1^2 p_4 p_6^3 p_7 ~,~ p_1^2 p_5 p_6^3 p_7 ~,~  p_2^2 p_3 p_4^2 p_7^2 ~,~ p_2^2 p_3 p_4 p_5 p_7^2 ~,~ p_2^2 p_3 p_5^2 p_7^2 ~,~  p_1 p_2 p_4 p_6^2 p_7^2 ~,~ 
&
\nn\\
&
p_1 p_2 p_5 p_6^2 p_7^2 ~,~ p_2^2 p_4 p_6 p_7^3 ~,~  p_2^2 p_5 p_6 p_7^3
~.~
&
\eea

\item The unrefined \textbf{Hilbert series} in terms of only $U(1)_R$ symmetry fugacities remains the same for the mesonic moduli spaces $\mathcal{M}^{mes}$ under the deformation of brane brick models corresponding to the birational transformation on the toric Calabi-Yau 4-folds. 
In our case, both Hilbert series refined under the $U(1)_R$ symmetry for the $P_{+-}(\textrm{SPP}/\mathbb{Z}_2)$ model and 
the $P^2_{+-}(\textrm{PdP}_3)$ model
are, 
\beal{esbt05}
g(\bar{t}_1, \dots, \bar{t}_4 ; P_{+-} (\textrm{SPP}/\IZ_2) ) = 
\frac{P(\bar{t}_1, \dots, \bar{t}_4)}{
(1- \bar{t}_1^2 \bar{t}_2 \bar{t}_3^4)^3 (1- \bar{t}_1^2 \bar{t}_2^2 \bar{t}_3^2 \bar{t}_4)^2 (1- \bar{t}_1^2 \bar{t}_2^3 \bar{t}_4^2)^2
}~,~
\eea
where the numerator is given by
\beal{esbt05b}
&&
P(\bar{t}_1, \dots, \bar{t}_4) =
1 + 3 \bar{t}_1^2 \bar{t}_2 \bar{t}_3^4 + 7 \bar{t}_1^2 \bar{t}_2^2 \bar{t}_3^2 \bar{t}_4 - 13 \bar{t}_1^4 \bar{t}_2^3 \bar{t}_3^6 \bar{t}_4 +  2 \bar{t}_1^6 \bar{t}_2^4 \bar{t}_3^{10} \bar{t}_4 + 2 \bar{t}_1^2 \bar{t}_2^3 \bar{t}_4^2 
\nn\\
&&
\hspace{0.5cm}
- 10 \bar{t}_1^4 \bar{t}_2^4 \bar{t}_3^4 \bar{t}_4^2 
-  4 \bar{t}_1^6 \bar{t}_2^5 \bar{t}_3^8 \bar{t}_4^2 + 4 \bar{t}_1^8 \bar{t}_2^6 \bar{t}_3^{12} \bar{t}_4^2 -  4 \bar{t}_1^4 \bar{t}_2^5 \bar{t}_3^2 \bar{t}_4^3 + 4 \bar{t}_1^6 \bar{t}_2^6 \bar{t}_3^6 \bar{t}_4^3 +  10 \bar{t}_1^8 \bar{t}_2^7 \bar{t}_3^{10} \bar{t}_4^3 
\nn\\
&&
\hspace{0.5cm}
- 2 \bar{t}_1^{10} \bar{t}_2^8 \bar{t}_3^{14} \bar{t}_4^3 
-  2 \bar{t}_1^6 \bar{t}_2^7 \bar{t}_3^4 \bar{t}_4^4  + 13 \bar{t}_1^8 \bar{t}_2^8 \bar{t}_3^8 \bar{t}_4^4 -  7 \bar{t}_1^{10} \bar{t}_2^9 \bar{t}_3^{12} \bar{t}_4^4 - 3 \bar{t}_1^{10} \bar{t}_2^{10} \bar{t}_3^{10} \bar{t}_4^5 -  \bar{t}_1^{12} \bar{t}_2^{11} \bar{t}_3^{14} \bar{t}_4^5
~,~
\nn\\
\eea
Here, the fugacities $\overline{t}_1, \dots, \overline{t}_4$ correspond to specific $U(1)_R$ charges $r_1, \dots, r_4$, as discussed in section \sref{sec:p2sppz2}.
The above indicates that the Hilbert series refined only under the $U(1)_R$ symmetry is also an invariant under birational transformations corresponding to non-mass relevant deformations of brane brick models.
This agrees generalizes the observations made in \cite{Ghim:2024asj,Ghim:2025zhs} in the context of birational transformations of toric Calabi-Yau 4-folds corresponding to mass deformations of brane brick models. 

\end{itemize}

Here, we see based on the example 
of the relevant deformation from the 
$P_{+-}(\textrm{SPP}/\mathbb{Z}_2)$ model 
to the model corresponding to $P^2_{+-}(\textrm{PdP}_3)$
in section \sref{sec:p2sppz2} that 
the invariants originally observed in \cite{Ghim:2024asj,Ghim:2025zhs} for mass deformations related to birational transformations of toric Calabi-Yau 4-folds
extend to relevant deformations that correspond to the same family of birational transformations.
These invariants are the number of generators of the mesonic moduli space
as well as the Hilbert series of the mesonic moduli space when it is refined only under the $U(1)_R$ symmetry.
It would be interesting to investigate further in the future the scope of birational transformations on toric Calabi-Yau 4-folds
in relation to corresponding deformations of brane brick models.
\\

\section{Conclusions} 

\label{section_conclusions}

In this paper we have extended the study of relevant deformations that connect $2d$ (0,2) gauge theories on D1-branes probing toric CY 4-folds in several directions. The geometric origin of these theories gives rise to interesting connections, offering new insights into both the gauge dynamics and the associated geometry. Below we summarize our key results and findings.

\begin{itemize}
\item We have expanded the analysis to include non-mass relevant deformations. Some of our examples, such as those in Sections \sref{sec:p2sppz2} and \sref{section_non_mass_to_mass_via_triality}, involve deformations consisting solely of plaquettes that are cubic or of higher order. Interestingly, such cases correspond to more than one CY 4-fold with the same quiver and different $J$- and $E$-terms.
\item The underlying geometry provides useful tools when field theoretic stools are still lacking. For instance, we have observed that an increase in the volume of the SE$_7$ base of the CY 4-fold correlates with RG flow towards the IR, thereby signaling the relevance of deformations.
\item Furthermore, leveraging divisor volumes together with the map between brick matchings (or, more generally, GLSM fields) and gauge theory fields, we have extracted the scaling dimensions of individual fields and of the corresponding terms in the Lagrangian. This approach enables a more precise determination of whether specific terms are relevant or irrelevant.
\item Geometry, through the map between gauge theory fields and GLSM fields, is also useful for identifying the terms necessary to realize a given deformation. All our results, together with those in \cite{Franco:2023tyf}, indicate that all terms responsible for a given deformation share the same extremal GLSM content. Conversely, to connect the gauge theories for two toric CY 4-folds, it is necessary to turn on all terms with a given extremal GLSM content. 
\item We investigated various aspects of the interplay between deformations and triality. In some cases, such as the example considered in Section \sref{section_non_mass_to_mass_via_triality}, non-mass relevant deformations are mapped to mass deformations in a triality dual theory.
\item In exploring the interplay between deformations and triality, we examined in greater detail examples such as the one in Section \sref{section_non_mass_to_mass_via_triality}, where matching deformations between two theories related by triality might naively appear to require turning on non-holomorphic deformations in one of them. A detailed analysis shows, however, that these situations are elegantly accounted for by the precise rules for transforming $J$- and $E$-terms under triality introduced in \cite{Franco:2017lpa}.
\item In Section \sref{section_irrelevant}, we showed that a deformation can yield the theory associated with another toric CY 4-fold, up to irrelevant terms. Moreover, the irrelevance of these extra terms can be established through geometric considerations. An analogous phenomenon has been observed for toric CY 3-folds and their associated 4d $\mathcal{N}=1$ gauge theories \cite{Cremonesi:2023psg}.
\item 
Finally, in Section \sref{section_polytope_mutation}, we presented evidence that when the Hilbert series of the mesonic moduli space is refined only under the $U(1)_R$ symmetry, it becomes invariant even under non-mass relevant deformations of the brane brick models corresponding to toric Calabi-Yau 4-folds related by a birational transformation. This extends the results of \cite{Ghim:2024asj,Ghim:2025zhs} to a broader class of deformations.

\end{itemize}

These results suggests new directions for further study, which we hope to revisit in the near future.

\section*{Acknowledgements}

The work of S. F., M. C. and G. G. has been supported by the U.S. National Science Foundation grants PHY-2112729 and PHY-2412479. 
D. G. was supported by JST PRESTO Grant Number JPMJPR2117 and the Royal Society under the International Collaboration Award Grant R2242058.
D. G. would like to thank University of Birmingham, Durham University and UNIST for their hospitality during part of this work.
R.-K. S. is supported by an Outstanding Young Scientist Grant (RS-2025-00516583) of the National Research Foundation of Korea (NRF).
He is also partly supported by the BK21 Program (``Next Generation Education Program for Mathematical Sciences'', 4299990414089) funded by the Ministry of Education in Korea and the National Research Foundation of Korea (NRF).
S. F. and R.-K. S. would like to thank the 2025 Simons Physics Summer Workshop and the Simons Center for Geometry and Physics for their hospitality during part of this work.

\newpage
\appendix

\section{$\bar{P}$-matrices for some of the models} \label{sec:pmdata}

\subsection{$\bar{P}$-matrices for $P_{+-} ( \textrm{PdP}_3 ) $}

The extended $\bar{P}$-matrix for Phase (a) of $P_{+-} ( \textrm{PdP}_3 ) $ is presented below:
\bea
\resizebox{0.72\hsize}{!}{$
\left(
\begin{array}{c|cccccc|cc|cc|cccccccccccccccccccccccc|c}
\; & p_1 & p_2 & p_3 & p_4 & p_5 & p_6 & s_1& s_2& r_1 & r_2 & q_1 & q_2 & q_3 & q_4 & q_5 & q_6 & q_7 & q_8 & q_9 & q_{10} & q_{11} & q_{12} & q_{13} & q_{14} & q_{15} & q_{16} & q_{17} & q_{18} & q_{19} & q_{20} & q_{21} & q_{22} & q_{23} & q_{24} & \Delta \\
\hline
X_{26} & 0 & 0 & 0 & 1 & 0 & 0 & 0 & 0 & 0 & 1 & 0 & 0 & 0 & 1 & 1 & 0 & 0 & 0 & 1 & 0 & 0 & 0 & 0 & 0 & 1 & 1 & 0 & 0 & 0 & 0 & 1 & 0 & 0 & 0 & 0.264637 \\
        X_{61} & 1 & 0 & 1 & 0 & 0 & 0 & 0 & 0 & 1 & 0 & 1 & 1 & 1 & 0 & 0 & 0 & 0 & 0 & 0 & 0 & 1 & 0 & 1 & 1 & 0 & 0 & 0 & 1 & 0 & 0 & 0 & 1 & 1 & 1 & 0.586522 \\
        X_{43} & 0 & 1 & 1 & 0 & 0 & 0 & 0 & 0 & 0 & 0 & 1 & 0 & 0 & 0 & 0 & 0 & 1 & 1 & 0 & 0 & 1 & 0 & 0 & 0 & 0 & 0 & 0 & 1 & 1 & 1 & 0 & 0 & 0 & 0 & 0.52778 \\
        X_{36} & 0 & 1 & 0 & 0 & 0 & 0 & 0 & 0 & 0 & 0 & 0 & 0 & 0 & 1 & 1 & 1 & 0 & 0 & 1 & 1 & 0 & 0 & 0 & 0 & 1 & 1 & 0 & 0 & 0 & 0 & 1 & 0 & 0 & 0 & 0.300436 \\
        X_{25} & 0 & 0 & 1 & 0 & 0 & 0 & 0 & 0 & 0 & 0 & 0 & 0 & 1 & 0 & 1 & 0 & 0 & 0 & 1 & 0 & 0 & 0 & 0 & 0 & 0 & 0 & 0 & 0 & 0 & 0 & 1 & 1 & 1 & 1 & 0.227344 \\
        X_{41} & 0 & 0 & 0 & 1 & 0 & 0 & 0 & 0 & 1 & 0 & 1 & 0 & 0 & 0 & 0 & 0 & 0 & 0 & 0 & 0 & 1 & 0 & 0 & 0 & 0 & 0 & 0 & 1 & 0 & 0 & 0 & 0 & 0 & 0 & 0.264637 \\
        X_{32} & 1 & 0 & 0 & 0 & 0 & 0 & 0 & 0 & 1 & 0 & 0 & 0 & 0 & 0 & 0 & 1 & 0 & 0 & 0 & 1 & 0 & 0 & 0 & 0 & 0 & 0 & 0 & 0 & 0 & 0 & 0 & 0 & 0 & 0 & 0.359178 \\
\end{array}
      \right)~.~$}
\eea      
For Phase (b) of $P_{+-} ( \textrm{PdP}_3 ) $ we have the following extended $\bar{P}$-matrix:
\bea
\resizebox{0.72\hsize}{!}{$
\left(
  \begin{array}{c|cccccc|cc|cc|ccccccccccccccccccccc|c}
  \;  & p_1 & p_2 & p_3 & p_4 & p_5 & p_6 & s_1 & s_2 & r_1 & r_2 & q_1 & q_2 & q_3 & q_4 & q_5 & q_6 & q_7 & q_8 & q_9 & q_{10} & q_{11} & q_{12} & q_{13} & q_{14} & q_{15} & q_{16} & q_{17} & q_{18} & q_{19} & q_{20} & q_{21} & \Delta\\
  \hline
X_{26} & 0 & 0 & 0 & 1 & 0 & 0 & 0 & 0 & 0 & 1 & 1 & 0 & 0 & 1 & 0 & 0 & 0 & 0 & 0 & 0 & 0 & 0 & 0 & 1 & 0 & 1 & 0 & 0 & 0 & 1 & 0 & 0.264637 \\
        X_{61} & 1 & 0 & 1 & 0 & 0 & 0 & 0 & 0 & 1 & 0 & 0 & 1 & 1 & 0 & 1 & 1 & 1 & 0 & 0 & 0 & 0 & 0 & 0 & 0 & 1 & 0 & 1 & 1 & 0 & 0 & 1 & 0.586522 \\
        X_{43} & 0 & 1 & 1 & 0 & 0 & 0 & 0 & 0 & 0 & 0 & 0 & 0 & 0 & 0 & 0 & 1 & 1 & 1 & 1 & 1 & 1 & 0 & 0 & 0 & 0 & 0 & 0 & 0 & 0 & 0 & 0 & 0.52778 \\
        X_{36} & 0 & 1 & 0 & 0 & 0 & 0 & 0 & 0 & 0 & 0 & 1 & 0 & 0 & 1 & 0 & 0 & 0 & 0 & 0 & 0 & 0 & 1 & 1 & 1 & 0 & 1 & 0 & 0 & 1 & 1 & 0 & 0.300436 \\
        X_{25} & 0 & 0 & 1 & 0 & 0 & 0 & 0 & 0 & 0 & 0 & 0 & 0 & 0 & 0 & 0 & 0 & 0 & 0 & 0 & 0 & 0 & 0 & 0 & 1 & 1 & 1 & 1 & 1 & 0 & 1 & 1 & 0.227344 \\
        X_{41} & 0 & 0 & 0 & 1 & 0 & 0 & 0 & 0 & 1 & 0 & 0 & 0 & 0 & 0 & 0 & 1 & 1 & 0 & 0 & 0 & 0 & 0 & 0 & 0 & 0 & 0 & 0 & 0 & 0 & 0 & 0 & 0.264637 \\
        X_{32} & 1 & 0 & 0 & 0 & 0 & 0 & 0 & 0 & 1 & 0 & 0 & 0 & 0 & 0 & 0 & 0 & 0 & 0 & 0 & 0 & 0 & 1 & 1 & 0 & 0 & 0 & 0 & 0 & 1 & 0 & 0 & 0.359178 \\
\end{array}
\right)~.~$}
\eea

\subsection{$\bar{P}$-matrices for $P^2_{+-} ( \textrm{dP}_3 ) $}

The extended $\bar{P}$-matrix for Phase (a) of $P^2_{+-} ( \textrm{dP}_3 )$ is presented below:
\bea
\resizebox{0.72\hsize}{!}{$
\left(
  \begin{array}{c|ccccccc|cc|cccccccccccccccccccccccc|c}
  \;  & p_2 & p_3& p_4 & p_5 & p_6 & p_7  & p_8  & r_1 & r_2 & q_1 & q_2 & q_3 & q_4 & q_5 & q_6 & q_7 & q_8 & q_9 & q_{10} & q_{11} & q_{12} & q_{13} & q_{14} & q_{15} & q_{16} & q_{17} & q_{18} & q_{19} & q_{20} & q_{21} & q_{22} & q_{23} & q_{24} & \Delta\\
  \hline
X_{26} & 1 & 0 & 0 & 0 & 0 & 0 & 0 & 0 & 0 & 0 & 0 & 0 & 1 & 1 & 0 & 0 & 0 & 1 & 0 & 0 & 0 & 0 & 0 & 1 & 1 & 0 & 0 & 0 & 0 & 1 & 0 & 0 & 0 & 0.124513 \\
        X_{61} & 0 & 1 & 0 & 0 & 0 & 1 & 0 & 0 & 0 & 1 & 1 & 1 & 0 & 0 & 0 & 0 & 0 & 0 & 0 & 1 & 0 & 1 & 1 & 0 & 0 & 0 & 1 & 0 & 0 & 0 & 1 & 1 & 1 & 0.372785 \\
        X_{43} & 1 & 1 & 0 & 0 & 0 & 1 & 0 & 0 & 0 & 1 & 0 & 0 & 0 & 0 & 0 & 1 & 1 & 0 & 0 & 1 & 0 & 0 & 0 & 0 & 0 & 0 & 1 & 1 & 1 & 0 & 0 & 0 & 0 & 0.497298 \\
        X_{36} & 0 & 0 & 1 & 0 & 0 & 0 & 1 & 0 & 0 & 0 & 0 & 0 & 1 & 1 & 1 & 0 & 0 & 1 & 1 & 0 & 0 & 0 & 0 & 1 & 1 & 0 & 0 & 0 & 0 & 1 & 0 & 0 & 0 & 0.313436 \\
        X_{25} & 0 & 0 & 0 & 0 & 0 & 1 & 0 & 0 & 0 & 0 & 0 & 1 & 0 & 1 & 0 & 0 & 0 & 1 & 0 & 0 & 0 & 0 & 0 & 0 & 0 & 0 & 0 & 0 & 0 & 1 & 1 & 1 & 1 & 0.183862 \\
        X_{41} & 0 & 0 & 1 & 0 & 0 & 0 & 0 & 0 & 0 & 1 & 0 & 0 & 0 & 0 & 0 & 0 & 0 & 0 & 0 & 1 & 0 & 0 & 0 & 0 & 0 & 0 & 1 & 0 & 0 & 0 & 0 & 0 & 0 & 0.124513 \\
        X_{32} & 0 & 1 & 0 & 0 & 0 & 0 & 0 & 0 & 0 & 0 & 0 & 0 & 0 & 0 & 1 & 0 & 0 & 0 & 1 & 0 & 0 & 0 & 0 & 0 & 0 & 0 & 0 & 0 & 0 & 0 & 0 & 0 & 0 & 0.188923 \\
\end{array}
\right)~.~$}
\eea

Below we present the extended $\bar{P}$-matrix for Phase (b) of $P^2_{+-} ( \textrm{dP}_3 ) $:
\bea
\resizebox{12.8 cm}{!}{$
\left(
  \begin{array}{c|ccccccc|cc|ccccccccccccccccccccc|c}
  \;  & p_2 & p_3 & p_4 & p_5 & p_6 & p_7 & p_8 & r_1 & r_2 & q_1 & q_2 & q_3 & q_4 & q_5 & q_6 & q_7 & q_8 & q_9 & q_{10} & q_{11} & q_{12} & q_{13} & q_{14} & q_{15} & q_{16} & q_{17} & q_{18} & q_{19} & q_{20} & q_{21} & \Delta\\
  \hline
X_{26} & 1 & 0 & 0 & 0 & 0 & 0 & 0 & 0 & 0 & 1 & 0 & 0 & 1 & 0 & 0 & 0 & 0 & 0 & 0 & 0 & 0 & 0 & 1 & 0 & 1 & 0 & 0 & 0 & 1 & 0 & 0.124513 \\
        X_{61} & 0 & 1 & 0 & 0 & 0 & 1 & 0 & 0 & 0 & 0 & 1 & 1 & 0 & 1 & 1 & 1 & 0 & 0 & 0 & 0 & 0 & 0 & 0 & 1 & 0 & 1 & 1 & 0 & 0 & 1 & 0.372785 \\
        X_{43} & 1 & 1 & 0 & 0 & 0 & 1 & 0 & 0 & 0 & 0 & 0 & 0 & 0 & 0 & 1 & 1 & 1 & 1 & 1 & 1 & 0 & 0 & 0 & 0 & 0 & 0 & 0 & 0 & 0 & 0 & 0.497298 \\
        X_{36} & 0 & 0 & 1 & 0 & 0 & 0 & 1 & 0 & 0 & 1 & 0 & 0 & 1 & 0 & 0 & 0 & 0 & 0 & 0 & 0 & 1 & 1 & 1 & 0 & 1 & 0 & 0 & 1 & 1 & 0 & 0.313436 \\
        X_{25} & 0 & 0 & 0 & 0 & 0 & 1 & 0 & 0 & 0 & 0 & 0 & 0 & 0 & 0 & 0 & 0 & 0 & 0 & 0 & 0 & 0 & 0 & 1 & 1 & 1 & 1 & 1 & 0 & 1 & 1 & 0.183862 \\
        X_{41} & 0 & 0 & 1 & 0 & 0 & 0 & 0 & 0 & 0 & 0 & 0 & 0 & 0 & 0 & 1 & 1 & 0 & 0 & 0 & 0 & 0 & 0 & 0 & 0 & 0 & 0 & 0 & 0 & 0 & 0 & 0.124513 \\
        X_{32} & 0 & 1 & 0 & 0 & 0 & 0 & 0 & 0 & 0 & 0 & 0 & 0 & 0 & 0 & 0 & 0 & 0 & 0 & 0 & 0 & 1 & 1 & 0 & 0 & 0 & 0 & 0 & 1 & 0 & 0 & 0.188923 \\
\end{array}
\right)~.~$}
\eea

\subsection{$\bar{P}$-matrices for $P_{+-} (\IC^3 / \IZ_2 \times \IZ_2)$ and $P_{+-}(\cC / \IZ_2)$.}

Below is the $\bar{P}$-matrix for Phase (a) of $P_{+-} (\IC^3 / \IZ_2 \times \IZ_2)$.
\bea
\resizebox{0.32\hsize}{!}{$
\left(
\begin{array}{c|ccccc|cc|cc|cc}
       \;  & p_1 & p_2 & p_3 & p_4 & p_5 & q_1 & q_2 & r_1 & r_2 & s_1 & s_2 \\
         \hline
        X_{31} & 1 & 0 & 0 & 0 & 0 & 0 & 0 & 0 & 1 & 0 & 1 \\
        X_{75} & 1 & 0 & 0 & 0 & 0 & 0 & 0 & 0 & 1 & 0 & 1 \\
        Y_{64} & 1 & 0 & 0 & 0 & 0 & 0 & 0 & 0 & 1 & 1 & 0 \\
        Z_{82} & 1 & 0 & 0 & 0 & 0 & 0 & 0 & 1 & 0 & 0 & 1 \\
        Y_{53} & 1 & 0 & 0 & 0 & 1 & 0 & 0 & 1 & 0 & 1 & 0 \\
        Z_{53} & 1 & 0 & 0 & 1 & 0 & 0 & 0 & 1 & 0 & 1 & 0 \\
        X_{21} & 0 & 1 & 0 & 0 & 0 & 0 & 1 & 0 & 1 & 0 & 0 \\
\end{array}\label{C3 3d print Phase A}
\right)~,~$}
\eea


Below is the $\bar{P}$-matrix for Phase (b) of $P_{+-} (\IC^3 / \IZ_2 \times \IZ_2)$.
\begin{align}
\resizebox{0.32\hsize}{!}{$
 \left(
\begin{array}{c|ccccc|cc|cc|cc}
       \;  & p_1 & p_2 & p_3 & p_4 & p_5 & q_1 & q_2 & r_1 & r_2 & s_1 & s_2 \\
         \hline
        X_{47} & 0 & 1 & 0 & 0 & 0 & 0 & 1 & 0 & 1 & 0 & 0 \\
        X_{57} & 1 & 0 & 0 & 0 & 0 & 0 & 0 & 0 & 1 & 0 & 1 \\
        X_{27} & 0 & 0 & 1 & 0 & 0 & 0 & 1 & 0 & 0 & 0 & 1 \\
        X_{31} & 1 & 0 & 0 & 0 & 0 & 0 & 0 & 1 & 0 & 1 & 0 \\
        Y_{64} & 1 & 0 & 0 & 0 & 0 & 0 & 0 & 1 & 0 & 0 & 1 \\
        Z_{82} & 1 & 0 & 0 & 0 & 0 & 0 & 0 & 0 & 1 & 1 & 0 \\
        X_{21} & 0 & 1 & 0 & 0 & 0 & 0 & 1 & 1 & 0 & 0 & 0 \\
\end{array}\label{C3 3d print Phase B}
\right)~,~$}
\end{align}

Below is the $\bar{P}$-matrix for Phase (a) of $P_{+-}(\cC / \IZ_2)$.
\bea
 \resizebox{0.32\hsize}{!}{$
 \left(\begin{array}{c|cccccc|cc|cc}
        \; & p_2 & p_3 & p_4 & p_5 & p_6 & p_7 & q_1 & q_2 & r_1 & r_2 \\
        \hline
        X_{21} & 0 & 1 & 0 & 0 & 0 & 0 & 0 & 0 & 1 & 0 \\
        X_{32} & 0 & 0 & 0 & 0 & 0 & 1 & 1 & 0 & 0 & 0 \\
        X_{34} & 1 & 0 & 0 & 0 & 0 & 0 & 0 & 0 & 1 & 0 \\
        X_{41} & 0 & 0 & 0 & 0 & 1 & 0 & 1 & 0 & 0 & 0 \\
        X_{65} & 0 & 1 & 0 & 0 & 0 & 0 & 0 & 0 & 1 & 0 \\
        X_{76} & 0 & 0 & 0 & 0 & 0 & 1 & 1 & 0 & 0 & 0 \\
        X_{78} & 1 & 0 & 0 & 0 & 0 & 0 & 0 & 0 & 1 & 0 \\
\end{array}\label{Conifold 3d print phase a P-matrix}
\right)~,~$}
\eea

\subsection{$\bar{P}$-matrices for model 1 and its deformation}

The extended $\bar{P}$-matrix for model 1 is presented below, continued to the next page:
\bea
\resizebox{0.32\hsize}{!}{$
\left(
\begin{array}{c|cccccc|ccc|ccc|cccccccc|c}
           \; & p_1 & p_2 & p_3 & p_4 & p_5 & p_6 & r_1 & r_2 & r_3 & l_1 & l_2 & l_3 & q_1 & q_2 & q_3 & q_4 & q_5 & q_6 & q_7 & q_8  & \Delta\\
          \hline
        X_{31} & 0 & 1 & 1 & 0 & 0 & 0 & 0 & 0 & 0 & 0 & 0 & 0 & 1 & 1 & 1 & 0 & 0 & 0 & 0 & 1 & 0.5494 \\
        X_{11,9} & 0 & 1 & 1 & 0 & 0 & 0 & 0 & 0 & 0 & 0 & 0 & 0 & 1 & 1 & 1 & 0 & 0 & 0 & 0 & 1 & 0.5494 \\
        X_{32} & 1 & 0 & 1 & 0 & 0 & 0 & 1 & 1 & 0 & 1 & 0 & 0 & 1 & 0 & 0 & 0 & 0 & 0 & 0 & 0 & 0.502203 \\
        X_{11,10} & 1 & 0 & 1 & 0 & 0 & 0 & 1 & 1 & 0 & 1 & 0 & 0 & 1 & 0 & 0 & 0 & 0 & 0 & 0 & 0 & 0.502203 \\
        X_{71} & 1 & 0 & 1 & 0 & 0 & 0 & 0 & 1 & 1 & 0 & 1 & 0 & 0 & 0 & 1 & 0 & 0 & 0 & 0 & 0 & 0.502203 \\
        X_{15,9} & 1 & 0 & 1 & 0 & 0 & 0 & 0 & 1 & 1 & 0 & 1 & 0 & 0 & 0 & 1 & 0 & 0 & 0 & 0 & 0 & 0.502203 \\
        X_{21} & 0 & 0 & 0 & 1 & 0 & 0 & 0 & 0 & 1 & 0 & 1 & 1 & 0 & 1 & 1 & 0 & 0 & 0 & 0 & 0 & 0.446566 \\
\end{array}
      \right)~,~$}
      \label{pm-m1-before}
\eea

\begin{align}
\resizebox{10 cm}{!}{$
\left(
\begin{array}{c|cccccccccccccccccccccccccccccccccccccccc}
        & t_1 & t_2 & t_3 & t_4 & t_5 & t_6 & t_7 & t_8 & t_{9} & t_{10} & t_{11} & t_{12} & t_{13} & t_{14} & t_{15} & t_{16} & t_{17} & t_{18} & t_{19} & t_{20} & t_{21} & t_{22} & t_{23} & t_{24} & t_{25} & t_{26} & t_{27} & t_{28} & t_{29} & t_{30} & t_{31} & t_{32} & t_{33} & t_{34} & t_{35} & t_{36} & t_{37} & t_{38} & t_{39} & t_{40} \\
         \hline
        X_{31} & 1 & 1 & 1 & 0 & 0 & 0 & 1 & 1 & 0 & 0 & 1 & 1 & 1 & 1 & 1 & 1 & 1 & 1 & 1 & 1 & 1 & 1 & 1 & 1 & 1 & 1 & 0 & 0 & 0 & 0 & 0 & 0 & 0 & 0 & 0 & 0 & 0 & 0 & 0 & 0 \\
        X_{11,9} & 1 & 1 & 1 & 0 & 0 & 0 & 1 & 1 & 0 & 0 & 0 & 1 & 1 & 1 & 1 & 1 & 0 & 0 & 1 & 0 & 1 & 1 & 1 & 1 & 1 & 0 & 1 & 1 & 1 & 1 & 1 & 1 & 1 & 1 & 1 & 1 & 0 & 0 & 0 & 0 \\
        X_{32} & 1 & 1 & 0 & 0 & 0 & 0 & 1 & 1 & 0 & 0 & 1 & 1 & 1 & 1 & 1 & 1 & 1 & 1 & 0 & 0 & 0 & 0 & 0 & 0 & 0 & 0 & 0 & 0 & 0 & 0 & 0 & 0 & 0 & 0 & 0 & 0 & 0 & 0 & 0 & 0 \\
        X_{11,10} & 1 & 1 & 0 & 0 & 0 & 0 & 1 & 1 & 0 & 0 & 0 & 1 & 1 & 1 & 1 & 1 & 0 & 0 & 1 & 0 & 1 & 1 & 1 & 1 & 0 & 0 & 1 & 1 & 1 & 1 & 1 & 1 & 1 & 1 & 0 & 0 & 0 & 0 & 0 & 0 \\
        X_{71} & 1 & 1 & 1 & 0 & 0 & 0 & 1 & 0 & 0 & 0 & 0 & 1 & 1 & 1 & 1 & 1 & 1 & 1 & 0 & 0 & 1 & 1 & 1 & 1 & 1 & 1 & 0 & 0 & 0 & 0 & 0 & 0 & 0 & 0 & 0 & 0 & 0 & 0 & 0 & 0 \\
        X_{15,9} & 1 & 1 & 1 & 0 & 0 & 0 & 1 & 0 & 0 & 0 & 0 & 0 & 0 & 1 & 1 & 1 & 0 & 0 & 0 & 0 & 0 & 1 & 1 & 1 & 0 & 0 & 0 & 0 & 1 & 1 & 1 & 1 & 1 & 1 & 1 & 1 & 0 & 0 & 0 & 0 \\
        X_{21} & 0 & 0 & 1 & 0 & 0 & 0 & 0 & 0 & 0 & 0 & 0 & 0 & 0 & 0 & 0 & 0 & 0 & 0 & 1 & 1 & 1 & 1 & 1 & 1 & 1 & 1 & 0 & 0 & 0 & 0 & 0 & 0 & 0 & 0 & 0 & 0 & 0 & 0 & 0 & 0 \\
\end{array}
      \right)~.~$}
\end{align}

Below we present the $\bar{P}$-matrix for the deformation of model 1.
\bea
\resizebox{0.34\hsize}{!}{$
\left(
\begin{array}{c|ccccccc|cc|cccccccc|c}
        \;  & p_1 & p_2 & p_3 & p_4 & p_5 & p_6 & p_7 & r_1 & r_2 & q_1 & q_2 & q_3 & q_4 & q_5 & q_6 & q_7 & q_8 & \Delta \\
         \hline
        X_{32} & 0 & 1 & 0 & 0 & 0 & 0 & 1 & 0 & 1 & 1 & 0 & 0 & 0 & 0 & 0 & 0 & 0 & 0.459721 \\
        X_{11,10} & 0 & 1 & 0 & 0 & 0 & 0 & 1 & 0 & 1 & 1 & 0 & 0 & 0 & 0 & 0 & 0 & 0 & 0.459721 \\
        X_{71} & 0 & 1 & 0 & 0 & 0 & 1 & 0 & 0 & 0 & 0 & 0 & 1 & 0 & 0 & 0 & 0 & 0 & 0.383693 \\
        X_{15,9} & 0 & 1 & 0 & 0 & 0 & 1 & 0 & 0 & 0 & 0 & 0 & 1 & 0 & 0 & 0 & 0 & 0 & 0.383693 \\
        X_{21} & 0 & 0 & 1 & 0 & 0 & 0 & 0 & 1 & 0 & 0 & 1 & 1 & 0 & 0 & 0 & 0 & 0 & 0.385513 \\
        X_{10,9} & 0 & 0 & 1 & 0 & 0 & 0 & 0 & 1 & 0 & 0 & 1 & 1 & 0 & 0 & 0 & 0 & 0 & 0.385513 \\
        X_{76} & 1 & 0 & 0 & 0 & 0 & 1 & 0 & 0 & 0 & 0 & 0 & 1 & 1 & 1 & 0 & 0 & 1 & 0.470853 \\
\end{array}
      \right)~,~$}
      \label{pm-m1-after}
\eea

\bea
\resizebox{0.62\hsize}{!}{$
\left(
\begin{array}{c|cccccccccccccccccccccccccccccccccccccccc}
      \; & t_1 & t_2 & t_3 & t_4 & t_5 & t_6 & t_7 & t_8 & t_{9} & t_{10} & t_{11} & t_{12} & t_{13} & t_{14} & t_{15} & t_{16} & t_{17} & t_{18} & t_{19} & t_{20} & t_{21} & t_{22} & t_{23} & t_{24} & t_{25} & t_{26} & t_{27} & t_{28} & t_{29} & t_{30} & t_{31} & t_{32} & t_{33} & t_{34} & t_{35} & t_{36} & t_{37} & t_{38} & t_{39} & t_{40} \\
     \hline
        X_{32}    & 1 & 1 & 0 & 0 & 0 & 0 & 1 & 1 & 0 & 0 & 1 & 1 & 1 & 1 & 1 & 1 & 1 & 1 & 0 & 0 & 0 & 0 & 0 & 0 & 0 & 0 & 0 & 0 & 0 & 0 & 0 & 0 & 0 & 0 & 0 & 0 & 0 & 0 & 0 & 0 \\
        X_{11,10} & 1 & 1 & 0 & 0 & 0 & 0 & 1 & 1 & 0 & 0 & 0 & 1 & 1 & 1 & 1 & 1 & 0 & 0 & 1 & 0 & 1 & 1 & 1 & 1 & 0 & 0 & 1 & 1 & 1 & 1 & 1 & 1 & 1 & 1 & 0 & 0 & 0 & 0 & 0 & 0 \\
        X_{71}    & 1 & 1 & 1 & 0 & 0 & 0 & 1 & 0 & 0 & 0 & 0 & 1 & 1 & 1 & 1 & 1 & 1 & 1 & 0 & 0 & 1 & 1 & 1 & 1 & 1 & 1 & 0 & 0 & 0 & 0 & 0 & 0 & 0 & 0 & 0 & 0 & 0 & 0 & 0 & 0 \\
        X_{15,9}  & 1 & 1 & 1 & 0 & 0 & 0 & 1 & 0 & 0 & 0 & 0 & 0 & 0 & 1 & 1 & 1 & 0 & 0 & 0 & 0 & 0 & 1 & 1 & 1 & 0 & 0 & 0 & 0 & 1 & 1 & 1 & 1 & 1 & 1 & 1 & 1 & 0 & 0 & 0 & 0 \\
        X_{21}    & 0 & 0 & 1 & 0 & 0 & 0 & 0 & 0 & 0 & 0 & 0 & 0 & 0 & 0 & 0 & 0 & 0 & 0 & 1 & 1 & 1 & 1 & 1 & 1 & 1 & 1 & 0 & 0 & 0 & 0 & 0 & 0 & 0 & 0 & 0 & 0 & 0 & 0 & 0 & 0 \\
        X_{10,9}  & 0 & 0 & 1 & 0 & 0 & 0 & 0 & 0 & 0 & 0 & 0 & 0 & 0 & 0 & 0 & 0 & 0 & 0 & 0 & 0 & 0 & 0 & 0 & 0 & 1 & 0 & 0 & 0 & 0 & 0 & 0 & 0 & 0 & 0 & 1 & 1 & 0 & 0 & 0 & 0 \\
        X_{76}    & 1 & 0 & 0 & 0 & 0 & 0 & 1 & 0 & 0 & 0 & 0 & 1 & 0 & 1 & 0 & 0 & 1 & 0 & 0 & 0 & 0 & 0 & 0 & 0 & 1 & 0 & 0 & 0 & 0 & 0 & 0 & 0 & 0 & 0 & 0 & 0 & 0 & 1 & 0 & 0 \\
\end{array}
      \right)~.~$}
\eea

\bibliographystyle{JHEP}
\bibliography{mybib}

\end{document}

%% file: pref.tex
\newcommand{\be}{\begin{equation}}
\newcommand{\ee}{\end{equation}}
\newcommand{\beq}{\begin{equation}}
\newcommand{\beql}[1]{\begin{equation}\label{#1}}
\newcommand{\eeq}{\end{equation}}
\newcommand{\ba}{\begin{array}}
\newcommand{\ea}{\end{array}}
\newcommand{\bea}{\begin{eqnarray}}
\newcommand{\beal}[1]{\begin{eqnarray}\label{#1}}
\newcommand{\eea}{\end{eqnarray}}
\newcommand{\ben}{\begin{enumerate}}
\newcommand{\een}{\end{enumerate}}
\newcommand{\bean}{\begin{eqnarray*}}
\newcommand{\eean}{\end{eqnarray*}}
\newcommand{\eref}[1]{(\ref{#1})}
\newcommand{\sref}[1]{\S\ref{#1}}
\newcommand{\tref}[1]{Table~\ref{#1}}
\newcommand{\nn}{\nonumber}

\newcommand{\fref}[1]{Figure \ref{#1}}
\newcommand{\btab}[1]{\begin{tabular}{#1}}
\newcommand{\etab}{\end{tabular}}

\newcommand{\comment}[1]{}

\newcommand{\IC}{\mathbb{C}}

\newcommand{\qed}{\nobreak \ifvmode \relax \else
      \ifdim\lastskip<1.5em \hskip-\lastskip
      \hskip1.5em plus0em minus0.5em \fi \nobreak
      \vrule height0.75em width0.5em depth0.25em\fi}

\newcommand{\cN}{{\cal N}}

\newcommand{\cC}{{\cal C}}

\newcommand{\IZ}{\mathbb{Z}}

\definecolor{darkspringgreen}{rgb}{0.09, 0.45, 0.27}
\definecolor{forestgreen}{rgb}{0.13, 0.55, 0.13}

%% file: main_final.bbl
\providecommand{\href}[2]{#2}\begingroup\raggedright\begin{thebibliography}{10}

\bibitem{Franco:2015tna}
S.~Franco, D.~Ghim, S.~Lee, R.-K.~Seong and D.~Yokoyama, \emph{{2d (0,2) Quiver
  Gauge Theories and D-Branes}},
  \href{https://doi.org/10.1007/JHEP09(2015)072}{\emph{JHEP} {\bfseries 09}
  (2015) 072} [\href{https://arxiv.org/abs/1506.03818}{{\ttfamily
  1506.03818}}].

\bibitem{Franco:2015tya}
S.~Franco, S.~Lee and R.-K.~Seong, \emph{{Brane Brick Models, Toric Calabi-Yau
  4-Folds and 2d (0,2) Quivers}},
  \href{https://doi.org/10.1007/JHEP02(2016)047}{\emph{JHEP} {\bfseries 02}
  (2016) 047} [\href{https://arxiv.org/abs/1510.01744}{{\ttfamily
  1510.01744}}].

\bibitem{Franco:2016nwv}
S.~Franco, S.~Lee and R.-K.~Seong, \emph{{Brane brick models and 2d (0, 2)
  triality}}, \href{https://doi.org/10.1007/JHEP05(2016)020}{\emph{JHEP}
  {\bfseries 05} (2016) 020}
  [\href{https://arxiv.org/abs/1602.01834}{{\ttfamily 1602.01834}}].

\bibitem{Franco:2016qxh}
S.~Franco, S.~Lee, R.-K.~Seong and C.~Vafa, \emph{{Brane Brick Models in the
  Mirror}}, \href{https://doi.org/10.1007/JHEP02(2017)106}{\emph{JHEP}
  {\bfseries 02} (2017) 106}
  [\href{https://arxiv.org/abs/1609.01723}{{\ttfamily 1609.01723}}].

\bibitem{Franco:2016fxm}
S.~Franco, S.~Lee and R.-K.~Seong, \emph{{Orbifold Reduction and 2d (0,2) Gauge
  Theories}}, \href{https://doi.org/10.1007/JHEP03(2017)016}{\emph{JHEP}
  {\bfseries 03} (2017) 016}
  [\href{https://arxiv.org/abs/1609.07144}{{\ttfamily 1609.07144}}].

\bibitem{Franco:2017cjj}
S.~Franco, D.~Ghim, S.~Lee and R.-K.~Seong, \emph{{Elliptic Genera of 2d (0,2)
  Gauge Theories from Brane Brick Models}},
  \href{https://doi.org/10.1007/JHEP06(2017)068}{\emph{JHEP} {\bfseries 06}
  (2017) 068} [\href{https://arxiv.org/abs/1702.02948}{{\ttfamily
  1702.02948}}].

\bibitem{Franco:2018qsc}
S.~Franco and A.~Hasan, \emph{{$3d$ printing of $2d$ $
  \mathcal{N}=\left(0,2\right) $ gauge theories}},
  \href{https://doi.org/10.1007/JHEP05(2018)082}{\emph{JHEP} {\bfseries 05}
  (2018) 082} [\href{https://arxiv.org/abs/1801.00799}{{\ttfamily
  1801.00799}}].

\bibitem{Franco:2019bmx}
S.~Franco and A.~Hasan, \emph{{Graded Quivers, Generalized Dimer Models and
  Toric Geometry}}, \href{https://doi.org/10.1007/JHEP11(2019)104}{\emph{JHEP}
  {\bfseries 11} (2019) 104}
  [\href{https://arxiv.org/abs/1904.07954}{{\ttfamily 1904.07954}}].

\bibitem{Franco:2020avj}
S.~Franco and A.~Hasan, \emph{{Calabi-Yau products: graded quivers for general
  toric Calabi-Yaus}},
  \href{https://doi.org/10.1007/JHEP02(2021)174}{\emph{JHEP} {\bfseries 02}
  (2021) 174} [\href{https://arxiv.org/abs/2004.13765}{{\ttfamily
  2004.13765}}].

\bibitem{Franco:2021elb}
S.~Franco and X.~Yu, \emph{{BFT$_2$: a General Class of $2d$$\mathcal{N}=(0,2)$
  Theories, 3-Manifolds and Toric Geometry}},
  \href{https://arxiv.org/abs/2107.00667}{{\ttfamily 2107.00667}}.

\bibitem{Franco:2022iap}
S.~Franco, \emph{{2d Supersymmetric Gauge Theories, D-branes and Trialities}},
  \href{https://arxiv.org/abs/2201.10987}{{\ttfamily 2201.10987}}.

\bibitem{Franco:2022gvl}
S.~Franco and R.-K.~Seong, \emph{{Fano 3-folds, reflexive polytopes and brane
  brick models}}, \href{https://doi.org/10.1007/JHEP08(2022)008}{\emph{JHEP}
  {\bfseries 08} (2022) 008}
  [\href{https://arxiv.org/abs/2203.15816}{{\ttfamily 2203.15816}}].

\bibitem{Franco:2022isw}
S.~Franco, D.~Ghim and R.-K.~Seong, \emph{{Brane brick models for the
  Sasaki-Einstein 7-manifolds
  Y$^{p,k}$({\ensuremath{\mathbb{C}}}{\ensuremath{\mathbb{P}}}$^{1}${\texttimes}
  {\ensuremath{\mathbb{C}}}{\ensuremath{\mathbb{P}}}$^{1}$) and
  Y$^{p,k}$({\ensuremath{\mathbb{C}}}{\ensuremath{\mathbb{P}}}$^{2}$)}},
  \href{https://doi.org/10.1007/JHEP03(2023)050}{\emph{JHEP} {\bfseries 03}
  (2023) 050} [\href{https://arxiv.org/abs/2212.02523}{{\ttfamily
  2212.02523}}].

\bibitem{Franco:2023tly}
S.~Franco, \emph{{4d crystal melting, toric Calabi-Yau 4-folds and brane brick
  models}}, \href{https://doi.org/10.1007/JHEP03(2024)091}{\emph{JHEP}
  {\bfseries 03} (2024) 091}
  [\href{https://arxiv.org/abs/2311.04404}{{\ttfamily 2311.04404}}].

\bibitem{Franco:2024lxs}
S.~Franco, \emph{{2d (0, 2) gauge theories from branes: Recent progress in
  brane brick models}},
  \href{https://doi.org/10.1142/S0217751X24460059}{\emph{Int. J. Mod. Phys. A}
  {\bfseries 39} (2024) 2446005}
  [\href{https://arxiv.org/abs/2402.06993}{{\ttfamily 2402.06993}}].

\bibitem{Franco:2023tyf}
S.~Franco, D.~Ghim, G.P.~Goulas and R.-K.~Seong, \emph{{Mass deformations of
  brane brick models}},
  \href{https://doi.org/10.1007/JHEP09(2023)176}{\emph{JHEP} {\bfseries 09}
  (2023) 176} [\href{https://arxiv.org/abs/2307.03220}{{\ttfamily
  2307.03220}}].

\bibitem{Klebanov:1998hh}
I.R.~Klebanov and E.~Witten, \emph{{Superconformal field theory on three-branes
  at a Calabi-Yau singularity}},
  \href{https://doi.org/10.1016/S0550-3213(98)00654-3}{\emph{Nucl.Phys.}
  {\bfseries B536} (1998) 199}
  [\href{https://arxiv.org/abs/hep-th/9807080}{{\ttfamily hep-th/9807080}}].

\bibitem{Franco:2005rj}
S.~Franco, A.~Hanany, K.D.~Kennaway, D.~Vegh and B.~Wecht, \emph{{Brane Dimers
  and Quiver Gauge Theories}},
  \href{https://doi.org/10.1088/1126-6708/2006/01/096}{\emph{JHEP} {\bfseries
  01} (2006) 096} [\href{https://arxiv.org/abs/hep-th/0504110}{{\ttfamily
  hep-th/0504110}}].

\bibitem{Franco:2005sm}
S.~Franco et~al., \emph{{Gauge theories from toric geometry and brane
  tilings}}, \href{https://doi.org/10.1088/1126-6708/2006/01/128}{\emph{JHEP}
  {\bfseries 01} (2006) 128}
  [\href{https://arxiv.org/abs/hep-th/0505211}{{\ttfamily hep-th/0505211}}].

\bibitem{Bianchi:2014qma}
M.~Bianchi, S.~Cremonesi, A.~Hanany, J.F.~Morales, D.~Ricci~Pacifici and
  R.-K.~Seong, \emph{{Mass-deformed Brane Tilings}},
  \href{https://doi.org/10.1007/JHEP10(2014)027}{\emph{JHEP} {\bfseries 10}
  (2014) 27} [\href{https://arxiv.org/abs/1408.1957}{{\ttfamily 1408.1957}}].

\bibitem{Cremonesi:2023psg}
S.~Cremonesi and J.~S{\'a}, \emph{{Zig-zag deformations of toric quiver gauge
  theories. Part I. Reflexive polytopes}},
  \href{https://doi.org/10.1007/JHEP05(2024)114}{\emph{JHEP} {\bfseries 05}
  (2024) 114} [\href{https://arxiv.org/abs/2312.13909}{{\ttfamily
  2312.13909}}].

\bibitem{Ghim:2024asj}
D.~Ghim, M.~Kho and R.-K.~Seong, \emph{{Combinatorial and algebraic mutations
  of toric Fano 3-folds and mass deformations of 2d(0,2) quiver gauge
  theories}}, \href{https://doi.org/10.1103/PhysRevD.110.086001}{\emph{Phys.
  Rev. D} {\bfseries 110} (2024) 086001}
  [\href{https://arxiv.org/abs/2407.19924}{{\ttfamily 2407.19924}}].

\bibitem{Ghim:2025zhs}
D.~Ghim, M.~Kho and R.-K.~Seong, \emph{{Birational transformations and 2d (0,
  2) quiver gauge theories beyond toric Fano 3-folds}},
  \href{https://doi.org/10.1007/JHEP06(2025)032}{\emph{JHEP} {\bfseries 06}
  (2025) 032} [\href{https://arxiv.org/abs/2502.08741}{{\ttfamily
  2502.08741}}].

\bibitem{Franco:2023flw}
S.~Franco and R.-K.~Seong, \emph{{Twin theories, polytope mutations and quivers
  for GTPs}}, \href{https://doi.org/10.1007/JHEP07(2023)034}{\emph{JHEP}
  {\bfseries 07} (2023) 034}
  [\href{https://arxiv.org/abs/2302.10951}{{\ttfamily 2302.10951}}].

\bibitem{Franco:2023mkw}
S.~Franco and D.~Rodriguez-Gomez, \emph{{Quiver tails and brane webs}},
  \href{https://doi.org/10.1007/JHEP10(2024)118}{\emph{JHEP} {\bfseries 10}
  (2024) 118} [\href{https://arxiv.org/abs/2310.10724}{{\ttfamily
  2310.10724}}].

\bibitem{Arias-Tamargo:2024fjt}
G.~Arias-Tamargo, S.~Franco and D.~Rodr\'\i{}guez-G\'omez, \emph{{The geometry
  of GTPs and 5d SCFTs}},
  \href{https://doi.org/10.1007/JHEP07(2024)159}{\emph{JHEP} {\bfseries 07}
  (2024) 159} [\href{https://arxiv.org/abs/2403.09776}{{\ttfamily
  2403.09776}}].

\bibitem{CarrenoBolla:2024fxy}
I.~Carre{\~n}o~Bolla, S.~Franco and D.~Rodr{\'\i}guez-G{\'o}mez, \emph{{The 5d
  tangram: brane webs, 7-branes and primitive T-cones}},
  \href{https://doi.org/10.1007/JHEP05(2025)175}{\emph{JHEP} {\bfseries 05}
  (2025) 175} [\href{https://arxiv.org/abs/2411.01510}{{\ttfamily
  2411.01510}}].

\bibitem{Gadde:2013lxa}
A.~Gadde, S.~Gukov and P.~Putrov, \emph{{(0, 2) trialities}},
  \href{https://doi.org/10.1007/JHEP03(2014)076}{\emph{JHEP} {\bfseries 03}
  (2014) 076} [\href{https://arxiv.org/abs/1310.0818}{{\ttfamily 1310.0818}}].

\bibitem{Seiberg:1994pq}
N.~Seiberg, \emph{{Electric - magnetic duality in supersymmetric nonAbelian
  gauge theories}},
  \href{https://doi.org/10.1016/0550-3213(94)00023-8}{\emph{Nucl. Phys.}
  {\bfseries B435} (1995) 129}
  [\href{https://arxiv.org/abs/hep-th/9411149}{{\ttfamily hep-th/9411149}}].

\bibitem{Franco:2016tcm}
S.~Franco, S.~Lee, R.-K.~Seong and C.~Vafa, \emph{{Quadrality for
  Supersymmetric Matrix Models}},
  \href{https://doi.org/10.1007/JHEP07(2017)053}{\emph{JHEP} {\bfseries 07}
  (2017) 053} [\href{https://arxiv.org/abs/1612.06859}{{\ttfamily
  1612.06859}}].

\bibitem{Franco:2017lpa}
S.~Franco and G.~Musiker, \emph{{Higher Cluster Categories and QFT Dualities}},
  \href{https://doi.org/10.1103/PhysRevD.98.046021}{\emph{Phys. Rev. D}
  {\bfseries 98} (2018) 046021}
  [\href{https://arxiv.org/abs/1711.01270}{{\ttfamily 1711.01270}}].

\bibitem{Closset:2018axq}
C.~Closset, S.~Franco, J.~Guo and A.~Hasan, \emph{{Graded quivers and B-branes
  at Calabi-Yau singularities}},
  \href{https://doi.org/10.1007/JHEP03(2019)053}{\emph{JHEP} {\bfseries 03}
  (2019) 053} [\href{https://arxiv.org/abs/1811.07016}{{\ttfamily
  1811.07016}}].

\bibitem{Intriligator:2003jj}
K.A.~Intriligator and B.~Wecht, \emph{{The Exact superconformal R symmetry
  maximizes a}},
  \href{https://doi.org/10.1016/S0550-3213(03)00459-0}{\emph{Nucl. Phys. B}
  {\bfseries 667} (2003) 183}
  [\href{https://arxiv.org/abs/hep-th/0304128}{{\ttfamily hep-th/0304128}}].

\bibitem{Gubser:1998vd}
S.S.~Gubser, \emph{{Einstein manifolds and conformal field theories}},
  \href{https://doi.org/10.1103/PhysRevD.59.025006}{\emph{Phys. Rev. D}
  {\bfseries 59} (1999) 025006}
  [\href{https://arxiv.org/abs/hep-th/9807164}{{\ttfamily hep-th/9807164}}].

\bibitem{Herzog:2002ih}
C.P.~Herzog, I.R.~Klebanov and P.~Ouyang, \emph{{D-branes on the conifold and
  N=1 gauge / gravity dualities}},  in \emph{{Les Houches Summer School:
  Session 76: Euro Summer School on Unity of Fundamental Physics: Gravity,
  Gauge Theory and Strings}}, pp.~189--223, 5, 2002
  [\href{https://arxiv.org/abs/hep-th/0205100}{{\ttfamily hep-th/0205100}}].

\bibitem{Benvenuti:2004dy}
S.~Benvenuti, S.~Franco, A.~Hanany, D.~Martelli and J.~Sparks, \emph{{An
  infinite family of superconformal quiver gauge theories with Sasaki-Einstein
  duals}}, \href{https://doi.org/10.1088/1126-6708/2005/06/064}{\emph{JHEP}
  {\bfseries 06} (2005) 064}
  [\href{https://arxiv.org/abs/hep-th/0411264}{{\ttfamily hep-th/0411264}}].

\bibitem{Gubser:1998fp}
S.S.~Gubser and I.R.~Klebanov, \emph{{Baryons and domain walls in an N=1
  superconformal gauge theory}},
  \href{https://doi.org/10.1103/PhysRevD.58.125025}{\emph{Phys. Rev. D}
  {\bfseries 58} (1998) 125025}
  [\href{https://arxiv.org/abs/hep-th/9808075}{{\ttfamily hep-th/9808075}}].

\bibitem{Martelli:2006yb}
D.~Martelli, J.~Sparks and S.-T.~Yau, \emph{{Sasaki-Einstein manifolds and
  volume minimisation}},
  \href{https://doi.org/10.1007/s00220-008-0479-4}{\emph{Commun. Math. Phys.}
  {\bfseries 280} (2008) 611}
  [\href{https://arxiv.org/abs/hep-th/0603021}{{\ttfamily hep-th/0603021}}].

\bibitem{Amariti:2012tj}
A.~Amariti and S.~Franco, \emph{{Free Energy vs Sasaki-Einstein Volume for
  Infinite Families of M2-Brane Theories}},
  \href{https://doi.org/10.1007/JHEP09(2012)034}{\emph{JHEP} {\bfseries 09}
  (2012) 034} [\href{https://arxiv.org/abs/1204.6040}{{\ttfamily 1204.6040}}].

\bibitem{Martelli:2005tp}
D.~Martelli, J.~Sparks and S.-T.~Yau, \emph{{The geometric dual of
  a-maximisation for toric Sasaki- Einstein manifolds}},
  \href{https://doi.org/10.1007/s00220-006-0087-0}{\emph{Commun. Math. Phys.}
  {\bfseries 268} (2006) 39}
  [\href{https://arxiv.org/abs/hep-th/0503183}{{\ttfamily hep-th/0503183}}].

\bibitem{Morrison:1998cs}
D.R.~Morrison and M.R.~Plesser, \emph{{Nonspherical horizons. 1.}},
  {\emph{Adv.Theor.Math.Phys.} {\bfseries 3} (1999) 1}
  [\href{https://arxiv.org/abs/hep-th/9810201}{{\ttfamily hep-th/9810201}}].

\bibitem{Beasley:2001zp}
C.E.~Beasley and M.R.~Plesser, \emph{{Toric duality is Seiberg duality}},
  \href{https://doi.org/10.1088/1126-6708/2001/12/001}{\emph{JHEP} {\bfseries
  0112} (2001) 001} [\href{https://arxiv.org/abs/hep-th/0109053}{{\ttfamily
  hep-th/0109053}}].

\bibitem{Feng:2001bn}
B.~Feng, A.~Hanany, Y.-H.~He and A.M.~Uranga, \emph{{Toric duality as Seiberg
  duality and brane diamonds}},
  \href{https://doi.org/10.1088/1126-6708/2001/12/035}{\emph{JHEP} {\bfseries
  12} (2001) 035} [\href{https://arxiv.org/abs/hep-th/0109063}{{\ttfamily
  hep-th/0109063}}].

\bibitem{Feng:2002zw}
B.~Feng, S.~Franco, A.~Hanany and Y.-H.~He, \emph{{Symmetries of toric
  duality}}, \href{https://doi.org/10.1088/1126-6708/2002/12/076}{\emph{JHEP}
  {\bfseries 12} (2002) 076}
  [\href{https://arxiv.org/abs/hep-th/0205144}{{\ttfamily hep-th/0205144}}].

\bibitem{Feng:2002fv}
B.~Feng, S.~Franco, A.~Hanany and Y.-H.~He, \emph{{Unhiggsing the del Pezzo}},
  {\emph{JHEP} {\bfseries 08} (2003) 058}
  [\href{https://arxiv.org/abs/hep-th/0209228}{{\ttfamily hep-th/0209228}}].

\bibitem{Hanany:2012hi}
A.~Hanany and R.-K.~Seong, \emph{{Brane Tilings and Reflexive Polygons}},
  \href{https://doi.org/10.1002/prop.201200008}{\emph{Fortsch.Phys.} {\bfseries
  60} (2012) 695} [\href{https://arxiv.org/abs/1201.2614}{{\ttfamily
  1201.2614}}].

\bibitem{2012arXiv1212.1785A}
M.~{Akhtar}, T.~{Coates}, S.~{Galkin} and A.M.~{Kasprzyk}, \emph{{Minkowski
  Polynomials and Mutations}},
  \href{https://doi.org/10.48550/arXiv.1212.1785}{\emph{arXiv e-prints} (2012)
  arXiv:1212.1785} [\href{https://arxiv.org/abs/1212.1785}{{\ttfamily
  1212.1785}}].

\end{thebibliography}\endgroup
